\documentclass[twocolumn,aps,pre,showpacs,showkey,superscriptaddress,floatfix]{revtex4-1}

\usepackage{amsmath,amssymb,amsthm}

\usepackage{hyperref}
\usepackage{amsthm}
\usepackage{graphicx}
\usepackage{subfigure}
\usepackage{color,bbm}
\usepackage[normalem]{ulem}

\newcommand{\bA     }{\mbox{\boldmath$A$}}

\newcommand{\bC     }{\mbox{\boldmath$C$}}
\newcommand{\bJ     }{\mbox{\boldmath$J$}}

\usepackage{color}

\begin{document}
\date{\today}
\title{\bf    Linear stability analysis for large dynamical systems \\ on  directed random graphs }

\author{Izaak Neri} 
\affiliation{Department of Mathematics, King’s College London, Strand, London, WC2R 2LS, UK}
\author{Fernando Lucas Metz}
 \affiliation{Instituto de Física, Universidade Federal do Rio Grande do Sul, Caixa Postal 15051, 91501-970 Porto Alegre, Brazil}
  \affiliation{London Mathematical Laboratory, 14 Buckingham Street, London WC2N 6DF, United Kingdom}
\begin{abstract}      
  We present a linear  stability analysis of   stationary states (or fixed points) in   large dynamical systems   defined on random  directed graphs with a prescribed  distribution of indegrees and outdegrees.         We obtain two remarkable results for such  dynamical systems: First, infinitely large systems on directed graphs can be stable even when the degree distribution has unbounded support; this result is surprising since their counterparts on nondirected graphs, i.e.~dynamical systems defined on nondirected random graphs  with a prescribed degree distribution with unbounded support, are  always unstable when system size is large enough.   Second,   we show that the phase transition between the stable and unstable is universal  in the sense  that it depends only on a  few  parameters, such as,  the mean degree and a degree correlation coefficient.   In addition, in the unstable regime we  characterize the nature of the destabilizing mode, which  also exhibits universal features.     These results follow   from an exact    theory for the leading eigenvalue of infinitely large graphs that are locally tree-like and oriented,  as well as,  the right and left eigenvectors associated with the leading eigenvalue.
We corroborate analytical results for infinitely large graphs    with numerical experiments on random graphs of finite size.        We discuss how the presented theory can be extended to graphs with diagonal disorder and to graphs that contain nondirected links.   Finally, we discuss the influence of   cycles and how they  can destabilize large dynamical systems when they induce  strong   feedback loops.     \end{abstract}
\maketitle

 \section{Introduction}            
 Scientists use  networks to depict   the causal interactions between the constituents of large dynamical systems \cite{barrat2008dynamical, newman2010networks, barthelemy2018spatial, dorogovtsev2013evolution, barabasi2016network}.   Currently, it is not well understood how  the   stability of a  large system is   affected by  the topology  of the underlying   interaction  network.   Relating system stability to network topology is important to understand, among others,   how systemic risk in financial  markets is governed by the topology of the  network of liabilities between financial institutions~\cite{gai2010contagion, haldane2011systemic, bardoscia2017pathways}; how the resilience of an ecosystem to external perturbations depends on the underlying  foodweb of trophic interactions~\cite{mccann2000diversity, dunne2002food, bascompte2009disentangling, bastolla2009architecture, allesina2012stability, coyte2015ecology}; and how networks of social interactions determine the spreading of rumours \cite{moreno2004dynamics, goltsev2012localization, weng2013virality}.        As these examples illustrate, in order to reduce risk and instability in  dynamical systems it is important  to identify  topological properties  of networks that     stabilize large  systems.  
 
In order to study the stability of large dynamical systems, we consider the linearized dynamics of a large,  complex dynamical  system in the vicinity of a stationary state or fixed point.  We model this dynamics with a set of randomly and linearly coupled   differential equations of the form 
   \begin{eqnarray}
\partial_t y_j(t) = \sum^n_{k=1}y_k(t) A_{kj},  \label{eq:linx}
 \end{eqnarray}  
   where  $t\geq 0$ is the time index, $\vec{y}^\dagger(t) = (y_1(t), y_2(t),\ldots, y_n(t))\in \mathbb{R}^n$ is a vector,
   and $\mathbf{A}$ is a random matrix that encodes for the underlying interaction network between the degrees of freedom.     We use the notation $\vec{y}(t)$ for column vectors and $\vec{y}^\dagger(t) = (y_1(t), y_2(t),\ldots, y_n(t))\in \mathbb{R}^n$ for their transpose (or conjugate transpose).      Models like Eq.~(\ref{eq:linx}) appear  when linearizing a set of nonlinearly coupled differential equations in the vicinity of a fixed point \cite{grobman1959homeomorphism, hartman1960lemma, may1972will} as occurs, for example, in the study of neural networks \cite{sompolinsky1988chaos, ahmadian2015properties, aljadeff2016low, kadmon2015transition, amir2016non} and   ecosystems \cite{may1972will, allesina2012stability, gibbs2018effect, haas2019subpopulations}.   The vector $\vec{y}$ describes then the deviation of the system from its fixed point, which is located at the origin, i.e., $\vec{y}=0$.        We will use the generic model, given by Eq.~(\ref{eq:linx}), to study how network topology affects the stability of stationary states.   
  
Since  we aim to develop a better understanding on how network topology affects system stability, we write Eq.~(\ref{eq:linx}) as
\begin{eqnarray}
\partial_t y_j(t) = \sum^n_{k=1; (k\neq j)}  y_k(t) \:C_{kj} J_{kj}    - d\: y_j(t) , \label{eq:linT}
\end{eqnarray}  
where  $d>0$ represents the  rate at which an isolated node relaxes to the stationary state, where  $C_{kj}\in\left\{0,1\right\}$ are the entries of the adjacency matrix of a directed graph, and  where $J_{kj}\in \mathbb{R}$ are the strengths of the couplings  between two nodes $k$ and $j$.      Note that 
Eq.~(\ref{eq:linT}) and Eq.~(\ref{eq:linx}) are related by 
\begin{eqnarray}
\mathbf{A} =-d \mathbf{1}_n +  \mathbf{J} \circ \mathbf{C}, \label{eq:model}
\end{eqnarray}
where 
\begin{eqnarray}
[\bJ\circ\bC]_{jk} = J_{jk}C_{jk} 
\end{eqnarray} 
and  $\mathbf{1}_n$ is the identity matrix.     The entries of the adjacency matrix $\mathbf{C}$ determine who interacts with whom, while the entries of the interaction matrix $\mathbf{J}$ denote the absolute strength of the interactions and whether these are inhibitory $J_{kj}<0$ or excitatory $J_{kj}>0$. 

 In absence of interactions between system constituents ($J_{jk}=0$), the fixed point  $\vec{y}=0$ is stable as $d>0$ . However, if the   constituents of the system interact strong enough, then a small perturbation  around the fixed point $\vec{y}=0$    will propagate through the network, grow in size, and   destabilize the system.      It is the underlying network topology, represented by the adjacency matrix $\mathbf{C}$, and the strength of the interactions, given by $\mathbf{J}$, that determine whether an initial perturbation will grow or  fade away.    For example, in an online social network,  the topology of the network  determines whether a rumour  spreads throughout the whole system or  only reaches  a couple of users.

The stability of the  fixed point  $\vec{y}=0$ in the dynamics given by Eq.~(\ref{eq:linx}) is governed by the sign of the real part of the {\it leading eigenvalue}  $\lambda_1(\mathbf{A})$, which is the eigenvalue with the largest real part.   As discussed in Appendix~\ref{AppNonDiag}, 
       if ${\rm Re}[\lambda_1(\mathbf{A})]>0$, then the fixed point is unstable and $\lim_{t\rightarrow \infty}|\vec{y}(t)| = \infty$.  Conversely, if ${\rm Re}[\lambda_1(\mathbf{A})]<0$, then the fixed point  is stable and  $\lim_{t\rightarrow \infty}|\vec{y}(t)| = 0$.      In addition, the left eigenvector associated with the leading eigenvalue determines the nature of the destabilizing mode.

To model dynamical systems on large networks, we consider that 
 $\mathbf{C}$ is the adjacency matrix of a   random  directed graph with a  prescribed degree distribution  $p_{K^{\rm in}, K^{\rm out}}(k, \ell)$ of indegrees $K^{\rm in}$ and outdegrees $K^{\rm out}$.    This  is  a  paradigmatic model for networked systems,  such as, the   World Wide Web  \cite{broder2000graph, pastor2007evolution},  neural networks \cite{brunel2000dynamics, amari2003handbook, sporns2010networks},  foodwebs \cite{dunne2002food}, and  online social networks \cite{mislove2007measurement, myers2014information}, and it is often called the configuration model~\cite{Molloy95,Molloy98,bollobas2001random, newman2010networks, dorogovtsev2013evolution} or the  uniform model \cite{dembo2010gibbs}.       The percolation properties of  random   directed graphs have been well understood, see Refs.~\cite{newman2001random, dorogovtsev2001giant, timar2017mapping}, and recently also spectral properties of random directed graphs have been thoroughly studied, see Refs.~\cite{rogers2009cavity, allesina2015predicting, neri2016eigenvalue, metz2018spectra, PhysRevResearch.2.023333}, but the properties of the leading eigenvalues of the adjacency matrices of  random directed graphs have not been studied so far.

In this paper,  we perform a linear stability analysis of fixed points in dynamical systems defined on  random  directed graphs.    To this aim,  we present a detailed analysis of the leading eigenvalue $\lambda_1(\mathbf{A})$ of the the adjacency matrices $\mathbf{A}$ of      random  directed graphs with a prescribed degree distribution and with randomly weighted links.      First,  building on Ref.~\cite{neri2016eigenvalue}, we derive exact analytical expressions for the typical value of $\lambda_1$ in the limit of infinitely large $n$.   In addition, we derive in this limit  exact expressions for the statistics of the entries of   right and  left eigenvectors associated with $\lambda_1$.    Second,  we use 
  these results  to  depict a phase diagram for the linear stability of fixed points in  dynamical systems defined on large  directed networks.     Third, the theoretical  results for infinitely large graphs are  compared with  numerical results for graphs of finite size, which include    random graphs   with power-law degree distributions.
   
Two implications of these results are surprising enough that they deserve further emphasis.    First, we  find that   dynamical systems on infinitely large,     random, and  directed graphs can be {\it stable}, even when the degree distribution has unbounded support.   This result is surprising because  dynamical systems on  random  nondirected   graphs with a degree distribution that has unbounded support are  unstable  if the system size is large enough.  Indeed,  the leading eigenvalue of an nondirected random graphs scales as  $\lambda_1 \sim \sqrt{k_{\rm max}}$ \cite{krivelevich2003largest, chung2004spectra, susca2019top}, where $k_{\rm max}$ is the expected largest  degree of the graph, and therefore the leading eigenvalue  of an nondirected graph diverges for large $n$. In contrast,  in this paper we obtain that the leading eigenvalue of a   random directed graph with a prescribed degree distribution is in general finite for $n\rightarrow \infty$, even when $k_{\rm max}$ diverges.     Hence, models on random directed graphs are significantly more stable than their counterparts on random nondirected graphs.

 Second, we obtain  a {\it universal} phase diagram for the stability of networked systems on  random directed graphs with a prescribed degree distribution.     Put in another way, we show that the leading eigenvalue of these random graphs only depends on a few system parameters, including the mean degree and a parameter that characterizes the correlations between indegrees and outdegrees.   
 
Both the {\it stability} and {\it universality} of dynamical systems defined on random directed graphs are rooted in a common fact: for large enough $n$, the local neighborhood of a randomly selected node is with probability one   a tree graph that contains only unidirectional links.   We call this the {\it locally tree-like and oriented property}.    Using the property, we derive a set of  recursion relations for the components of   right  and left eigenvectors associated with the leading eigenvalue.    These recursion relations have first been derived in Ref.~\cite{neri2016eigenvalue} using the cavity method  \cite{rogers2008cavity, rogers2009cavity, rogers2010spectral, metz2011spectra, bolle2013spectra, metz2018spectra}, a method borrowed from the statistical physics of spin glasses \cite{mezard2001bethe, mezard2003cavity}.    In the present paper, we  present an alternative derivation of the recursion relations based on the {\it Schur} formula \cite{bordenave2010resolvent},  which we believe is simpler to understand and thus more insightful.

The outline of the paper is the following.  In  Sec.~\ref{eq:modelDef}, we define the random matrices  and  spectral quantities we  study in this paper.     In Sec.~\ref{eq:theory}, we present  an overview of  the theoretical results  derived in this paper.
 In Sec.~\ref{sec:app}, we apply these theoretical results to a linear stability  analysis of stationary states in networked systems.   
   In Sec.~\ref{sec:examples},   we compare   theoretical results for infinitely large matrices with numerical  data for   matrices of finite size.   In Sec.~\ref{sec:ext}, we discuss   extensions of the theory presented in  Sec.~\ref{eq:theory}  to the cases of adjacency matrices with diagonal disorder and adjacency matrices of random graphs that contain  nondirected links.  Lastly,  in Sec.~\ref{sec:discu}, we present a discussion of the main results.    A detailed description of   mathematical derivations  are presented in the appendices.   In Appendix~\ref{AppNonDiag},  we show that a linear set of  randomly coupled differential equations, of the form given by Eq.~(\ref{eq:linx}), is stable if and only if all the eigenvalues of $\mathbf{A}$ are negative.
 Appendix~\ref{sec:finite} details the algorithm we use to generate graphs with a prescribed degree distribution, and in Appendix~\ref{OrientedRing}, we discuss properties of oriented ring graphs.    In Appendix~\ref{App:specConn}, we show that the algebraic multiplicity of the zero eigenvalue of a  random directed  graph is related to the size of its strongly connected component.    Lastly, in Appendices~\ref{sec:der}-\ref{sec:der2}, we  derive  recursion relations for the entries of right and left eigenvectors of random and directed graphs, which are  based on the Schur formula.     
 
\subsection{Notation}
We use lower case symbols for deterministic variables, e.g., $x$ and $y$.  We  write (column) vectors as $\vec{x}$ and $\vec{y}$, while for adjoint row vectors we write  $\vec{x}^\dagger$ and $\vec{y}^\dagger$.     The inproduct $\vec{x}\cdot \vec{y} = \vec{x}^\dagger\vec{y} =  \sum^{n}_{k=1}x^\ast_ky_k$, where $x^\ast_k$ is the complex conjugate of $x_k$.     
  Matrices are written in boldface, e.g., $\mathbf{x}$ and $\mathbf{y}$.  
We write random variables in upper case, e.g., $X$ and $Y$. The probability distribution of a random variable $X$ is denoted by $p_X(x)$.
There are a few exceptions to the use of upper case letters to represent random quantities. For example, we use the  notation $\lambda_j(\mathbf{A})$ to denote the $j$-th eigenvalue of a random matrix $\mathbf{A}$, and we write   $p_X(x;\mathbf{A})$ for the probability distribution of a random variable $X$ that depends on the matrix $\mathbf{A}$. We denote averages with respect to the distribution $p_{\mathbf{A}}(\mathbf{a})$ by $\langle \cdot\rangle$.   We denote the identity matrix by $\mathbf{1}_n$ and  we use $\left\{1,2,\ldots,n\right\} =  [n]$.   We write $\int_{\mathbb{R}} {\rm d}x f(x)$ for an integral over the real line and  $\int_{\mathbb{C}} {\rm d}^2z f(z) = \int {\rm d}x{\rm d}y f(x+{\rm i}y)$ for an integral over the complex plane.   We denote the Dirac distribution over the real line by  $\delta(x)$ and we denote the Dirac distribution over the complex plane by $\delta(z) = \delta(x)\delta(y)$, where $z=x+{\rm i}y\in\mathbb{C}$.

 \section{System setup and definitions}   \label{eq:modelDef}
 In this section, we define the  random matrices   and the spectral properties    we study in this paper.   

\subsection{Adjacency matrices of random directed graphs with a prescribed degree distribution}\label{sec:modeldef} 
We consider random matrices $\mathbf{A}$, as defined by Eq.~(\ref{eq:model}), where $\bJ$ is a square  matrix of size $n$ with real entries $J_{jk}\in \mathbb{R}$ that are i.i.d.~random variables drawn from a distribution $p_J$, and  where $\bC$ is the adjacency matrix of a random and directed graph $\mathcal{G}$ of size  $n$ with a prescribed degree distribution  $p_{K^{\rm in}, K^{\rm out}}(k, \ell)$  of indegrees $K^{\rm in}$ and outdegrees $K^{\rm out}$ \cite{bollobas2001random, newman2010networks, dorogovtsev2013evolution};  note that we call the number of vertices of a graph its size and not the number of links.  
 
For a simple graph $\mathcal{G}$ the entries of the adjacency matrix satisfy $C_{jk}\in\left\{0,1\right\}$ and $C_{jj} = 0$. 
We use the convention that  $C_{jk}=1$ if the graph $\mathcal{G}$ has a directed edge  from node $j$ to node $k$. Therefore, the
indegree $K^{\rm in}_j$ of the $j$-th node equals the number of nonzero elements in the $j$-th column of $\bC$, 
\begin{eqnarray}
K^{\rm in}_j := \sum^n_{k=1}C_{kj},
\end{eqnarray}
and the outdegree $K^{\rm out}_j$  of the $j$-th node equals the number of nonzero elements in the $j$-th row,
\begin{eqnarray}
 K^{\rm out}_j := \sum^n_{k=1}C_{jk}.
\end{eqnarray}   
The inneighborhood $\partial^{\rm in}_j$ and the outneighbourhod $\partial^{\rm out}_j$ of the   $j$-th node are defined by
\begin{eqnarray}
  \partial^{\rm in}_j :=  \left\{k\in[n]:  C_{kj}=1\right\}  \label{eq:pin} 
  \end{eqnarray} 
and
\begin{eqnarray}
  \quad \partial^{\rm out}_j :=  \left\{k\in[n]:  C_{jk}=1\right\} ,\label{eq:pout}
\end{eqnarray} 
respectively, 
and 
\begin{eqnarray}
\partial_j := \partial^{\rm in}_j  \cup \partial^{\rm out}_j \label{eq:neighbour}
\end{eqnarray}
is the neighborhood of node $j$.

We say that  $\mathcal{G}$ is a random graph with  a  prescribed degree distribution $p_{K^{\rm in}, K^{\rm out}}(k, \ell)$ if the following properties hold:~(i) the degrees  $(K^{\rm in}_j,  K^{\rm out}_j)$  are i.i.d.~random variables with a joint probability distribution $p_{K^{\rm in}, K^{\rm out}}(k, \ell)$ and with the additional constraint $\sum^{n}_{j=1}K^{\rm in}_j = \sum^n_{j=1}K^{\rm out}_j$; (ii) given a certain degree sequence $\left\{K^{\rm in}_j,  K^{\rm out}_j\right\}^n_{j=1}$, the nodes are connected randomly and hence the edges of $\mathcal{G}$ are generated  by the configuration model  \cite{bollobas2001random, newman2010networks, dorogovtsev2013evolution}.   In the Appendix~\ref{sec:finite}, we describe in  detail the   algorithm we use to sample random graphs with a prescribed degree distribution.     

 In the specific case when $J_{jk} = 1$ and $d=0$, $\mathbf{A}$ is the  adjacency matrix of a  random  directed   graph.  
  The variables $J_{jk}$ are the weights associated with the  links of the graph represented by the adjacency matrix $\bC$, and hence for $J_{jk} \neq 1$  the  matrix $\mathbf{A}$  is the adjacency matrix of a weighted graph.     The constant parameter $d$ affects the spectral properties of $\mathbf{A}$   in a trivial manner, but plays an important role in a stability analysis of dynamical systems.

\subsection{Ensemble parameters}\label{def:EnsembleParam}
The random matrix ensemble, defined by Eq.~(\ref{eq:model}), depends on the following parameters: the distribution $p_J$ of weights $J_{ij}$,   the joint  distribution $p_{K^{\rm in}, K^{\rm out}}$ of indegrees and outdegrees,  the real number $d$, and the size~$n$.   

 We  often use the  moments of  $p_J$ and $p_{K^{\rm in}, K^{\rm out}}$ to specify the model of interest. 
  The $m$-th  moment of $p_J$ is defined by 
  \begin{eqnarray}
\langle J^m\rangle := \int^{\infty}_{-\infty}{\rm d}x \:x^m\: p_J(x),
\end{eqnarray}
and the $(m,o)$-th moment of $p_{K^{\rm in}, K^{\rm out}}$ is given by
  \begin{eqnarray}
\langle \left(K^{\rm in}\right)^m \left(K^{\rm out}\right)^o \rangle  := \sum^{\infty}_{k=0}\sum^{\infty}_{\ell=0}p_{K^{\rm in}, K^{\rm out}}\left(k,\ell \right)k^m \ell^o .\nonumber\\
  \end{eqnarray}
 Among those, important parameters are the {\it mean degree} 
  \begin{eqnarray}
c := \langle K^{\rm in}\rangle =  \langle K^{\rm out}\rangle  \label{eq:meanDef}
\end{eqnarray}    
and the {\it degree correlation coefficient}  
\begin{eqnarray}
\rho :=  \frac{\langle K^{\rm in}K^{\rm out}\rangle - c^2 }{c^2} . \label{eq:assort}
\end{eqnarray}

The mean degree is equal to the average number of edges that enter or leave a uniformly and randomly selected  vertex in the graph.   
The parameter $c \langle J \rangle $ is  the average interaction  strength felt by a degree of freedom in the dynamical system governed by Eq.~(\ref{eq:linT}).
The degree correlation coefficient $\rho$ characterizes the correlations between indegrees and outdegrees of vertices in the graph.    If $\langle K^{\rm in}_jK^{\rm out}_j\rangle = \langle K^{\rm in}_j\rangle \langle K^{\rm out}_j\rangle$, then $\rho = 0$, which means that indegrees and outdegrees  are uncorrelated. If  $\rho>0$ ($\rho<0$), then  indegrees and outdegrees are positively (negatively) correlated.

\begin{figure}[t]
\centering
{\includegraphics[width=0.4\textwidth]{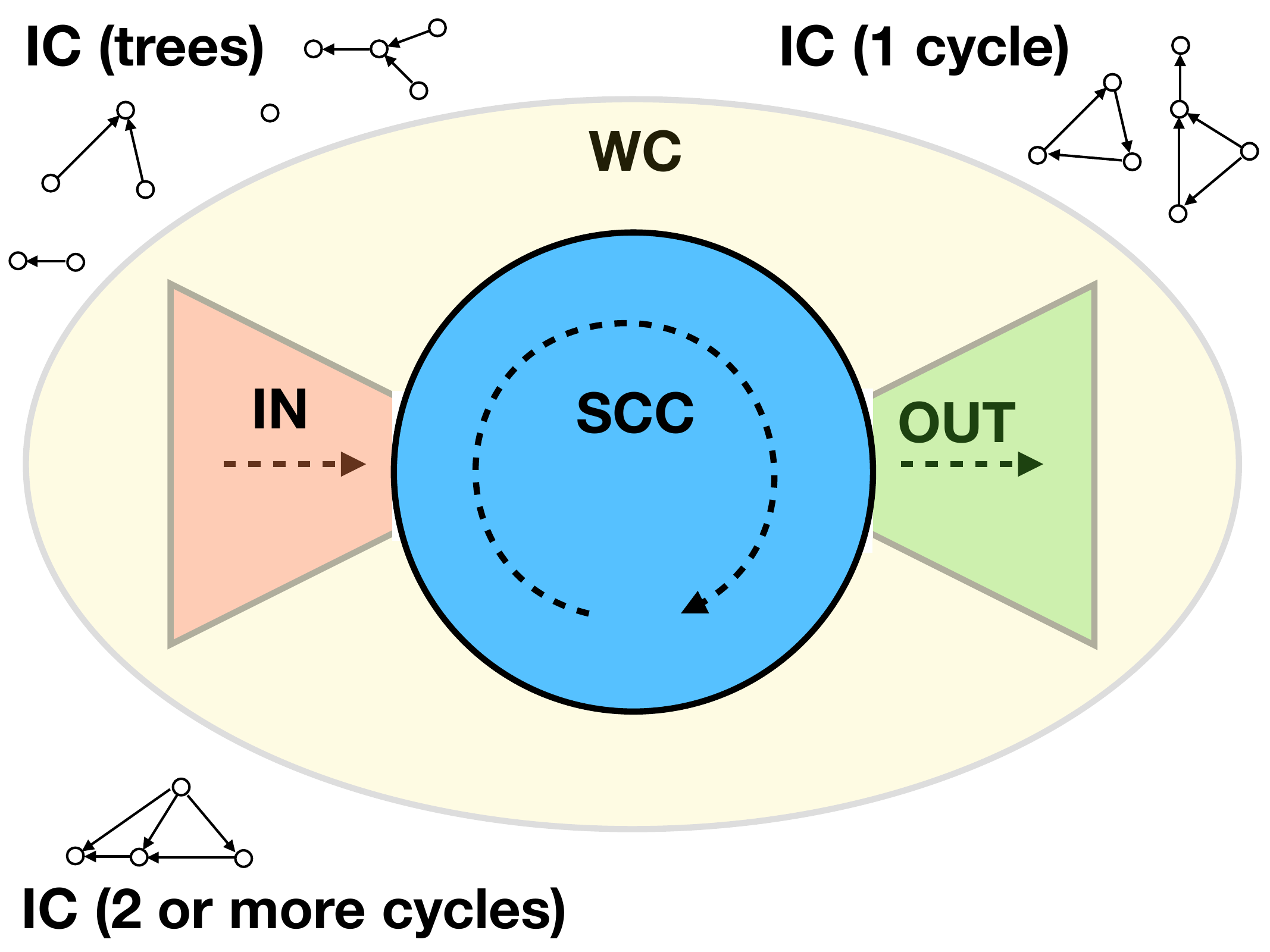}} 
\caption{{\it Topology of directed graphs.} Graphical illustration of  the connected components of directed graphs   (bow-tie diagram, see also Refs.~\cite{broder2000graph, dorogovtsev2001giant, timar2017mapping}): largest strongly connected component (SCC), largest incomponent (IN), largest outcomponent (OUT), largest weakly connected component (WC), and isolated components (IC), which consist of isolated trees and cycles.     
    } \label{fig:bowtie}
\end{figure}

  \subsection{Topology of  directed graphs}\label{Sec:ConComp}  
  We discuss   properties of the topology of random, directed graphs with a prescribed degree distribution that will be relevant to understand  their spectra, namely,   connected components,    percolation transitions,  the  locally tree-like and oriented structure, and   oriented rings.  
  
  \subsubsection{Connected components of directed graphs}\label{sec:concomp}
Connected components are subgraphs   that characterize the topology of a directed graph.   In particular, the connected components determine which  nodes in the graph are affected by a local perturbation.   

The  topology of a directed graph can be depicted with  a bow-tie diagram, see Fig.~\ref{fig:bowtie} and Refs.~\cite{broder2000graph, dorogovtsev2001giant, timar2017mapping,Kryven2016}.         The bow-tie diagram depicts the following subgraphs of a directed graph: the largest strongly connected component (SCC), the  incomponent (IN), and the  outcomponent (OUT).   Besides these three components, directed graphs also have a largest weakly connected component (WC) and isolated components (IC), also depicted in Fig.~\ref{fig:bowtie}.   Finally, directed graphs  contain  tendrils~\cite{dorogovtsev2001giant, timar2017mapping}.   Since tendrils play a minor role in   the spectral properties of directed graphs,  we  omit them in Fig.~\ref{fig:bowtie}.     

We  present definitions of the abovementioned subgraphs.  
 The SCC  is the largest subgraph that is strongly connected.   A subgraph is {\it strongly connected} if   for each pair of vertices in the subgraph, say $j$ and $k$, the following two conditions are met:  (a) there exist at least one path starting in $j$ and ending in $k$   (b) there exist  at least one path starting in $k$ and ending in $j$.      The  IN consists of all nodes  that can reach the strongly connected component  and the OUT consist of all nodes that can be reached from the strongly connected component (by following the edges of the directed graph).   The WC is the largest connected component obtained by ignoring the directionality of edges.  The tendrils consist of all vertices that belong to the weakly connected component, but do not belong to the incomponent and outcomponent.    Finally, the IC are  connected subgraphs that are disconnected from the  largest weakly connected component.

  \subsubsection{Size  of the connected components of random directed graphs with a prescribed degree distribution}\label{sec:sizeConn}
   For random directed graphs with a prescribed degree distribution $p_{K^{\rm in}, K^{\rm out}}(k, \ell)$, the relative sizes of the connected components  are deterministic  in the limit of large $n$.    We denote the limiting value of the  relative size of the SCC by $s_{\rm sc}$ (i.e., the fraction of nodes that belong to the SCC), and analogously, we use  $s_{\rm in}$,  $s_{\rm out}$,  $s_{\rm wc}$, $s_{\rm t}$  and $s_{\rm ic}$, for the limiting values of the  relative sizes of the incomponent, outcomponent, largest weakly connected component, tendrils, and isolated components, respectively.

   We say that a  random graph has a giant SCC when  $s_{\rm sc}>0$  and, analogously, we say  that a random graph has a giant IN, OUT, or WC when, respectively, $s_{\rm in}>0$,  $s_{\rm out}>0$,  or $s_{\rm wc}>0$.    
 
  For small enough values of $c(\rho+1)$, it holds that $s_{\rm sc}=0$ and $s_{\rm wc} = 0$,  whereas for large enough values of    $c(\rho+1)$,   it holds that $s_{\rm sc}>0$ and $s_{\rm wc} > 0$.    The  percolation transitions associated with a giant SCC and a giant WC  take place at the threshold values of  $c(\rho+1)$ for which the quantities $s_{\rm sc}$ and $s_{\rm wc}$  vanish, respectively.  Since by definition $s_{\rm in}\geq s_{\rm sc}$ and $s_{\rm out}\geq s_{\rm sc}$, and $s_{\rm in} = s_{\rm out}=0$ if $s_{\rm sc}=0$,  the percolation transition associated with the IN and OUT is identical to the percolation transition associated with the SCC.    Hence, in directed graphs there exist two percolation transitions, namely a transition associated with the SCC and one associated with the WC.

In    Ref.~\cite{dorogovtsev2001giant}, an exact set of  equations have been derived for the relative sizes of the various connected components in directed graphs.   It was found that
 \begin{eqnarray}
s_{\rm out} &=& 1- \sum^{\infty}_{k=0}a^{k}\sum^{\infty}_{\ell=0} p_{K^{\rm in},K^{\rm out}}(k,\ell), \label{eq:sin} 
\end{eqnarray}   
and 
\begin{eqnarray}
s_{\rm in} &=& 1- \sum^{\infty}_{\ell=0} b^{\ell} \sum^{\infty}_{k=0} p_{K^{\rm in},K^{\rm out}}(k,\ell), \label{eq:sout}
\end{eqnarray}   
 where $a$ and $b$ are the smallest nonnegative solutions to the equations 
\begin{eqnarray}
a &=& \sum^{\infty}_{k=0}a^{k}\sum^{\infty}_{\ell=0}\frac{\ell \: p_{K^{\rm in},K^{\rm out}}(k,\ell)}{c},  \label{eq:a} 
\end{eqnarray}   
and
\begin{eqnarray}
b &=&\sum^{\infty}_{\ell=0} b^{\ell} \sum^{\infty}_{k=0}\frac{k \: p_{K^{\rm in},K^{\rm out}}(k,\ell)}{c}. \label{eq:bb}
\end{eqnarray} 
The size of the  SCC  is given by 
\begin{eqnarray}
s_{\rm sc} =   s_{\rm in} +s_{\rm out}  + s_{\rm t} - s_{\rm wc} ,  \label{eq:sc}
\end{eqnarray}
where 
  \begin{eqnarray}
s_{\rm t} - s_{\rm wc} =   \sum^{\infty}_{k=0}\sum^{\infty}_{\ell=0}p_{K^{\rm in},K^{\rm out}}(k,\ell)\:a^{k} b^{\ell} -1 .
\end{eqnarray}

The  percolation transition of the SCC happens when $s_{\rm sc}$ turns positive, which happens when
\begin{eqnarray}
\sum^{\infty}_{k=0}\sum^{\infty}_{\ell=0}\frac{k\:\ell \: p_{K^{\rm in},K^{\rm out}}(k,\ell)}{c} =1. \label{eq:percCond}
\end{eqnarray}  
Using in  Eq.~(\ref{eq:percCond}) the definitions (\ref{eq:meanDef}) and (\ref{eq:assort}) for, respectively,  the mean degree $c$ and the degree correlation coefficient $\rho$, we obtain  that at  the critical connectivity
 \begin{eqnarray}
c= \frac{1}{1+\rho} \label{eq:stronglyperco}
\end{eqnarray}
a  giant SCC  emerges in a directed random graph with a prescribed degree distribution.

Equation (\ref{eq:stronglyperco}) implies that random graphs with positively correlated indegrees and outdegrees percolate at lower connectivities than random graphs with negatively correlated indegrees and outdegrees.

  \subsubsection{Oriented and locally tree-like structure} 
  If the mean degree $c$ is finite, then random graphs with a prescribed degree distribution are locally tree-like and oriented.   
 This means that for large enough $n$, the finite neighborhood of a randomly selected node is with probability one  an oriented tree \cite{aldous2004objective}.  We say that a graph is a tree if it is connected and does not contain a cycle and we say that a graph is oriented if all its edges are unidirectional, i.e.,  $C_{ij}C_{ji} = 0$ for each pair $(i,j)$.   For a precise mathematical definition of locally tree-like graphs, we refer to the section 2.1 of Ref.~\cite{dembo2010gibbs}.

  \subsubsection{Oriented rings} \label{Sec:orientedRing}
  Since random  directed graphs with a prescribed degree distribution are locally tree-like,  one may think that cycles of finite length are not important to describe their spectral properties in  the limit of large $n$.      However, this is only partly true since in the limit $n\rightarrow \infty$ there nevertheless  exists a finite number of cycles of finite length $\ell$, and these cycles may affect the value of the leading eigenvalue. 
     
We focus on subgraphs that are oriented rings since  only their contribution matters to the spectrum of $\mathbf{A}$.
 An oriented ring of length $\ell$ is an $\ell$-tuple of nodes $i_1,i_2,\ldots,i_{\ell}$  for which 
\begin{eqnarray}
A_{i_1i_2}A_{i_2i_3}\ldots A_{i_{\ell-1}i_{\ell}}A_{i_{\ell}i_1} \neq 0. \label{eq:orientedRingDef}
\end{eqnarray}
In the limit $n \rightarrow \infty$, the average number of oriented rings of length $\ell$ in a random  directed graph with a prescribed degree distribution is given by (see Appendix~\ref{OrientedRing}) 
\begin{eqnarray}
\langle N(\ell)\rangle   = \frac{1}{\ell}[c (\rho +1)]^{\ell}, \label{eq:cyclesrxx}
\end{eqnarray}  
and the total number of oriented rings of finite length reads
 \begin{eqnarray}
\langle N \rangle = \sum^{\infty}_{\ell=2}\langle N(\ell)\rangle  = -\ln[1-c(\rho+1)]- c(\rho+1) .  \nonumber
\end{eqnarray} 
 Note that $\langle N \rangle$ diverges for $c(\rho+1)\rightarrow 1$.
 
The distribution of $N(\ell)$ is Poissonian with mean  $\langle N(\ell)\rangle$ \cite{wormald1981asymptotic}, and therefore the probability $p_+$ that there exists at least one oriented ring of length $\ell \geq 2$ is given by 
 \begin{eqnarray}
 p_+ =1- e^{-\langle N \rangle } =1- [1-c(\rho+1)]e^{c(\rho+1)}.\label{eq:p+}
 \end{eqnarray}
Note that $p_+ \rightarrow 1$ when  $c (\rho +1)\rightarrow 1$ and $p_+ \rightarrow 0$ when $c (\rho +1)\rightarrow 0$.

 \subsection{Spectral observables}\label{specobs} 
 
 \subsubsection{Finite matrices}
  The eigenvalues  $\left\{\lambda_\alpha(\bA)\right\}_{\alpha \in [n]}$  are the complex roots of  the algebraic equation  \cite{horn1985}
  \begin{equation}
  {\rm det}(\bA - \lambda \mathbf{1}_n) = 0 .
  \end{equation}    
      We sort the eigenvalues in decreasing order, so that   
        \begin{equation}
      {\rm Re}[\lambda_1(\bA )]\geq {\rm Re}[\lambda_2(\bA )]\ldots \geq {\rm Re}[\lambda_n(\bA )].
        \end{equation}    
       If an eigenvalue is degenerate, then it appears more than once in the sequence.  
We call  $\lambda_{1}$  the {\it leading} eigenvalue of $\mathbf{A}$  and 
 $\lambda_2$   the  {\it subleading} eigenvalue.

      A right eigenvector  $\vec{R}(\bA)$ and a  left eigenvector  $\vec{L}(\bA)$ associated with  an eigenvalue $\lambda_{\alpha}$ are nonzero vectors that fulfil   
 \begin{eqnarray}
 \bA\, \vec{R} = \lambda_{\alpha}\,  \vec{R},\quad {\rm and} \quad  \vec{L}^\dagger\, \bA  = \lambda_{\alpha}\, \vec{L}^\dagger. \label{eq:eigvDef}
 \end{eqnarray}   
We use the notation $R_{j}$ and $L_{j}$ for the components or entries of the 
right and left eigenvectors,  respectively, where $j\in[n]$.            

The number $m$ of linearly independent right eigenvectors (or left eigenvectors) is smaller or equal than the size of the matrix and  greater or equal than the number of  eigenvalues of $\mathbf{A}$.     If $m=n$, then the matrix is diagonalizable.   

Right and left eigenvectors of   $\mathbf{A}$ can be chosen   biorthonormal, 
  \begin{eqnarray}
 \vec{L}_{\beta}\cdot \vec{R}_{\alpha} = \delta_{\alpha\beta},     \label{eq:norm}
 \end{eqnarray}     
 where   $\alpha,\beta \in  [m]$ is a label to identify  the $m$ linearly independent  right (left) eigenvectors.   
 Biorthonormality is not sufficient to uniquely characterize right and left eigenvectors since  they
 can be rescaled as $c_{\alpha}\vec{R}_{\alpha}$ and $c^{-1}_{\alpha}\vec{L}_{\alpha}$, with $c_{\alpha}\in\mathbb{C}$.       
  In order to uniquely define the right and left eigenvectors,  
  we take the convention that  
     \begin{eqnarray}
{\rm Im}\left[\sum^n_{j=1}R_{\alpha,j} \right] = 0,\quad  {\rm Re}\left[\sum^n_{j=1}R_{\alpha,j} \right] \geq 0, \label{eq:convR}
 \end{eqnarray} 
 and  we set 
      \begin{eqnarray}
\sum^n_{j=1}|R_{\alpha,j}|^2 = n.   \label{eq:convR2}
   \end{eqnarray} 
   The relation (\ref{eq:convR}) specifies the  argument  of $c_{\alpha}$ and the relation (\ref{eq:convR2}) specifies its  norm.  
When using the conventions (\ref{eq:norm})-(\ref{eq:convR2}),  the   norm $\sum^n_{j=1}|L_{\alpha,j}|^2$ and the  argument of  $\sum^n_{j=1}L_{\alpha,j} $  are functions of the entries of~$\bA$.

   \subsubsection{Infinitely large matrices}
In order to characterize properties of random matrices in the limit of $n\rightarrow \infty$, we use sets and distributions. The spectrum of $\mathbf{A}$ is the  set
  \begin{eqnarray}
  \sigma(\bA) := \left\{\lambda \in \mathbb{C}:{\rm det}\left(\bA - \lambda \mathbf{1}_n\right) = 0\right\}
  \end{eqnarray} 
of  eigenvalues of  $\bA$. For finite  $n$, $\sigma(\bA)$ is discrete.   For large $n,$ the closure  of the spectrum  $\sigma(\bA)$  converges to the limit
    \begin{eqnarray}
\overline{ \lim_{n\rightarrow \infty}\sigma(\bA)} =   \sigma  \cup \Gamma, \label{eq:lim}
    \end{eqnarray} 
    where $\sigma$ is a deterministic set and $\Gamma$ is a random set. The deterministic spectrum 
    \begin{equation}  
    \sigma = \sigma_{\rm c}\cup \sigma_{\rm d}
    \end{equation}
       consists of  a continuous part $\sigma_{\rm c}$ and a discrete part $\sigma_{\rm d}$. 
       
       The continuous part 
       \begin{equation}
       \sigma_{\rm c} = \sigma_{\rm sc}\cup \sigma_{\rm ac}
           \end{equation}
consists of  a set $\sigma_{\rm ac}$ of nonzero Lebesgue measure,  which we call the absolutely continuous part, and a set 
    $\sigma_{\rm sc}$ of zero Lebesgue measure, which we call the singular continuous part.     We will be interested in the boundary $\partial \sigma_{\rm ac}$ of the set $\sigma_{\rm ac}$ and use the notation 
    \begin{equation}
    \lambda_{\rm b} \in \partial \sigma_{\rm ac}
        \end{equation}
     for eigenvalues located at the boundary of  $\sigma_{\rm ac}$.  
    
    The discrete part of the spectrum consists  of   deterministic {\it outlier eigenvalues}, which we denote by $\lambda_{\rm isol}$.  
    We say that  $\lambda_{\rm isol}\in \sigma$ is an  outlier eigenvalue --- sometimes also called an  isolated eigenvalue --- if there exists an $\epsilon>0$, such that,   
   \begin{equation}
   \sigma  \cap  \left\{\lambda \in \mathbb{C}: |\lambda_{\rm isol} - \lambda|<\epsilon\right\}= \left\{\lambda_{\rm isol}\right\}.
   \end{equation}  
       In the examples considered in this paper, there will be maximal one deterministic outlier eigenvalue.

Lastly, the limiting spectrum in Eq.~(\ref{eq:lim}) may contain a random set $\Gamma$ that consists of stochastic (outlier) eigenvalues.

The spectral distribution 
\begin{eqnarray}
\mu(\lambda;\mathbf{A}) =  \frac{1}{n}\sum^n_{\alpha=1} \delta\left(\lambda-\lambda_\alpha\left(\mathbf{A}\right)\right)
\end{eqnarray}  
  denotes the relative number of eigenvalues that occupy a certain region of the complex plane, and we denote its asymptotic expression by 
  \begin{eqnarray}
    \mu(\lambda) = \lim_{n\rightarrow \infty}\mu(\lambda;\mathbf{A}). \label{eq:mu}
  \end{eqnarray} 
The support of the distribution is the closure of  the set $\left\{\lambda\in\mathbb{C}: \mu(\lambda)\neq 0\right\}$.  Since in general $\mu(\lambda_{\rm isol}) = 0$, the outliers do not belong to the support of $\mu$, and therefore the support of $\mu$ is a subset of $\sigma$.

 We are also  interested  in the statistics of the components of right and left eigenvectors.   Let $\vec{R}$ ($\vec{L}$) be the right  (left) eigenvector associated with an eigenvalue $\lambda$.   We define the random variable $R$  ($L$) as a uniformly randomly sampled entry of the eigenvector.       If $R$ and $L$ refer to an outlier, then we use the notation $R_{\rm isol}$ and $L_{\rm isol}$; if $R$ and $L$ refer to an eigenvalue located at the boundary of  $\sigma_{\rm ac}$, then we use  $R_{\rm b}$ and $L_{\rm b}$.

The distributions of  the random variables $R$ and $L$ are defined by
 \begin{eqnarray}
p_{R}(r|\mathbf{A}) &= \frac{1}{n}\sum^n_{i=1}\delta(r-R_{ i})
 \label{eq:DefEigv}
\end{eqnarray} 
and 
 \begin{eqnarray}
p_{L}(l|\mathbf{A}) &= \frac{1}{n}\sum^n_{i=1}\delta(l-L_{ i}), \label{eq:DefEigv2}
\end{eqnarray} 
respectively,
where $\delta(z)$ is the Dirac-delta distribution in the complex plane.   In the limit $n\rightarrow \infty$,
the distributions $p_{R}(r|\mathbf{A})$ and  $p_{L}(l|\mathbf{A})$ often  converge to deterministic limits 
\begin{eqnarray}
p_R(r) = \lim_{n\rightarrow \infty}p_{R}(r|\mathbf{A}),\quad  p_L(l) = \lim_{n\rightarrow \infty} p_{L}(l|\mathbf{A}). \label{eq:asymptpR}
\end{eqnarray}
We  denote the moments of the limiting distributions $p_{R}(r)$ and $p_{L}(l)$ by  
\begin{eqnarray}
\langle R^m \rangle  = \int {\rm d}^2r\: p_{R}(r) r^m , \quad {\rm and}  \quad \langle L^m\rangle  = \int {\rm d}^2l\: p_{L}(r) l^m,  \nonumber\\
\end{eqnarray}
where ${\rm d}^2r = {\rm d}{\rm Re}(r) {\rm d}{\rm Im}(r)$ and ${\rm d}^2l = {\rm d}{\rm Re}(l) {\rm d}{\rm Im}(l)$.

We say that a spectral quantity of a random directed graph  is {\it universal} if  it converges for $n \rightarrow \infty$ to a deterministic limit that only depends on the first  few moments of  the distributions $p_J$ and $p_{K^{\rm in}, K^{\rm out}}$.

\section{Spectral properties of infinitely large random and directed graphs} \label{eq:theory} 
In this section, we present  the main theoretical results in the limit of large~$n$ for  the spectral properties  of   adjacency matrices of random directed graphs with a
prescribed degree distribution [as defined in Eq.~(\ref{eq:model})].      

The giant SCC  plays an important role in   the spectrum of random directed graphs.   Let us therefore recollect that for directed random graphs with a prescribed degree distribution 
\begin{eqnarray}
s_{\rm sc} =0  \quad {\rm if }  \quad c(\rho+1)\leq 1,
\end{eqnarray}
and 
\begin{eqnarray}
s_{\rm sc} >0   \quad {\rm if }  \quad c(\rho+1)> 1.
\end{eqnarray}

This section is organized as follows.  
First, we discuss in Sec.~\ref{sec:specDistri}  how the  spectral distribution $\mu(\lambda)$ depends on the size of the SCC. 
     Second, we discuss in Sec.~\ref{sec:spectrum} how the  deterministic part  $\sigma$ of the spectrum   is governed by  the SCC.   In particular, we show that if  $c(\rho+1)>1$, then  $\sigma$ contains a  continuous part $\sigma_{\rm ac}$ and (possibly) a deterministic  outlier $\lambda_{\rm isol}$, both determined by the SCC.  On the other hand,  if $c(\rho+1)<1,$ then the spectrum $\sigma = \left\{-d\right\}$.     In   Sec.~\ref{sec:spectrum}  we also discuss how the  nondeterministic part  $\Gamma$  of the spectrum is determined by oriented ring graphs.   
Third, in Sec.~\ref{sec:rec}, we present recursion relations in the distribution of entries of  right eigenvectors associated with   deterministic outliers  $\lambda_{\rm isol}$ or with  eigenvalues $\lambda_{\rm b}$ located at the boundary of  $\sigma_{\rm ac}$.     Subsequently, we use in Secs.~\ref{sec:bound} and \ref{outliersub} these recursive distributional equations to derive analytical results for  the  boundary of   $\sigma_{\rm ac}$ and the deterministic outliers $\lambda_{\rm isol}$, respectively.    In Sec.~\ref{sec:lead}, we present results for the leading eigenvalue $\lambda_1$. We obtain exact analytical expressions for the typical value of the leading eigenvalue  $\lambda_1$ in the regime where $c(\rho+1)>1$, while for  $c(\rho+1)<1$ we show that the leading eigenvalue is governed by oriented ring graphs.     Lastly, in Sec.~\ref{Sec:gap}, we discuss the spectral gap, and in Sec.~\ref{sec:Perron}, we comment on the relation between the derived results and the Perron-Frobenius theorem~\cite{horn2012matrix}.     

We focus on right eigenvectors since  the  left eigenvectors of $\mathbf{A}$ are simply the right eigenvectors of~$\mathbf{A}^T$.  Therefore, results for left eigenvectors
can be obtained from the expressions for right eigenvectors through the substitutions "$R\rightarrow L$" and "${\rm in}\leftrightarrow  {\rm out}$".

\subsection{Spectral distribution}\label{sec:specDistri}
We discuss how the  spectral distribution $\mu(\lambda)$ of an adjacency matrix of a random directed graph depends on the size of its  connected components.
In Appendix~\ref{App:specConn}, we show that
the spectral distribution  $\mu(\lambda)$ 
  takes the form   
  \begin{eqnarray}
\mu(\lambda) = (1-s_{\rm sc})\delta(\lambda+d) +s_{\rm sc} \:\tilde{\mu}(\lambda), \label{eq:muOrientedxx}
  \end{eqnarray} 
  where $ \tilde{\mu}(\lambda)$ is a normalized  distribution associated with the SCC and supported on $\sigma_{\rm ac}$; see Fig.~9 of Ref.~\cite{metz2018spectra} for an example of $\tilde{\mu}(\lambda)$ in the case of directed Erd\H{o}s-R\'{e}nyi ensembles.      

  Eq.~(\ref{eq:muOrientedxx})  implies that the algebraic multiplicity of the $-d$-eigenvalue  is equal to 
  \begin{equation}
  n(1-s_{\rm sc})(1+o_n(1)).      
  \end{equation} 
  
  The high degeneracy of the  $-d$-eigenvalue follows from the fact that  (i) random, directed graphs with a prescribed degree distribution are locally tree-like and oriented  and (ii) an oriented tree graph has only zero eigenvalues, and in the present case where the diagonal elements are all set equal to $-d$,  all eigenvalues of an oriented tree graph are  equal to $-d$.     Hence, a random directed graph develops  eigenvalues  that differ from $-d$  trough the presence of oriented rings, which are defined by Eq.~(\ref{eq:orientedRingDef}) in Sec.~\ref{Sec:orientedRing}.     

\subsection{Spectrum}\label{sec:spectrum}
The spectrum $\sigma\cup \Gamma$ of a  random  directed graph in the limit of infinitely large $n$ is determined by three topological components, namely the SCC,  nodes that do not belong to the SCC, and oriented rings of finite length.

If $c(\rho+1)>1$, then   the deterministic part $\sigma$ of the spectrum consists of a continuous set $\sigma_{\rm ac}$  and (possibly) an outlier $\lambda_{\rm isol}$, both determined by the SCC.       

On the other hand, if $c(\rho+1)<1$, then  $\sigma = \left\{-d\right\}$. 

In addition, due to the presence of cycles of finite length,   random and directed graphs can contain stochastic outliers.        Stochastic outliers appear in the spectrum due to the presence of oriented rings in the directed random graph.      As shown in the Appendix~\ref{OrientedRing}, the eigenvalues of an oriented ring of length $\ell$ are located on a circle of radius 
\begin{eqnarray}
\gamma = \left(\prod^n_{j=1}|J_j|\right)^{1/\ell}, \label{eq:radiusRing}
\end{eqnarray} 
where $J_j$ are the random weights attributed to the ring graph.      If $c(\rho+1)<1$, then these eigenvalues appear as outliers in the spectrum.    On the other hand if $c(\rho+1)>1$, then the eigenvalues of an oriented rings  form  stochastic outliers  only when 
$\gamma$ is large enough, so that they do not belong to $\sigma_{\rm ac}$.    As a consequence, unweighted  graphs, i.e., with $J_{ij}=1$, do not contain stochastic outliers when   $c(\rho+1)>1$.      However, if the graph has weighted links, then  stochastic outlier eigenvalues exist, even though the probability to observe them is in general small. 

   \subsection{Recursive distributional equations for right eigenvectors}  \label{sec:rec}
   In Appendix~\ref{sec:der}, we derive  a set of recursive distributional equations for the asymptotic distributions $p_R$ as defined in Eq.~(\ref{eq:asymptpR}) for  the right eigenvectors associated with deterministic eigenvalue outliers $\lambda_{\rm isol}$ and  with   eigenvalues located at the boundary of the continuous part  $\sigma_{\rm ac}$.    In particular, we show that the  distribution $p_R$    solves the recursive distributional equation
 \begin{eqnarray}
 \lefteqn{p_R(r) = \sum^{\infty}_{k=0} \sum^{\infty}_{\ell=0}p_{K^{\rm in}, K^{\rm out}}(k,\ell)    }&& 
 \nonumber\\ 
&& \int \prod^{\ell}_{j=1}{\rm d}^2 r_j  q_R(r_j) \int \prod^{\ell}_{j=1}{\rm d} x_j  p_J(x_j) \delta \left[r - \frac{\sum^{\ell}_{j=1}x_j r_j }{\lambda+d}\right], \nonumber\\ \label{eq:pRecR}
 \end{eqnarray} 
 where $q_R$  is  a distribution that solves 
  \begin{eqnarray}
 \lefteqn{q_R(r) = \sum^{\infty}_{k=0} \sum^{\infty}_{\ell=0}p_{K^{\rm in}, K^{\rm out}}(k,\ell)     \frac{k}{c} }&& 
 \nonumber\\ 
&& \int \prod^{\ell}_{j=1}{\rm d}^2 r_j  q_R(r_j) \int \prod^{\ell}_{j=1}{\rm d} x_j  p_J(x_j) \delta \left[r - \frac{\sum^{\ell}_{j=1}x_j r_j }{\lambda+d}\right].\label{eq:qRecR} \nonumber\\
 \end{eqnarray} 
When 
    \begin{eqnarray}
  p_{K^{\rm in}, K^{\rm out}}(k, \ell) = p_{K^{\rm in}}\left(k\right)p_{K^{\rm out}}\left(\ell \right),
   \end{eqnarray}
   it holds that  $p_R(r)=q_R(r)$ and 
 we  recover the results from Ref.~\cite{neri2016eigenvalue}.

 The relations (\ref{eq:pRecR}) and (\ref{eq:qRecR})    admit, for any value of $\lambda$, the {\it trivial} solution 
 \begin{eqnarray} 
 p_{R}(r) = \delta(r), \label{eq:trivial} 
 \end{eqnarray}
 which cannot be associated with a  right eigenvector of the random matrix $\mathbf{A}$.    
 However, the relations  (\ref{eq:pRecR}) and (\ref{eq:qRecR})  also admit   {\it normalizable} solutions  for which  there exist a positive number $\alpha>0$ so that 
 \begin{eqnarray}
 \int {\rm d}^2 r \: p_R(r)|r|^\alpha \in (0,\infty).
 \end{eqnarray}   
 These normalizable solutions are associated with  right eigenvectors of the random matrix $\mathbf{A}$.

 As a consequence, we  can obtain explicit expressions for the outliers $\lambda_{\rm isol}$  and the eigenvalues $\lambda_{\rm b} \in \partial \sigma_{\rm ac}$ by identifying values of $\lambda$ for which the relations (\ref{eq:pRecR}) and (\ref{eq:qRecR}) admit normalizable solutions.  This is the program that we pursue in Appendix~\ref{sec:SolRec}, while we present the main results of those derivations in the next subsections.

  \subsection{Eigenvalues at the boundary  of the continuous part of the spectrum}  \label{sec:bound}
The spectrum $\sigma$  contains a continuous part $\sigma_{\rm ac}$ if $c(\rho+1)>1$, as we have shown in Sec.~\ref{sec:specDistri}.        For values  $\lambda =\lambda_{\rm b}\in\partial\sigma_{\rm ac}$ located at the boundary of $\sigma_{\rm ac}$, the relations (\ref{eq:pRecR}) and (\ref{eq:qRecR}) admit a normalizable solution.  
Using this criterion, we obtain in Appendix~\ref{sec:SolRec} that
   \begin{eqnarray}
  |\lambda_{\rm b}+d|^2 =  c (\rho + 1) \langle J^2  \rangle  .\label{eq:boundary}
  \end{eqnarray}
  
  The relations  (\ref{eq:pRecR}) and (\ref{eq:qRecR}) provide us also with the statistics of right eigenvectors $\vec{R}_{\rm b}$ associated with eigenvalues $\lambda_{\rm b}$.     We distinguish between 
   the cases where $\lambda_{\rm b}\notin \mathbb{R}$ and $\lambda_b \in \mathbb{R}$.   In the former case,  the components $R_{\rm b}$ are complex-valued random variables with
  \begin{eqnarray}
\langle R_{\rm b}\rangle  &=  \langle R^2_{\rm b}\rangle  &= 0. \label{eq:Rb1}
\end{eqnarray}
On the other hand, if $\lambda_b \in \mathbb{R}$, then the  components are real-valued random variables with 
  \begin{eqnarray}
\langle R_{\rm b}\rangle  &= 0 , 
\quad \langle R^2_{\rm b}\rangle &= 1 . \label{eq:Rb2}
\end{eqnarray}   

In addition to these results, we show in Appendix~\ref{Sec:ZeroValuedR} that the distribution $p_{R_{{\rm b}}}(r)$ contains a delta peak at the origin due to all nodes that do not belong to the giant outcomponent,  i.e.,
\begin{eqnarray}
p_{R_{{\rm b}}}(r) &=& (1-s_{\rm in})\delta(r) + s_{\rm in}\:\tilde{p}_{R_{{\rm b}}}(r), \label{eq:pRb}
\end{eqnarray} 
where $s_{\rm in}$ is the size of the giant incomponent given by Eq.~(\ref{eq:sout}), and $\tilde{p}_{R_{{\rm b}}}(r)$ is a  normalized distribution.

\subsection{Outlier eigenvalue} \label{outliersub}  
There exists a second type of normalizable solutions to the Eqs.~(\ref{eq:pRecR}) and (\ref{eq:qRecR}), which are associated with deterministic outlier eigenvalues $\lambda_{\rm isol}$.
If  $c (\rho + 1)>1$ and  $\langle J^2\rangle <c  (\rho + 1)|\langle J\rangle|$, then there exists a deterministic eigenvalue outlier located at
\begin{eqnarray}
\lambda_{\rm isol}  = -d+  c (\rho + 1) \langle J \rangle  \label{eq:outlier}. 
\end{eqnarray} 
Reference~\cite{restrepo2007approximating}  observes that Eq.~(\ref{eq:outlier}) describes well the  largest eigenvalue of unweighted adjacency matrices of  random graphs with a prescribed degree distribution.   In Appendix~\ref{sec:SolRec}, we show that Eq.~(\ref{eq:outlier}) is in fact an exact expression for the deterministic outlier.  

The   entries of the eigenvector $\vec{R}_{\rm isol}$ are  real, and  the first moment of $R_{\rm isol}$  satisfies
\begin{eqnarray}
\frac{\langle R_{\rm isol}\rangle^2 }{\langle R^2_{\rm isol}\rangle} = \frac{c^3 (\rho + 1) [c(\rho + 1)\langle J\rangle^2- \langle J^2\rangle]}{    c^2(\rho+1)^2\langle J\rangle^2 [\langle (K^{\rm out})^2\rangle -c]  + \langle J^2\rangle \rho^{\rm out}_2   } , \nonumber\\
\label{eq:R}
\end{eqnarray}
where   
\begin{eqnarray}
\rho^{\rm out}_2 = \langle K^{\rm in}(K^{\rm out})^2\rangle - c(1+\rho) \langle (K^{\rm out})^2\rangle.
\end{eqnarray}
For uncorrelated indegrees and outdegrees it holds that $\rho=0$ and $ \rho^{\rm out}_2 = 0$, and   we recover  the  results in Ref.~\cite{neri2016eigenvalue}.   

Analogous to Eq.~(\ref{eq:pRb}), the distribution $p_{R_{{\rm isol}}}$  takes  the form 
\begin{eqnarray}
p_{R_{{\rm isol}}}(r) &=& (1-s_{\rm in})\delta(r) + s_{\rm in}\tilde{p}_{R_{{\rm isol}}}(r),  \label{eq:pRIsol}
\end{eqnarray}  
where $s_{\rm in}$ is the size of the giant incomponent (\ref{eq:sout}) and $\tilde{p}_{R_{{\rm isol}}}(r)$ is a  normalized distribution (see Appendix~\ref{Sec:ZeroValuedR}).

\subsection{Leading eigenvalue}  \label{sec:lead} 
We discuss the implications of the results derived   in Secs.~\ref{sec:bound} and \ref{outliersub}  for the leading eigenvalue $\lambda_1$ of  random graphs with a prescribed degree distribution $p_{K^{\rm in},K^{\rm out}}$.       

\subsubsection{Distribution of $\lambda_1$}
The theory in Secs.~\ref{outliersub} and \ref{sec:bound}  provides exact expressions for the boundary $\partial \sigma_{\rm ac}$ of the continuous part of the spectrum, which is given by the eigenvalues $\lambda_{\rm b}$ in Eq.~(\ref{eq:boundary}), and the deterministic eigenvalue outlier $\lambda_{\rm isol}$, which is given by Eq.~(\ref{eq:outlier}), in random directed graphs that are infinitely large.    The question remains  how  the leading eigenvalue $\lambda_1$ is related to $\lambda_{\rm b}$ and $\lambda_{\rm isol}$.

 If we neglect the contributions from cycles of finite length $\ell$, then the leading eigenvalue of an infinitely large random directed graph  is given by
\begin{eqnarray}
 \lambda^\ast = \left\{ \begin{array}{ccc}  {\rm max}\left\{\lambda_{\rm isol}, |\lambda_{\rm b}+d|-d\right\} &{\rm if}& c(\rho+1)\geq1, \\ -d & {\rm if} & c(\rho+1)<1, \end{array} \right. \nonumber\\ \label{eq:lamddaAst}
\end{eqnarray} 
where $\lambda_{\rm isol}$ and $\lambda_{\rm b}$ are given by Eqs.~(\ref{eq:outlier}) and (\ref{eq:boundary}), respectively.   Hence, if a random directed graph contains no cycles of small length $\ell$ in the limit $n\rightarrow \infty$, then Eq.~(\ref{eq:lamddaAst}) is exact.   However, as we have discussed in Sec.~\ref{Sec:orientedRing}, random directed graphs with a prescribed degree distribution $p_{K^{\rm in},K^{\rm out}}$ typically contain a finite number of cycles of a given  length $\ell$, even in the limit $n\rightarrow \infty$, and therefore we need to discuss how these cycles will affect~$\lambda_1$.

      \begin{figure}[t]
\centering
\includegraphics[width=0.4\textwidth]{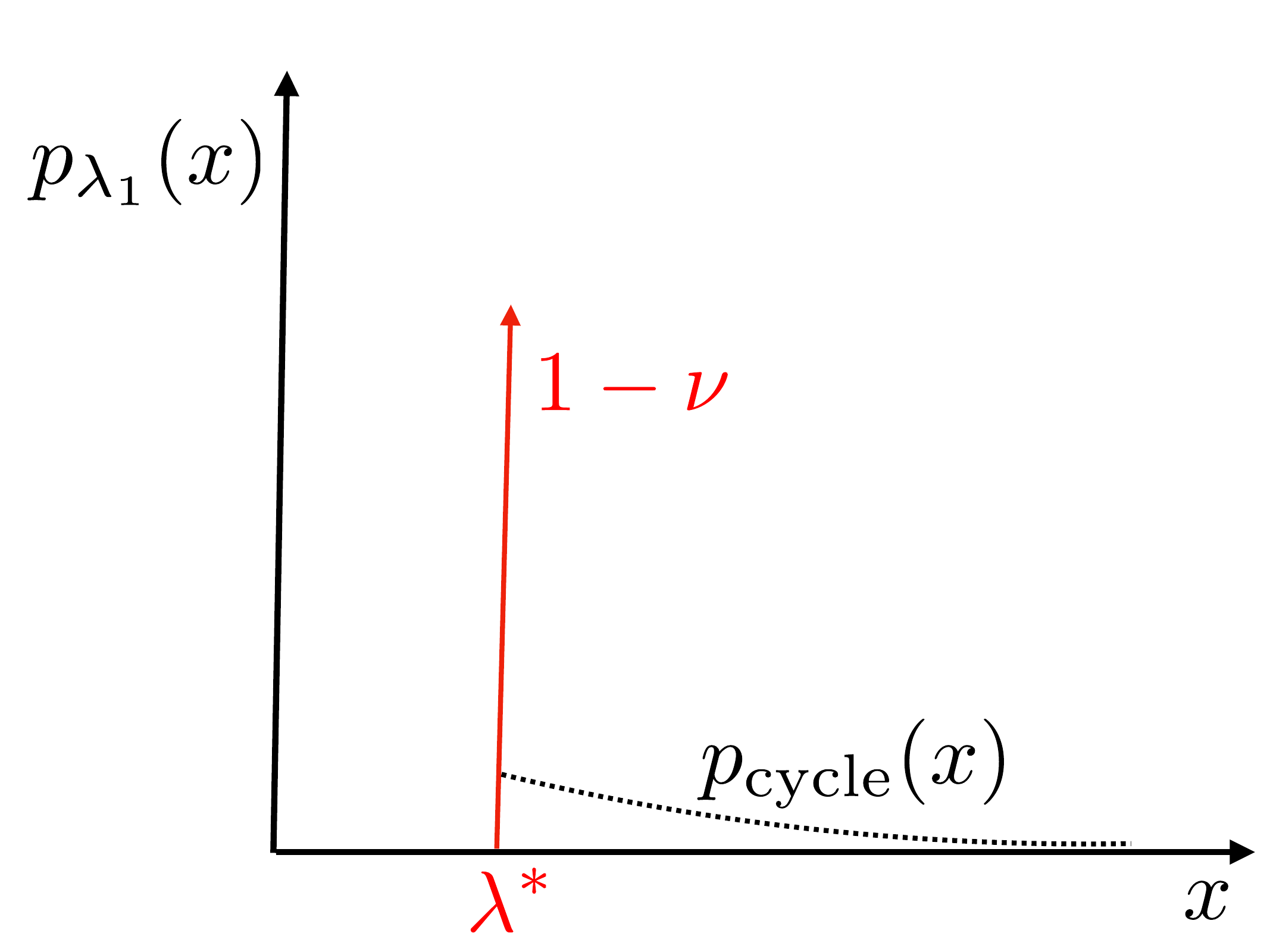}
\caption{ {\it Distribution of the leading eigenvalue.} Sketch of the distribution $p_{\lambda_1}$ of the leading eigenvalue $\lambda_1$  of random matrices $\mathbf{A}$, as defined in Sec.~\ref{eq:modelDef}, in the regime $c(\rho+1)>1$.    The distribution consists of a  delta distribution at the typical value $\lambda^\ast$ given by Eq.~(\ref{eq:lamddaAst}) and  a continuous distribution $p_{\rm cycle}$ with a total weight $\nu\approx 0$.  } \label{fig:IllustrationSketch}
\end{figure}  
 
   Cycles that are oriented rings may contribute stochastic outlier eigenvalues to the spectrum, see Sec.~\ref{Sec:orientedRing}.   As a consequence, $\lambda_1$ is not a self-averaging  quantity but is instead  a random variable with a distribution
\begin{eqnarray}
p_{\lambda_1}(x) := \lim_{n\rightarrow \infty} \Big\langle  \delta(x-\lambda_1(\mathbf{A}))\Big\rangle
\end{eqnarray}
of nonzero variance.  

The distribution $p_{\lambda_1}$ takes the form 
\begin{eqnarray}
p_{\lambda_1}(x) =(1-\nu) \delta(x- \lambda^\ast)  +  \nu \: p_{\rm cycle}(x), \label{eq:plambda1}
\end{eqnarray}
where  $\nu$ is the probability that the leading eigenvalue is a stochastic outlier contributed by  an oriented ring,  and    $p_{\rm cycle}(x)$ is the distribution of those stochastic outliers that are leading eigenvalues.   Note that the  distribution    $p_{\rm cycle}(x)$  is supported on the half line  $[\lambda^\ast,\infty)$, see Fig.~\ref{fig:IllustrationSketch} for a sketch of $p_{\lambda_1}$.  

Since for $c(\rho+1)<1$ it holds that  $\lambda^\ast = -d$,   oriented rings will play an important role in $p_{\lambda_1}(x)$ when $\mathbf{A}$ does not have a giant SCC.   On the other hand, if $\mathbf{A}$ has a giant SCC, i.e.~$c(\rho+1)>1$, then it will be unlikely that the leading eigenvalue is a stochastic outlier.     We show this explicitly in the next subsection for unweighted graphs, and subsequently we discuss the case of weighted graphs.

\subsubsection{Unweighted graphs}\label{sec:unweightedGraphs}
We consider adjacency matrices $\mathbf{A}$   of  unweighted graphs, 
 such that $J_{ij}=1$ for all values of $i$ and $j$.  In this case, we obtain  an exact expression  for $p_{\lambda_1}(x)$.   Indeed, since the eigenvalues of oriented rings with $J_{ij}=1$ are located on a circle of radius $1$ centred around $-d$, see Eq.~(\ref{eq:radiusRing}), it holds that
\begin{equation}
\nu = \left\{ \begin{array}{ccc} 0 &{\rm if}& c(\rho+1) \geq 1, \\ p_+ &{\rm if}& c(\rho+1)<1,\end{array} \right. \label{eq:nu}
\end{equation}  
 where $p_+$ is the probability that the graph contains at least one oriented ring graph, given by Eq.~(\ref{eq:p+}).   Moreover, it holds that
\begin{equation}
p_{\rm cycle}(x) = \delta(x-1+d),
\end{equation}
and that
\begin{eqnarray}
 \lambda^\ast = \left\{ \begin{array}{ccc}  -d +c(\rho+1) &{\rm if}& c(\rho+1)\geq1, \\ 0 & {\rm if} & c(\rho+1)<1.  \end{array} \right.  \label{eq:lamddaAstxx}
\end{eqnarray}
Using Eqs.~(\ref{eq:nu}-\ref{eq:lamddaAstxx}) in Eq.~(\ref{eq:plambda1}), 
we obtain that 
\begin{eqnarray}
\lefteqn{p_{\lambda_1}(x) }&& \nonumber\ \\ 
& =&  \left\{\begin{array}{ccc} \delta(x +d -c [\rho + 1]) &{\rm if}& c(\rho+1) \geq 1, \\ (1-p_+)\delta(x+d) + p_+ \delta(x-1+d)  &{\rm if}& c(\rho+1) < 1 . \end{array} \right. \nonumber\\ \label{eq:distrilambda1Unw}
\end{eqnarray}     
Hence, the leading eigenvalue of an unweighted random directed graph is deterministic and given by the value $\lambda^\ast$  if the graph contains a giant SCC.    On the other hand, if there is no giant SCC, then with probability $p_+$ an oriented ring determine the leading eigenvalue.

From Eq.~(\ref{eq:distrilambda1Unw}), we  obtain the average leading eigenvalue, which is given by 
\begin{eqnarray}
\langle  \lambda_1 \rangle  = \left\{ \begin{array}{ccc}   -d + c (\rho + 1)  &  {\rm if}& c (\rho + 1)\geq 1, \\ -d + p_+& {\rm if}& c(\rho+1)<1,\end{array} \right. \label{eq:meanlambda1}
\end{eqnarray} 
and its variance  
\begin{eqnarray}
 {\rm var}[\lambda_1]  
& =&  \left\{\begin{array}{ccc}  0 &{\rm if}& c(\rho+1) \geq 1, \\ 
p_+(1-p_+) &{\rm if}& c(\rho+1) < 1, 
 \end{array}\right. \label{eq:Var}
\end{eqnarray}
where   $p_+$ is given by Eq.~(\ref{eq:p+}).   Note that  ${\rm var}[\lambda_1]  =0$ if $\mathbf{A}$ has a giant SCC, and the leading eigenvalue is thus self-averaging in this regime, while   ${\rm var}[\lambda_1]  > 0$ if  $\mathbf{A}$  does not have a giant SCC, and the leading eigenvalue is thus not self-averaging in this regime.   

In the next section, we discuss how these results  extend to the case of weighted graphs for which the $J_{ij}$ are drawn from a nontrivial distribution $p_J$. 

\subsubsection{Weighted graphs}\label{sec:weighted}
 In  the general case of weighted graphs, it is  difficult to obtain exact expressions for $\nu$ and $p_{\rm cycle}(x)$.    However, we can discuss  the qualitative features of 
$p_{\lambda_1}(x)$  in the two regimes $c(\rho+1)<1$ and  $c(\rho+1)>1$.    

 If $c(\rho+1)>1$, then   
\begin{equation}
\nu\approx 0,  \label{eq:Delta} 
\end{equation}
since it is unlikely that an oriented ring  contributes an eigenvalue to the spectrum that is  larger than $\lambda^\ast$; this would require that $\gamma$, given by Eq.~(\ref{eq:radiusRing}), is larger than $\lambda^\ast$.      Therefore,  if the graph has a giant SCC, then the variance of $\lambda_1$ will be small and the typical value of $\lambda_1$  is given by $\lambda^\ast$ in Eq.~(\ref{eq:lamddaAst}).       As a consequence, if the graph has a giant SCC, then 
\begin{eqnarray}
 \langle \lambda_1\rangle  \approx \lambda^\ast  =  \left\{\begin{array}{ccc} -d +c (\rho + 1) \langle J \rangle &{\rm if}&  \langle J \rangle > \sqrt{\frac{\langle J^2 \rangle}{c(\rho + 1)}}, \\ -d+ \sqrt{c (\rho + 1) \langle J^2 \rangle}&{\rm if}& \langle J \rangle \leq    \sqrt{\frac{\langle J^2 \rangle}{c (\rho + 1)}}.\end{array}\right.  \nonumber\\\label{eq:lambda1}
\end{eqnarray} 
since $\lambda^\ast$ is the typical value of $\lambda_1$.  

   On the other hand, when $c(\rho+1)<1$, then $\lambda^\ast = 0$, and therefore the leading eigenvalue is with a probability
 \begin{eqnarray}
 \nu = p_+,  
 \end{eqnarray} 
a stochastic outlier coming from an oriented ring graph.      Hence, in the absence of a SCC, the variance of $p_{\lambda_1}(x)$ is large.

\subsubsection{Right eigenvector associated with $\lambda_1$}
We  derive exact expressions for the first moment  $\langle R_{\rm 1}\rangle$  of  eigenvectors associated with the leading eigenvalue $\lambda_1$.   

We first consider the case $c(\rho+1)>1$.   Assuming that    the leading eigenvalue takes its typical value $\lambda^\ast$,  given by either the outlier $ \lambda_{\rm isol}$ or  the maximum value of ${\rm Re}[\lambda_{\rm b}]$, see Eq.~(\ref{eq:lambda1}), we obtain that    
 \begin{eqnarray}
   \frac{\langle R_{\rm 1}\rangle^2}{\langle |R_{\rm 1}|^2\rangle} = \left\{\begin{array}{ccc}  \langle R_{\rm isol}\rangle^2/\langle |R_{\rm isol}|^2\rangle  &{\rm if}&
   \langle J \rangle > \sqrt{\frac{\langle J^2 \rangle}{c(\rho + 1)}}, \\ 0&{\rm if}& \langle J \rangle \leq
   \sqrt{\frac{\langle J^2 \rangle}{c (\rho + 1)}},\end{array}\right.\nonumber\\ \label{eq:R1}
\end{eqnarray} 
where $ \langle R_{\rm isol}\rangle^2/\langle |R_{\rm isol}|^2\rangle$ is given by Eq.~(\ref{eq:R}).

On the other hand, if  $c(\rho+1)<1$, then the right eigenvector of $\lambda_1$ will be localized on a  finite number of nodes and 
 \begin{eqnarray}
   \frac{\langle R_{\rm 1}\rangle^2}{\langle |R_{\rm 1}|^2\rangle} =  0.
   \end{eqnarray} 

Interestingly, we observe in Eq.~(\ref{eq:R1}) that $\langle R_{\rm 1} \rangle$  behaves as an order-parameter  of a phase transition between a ferromagnetic phase   ($\langle R_1 \rangle>0$)  and a spin glass phase ($\langle R_1 \rangle=0$). 
    A similar type of  behaviour has been found in sparse symmetric random matrices \cite{kabashima2010cavity, kabashima2012first, takahashi2014fat, susca2019top}.  
        The analogy between  $\langle R_1 \rangle$ and the  order parameter  of a ferromagnetic  phase  can be made explicit. Indeed, the leading right eigenvector $\vec{R}_1$ is the  stationary state of a spherical model defined on the graph represented by the  adjacency matrix  $\mathbf{A}$, see  equations (45) till (52) in  Ref.~\cite{metz2018spectra}.     The spherical model at zero temperature exhibits either a ferromagnetic phase or a spin-glass phase, see Ref.~\cite{kosterlitz1976spherical}, and $\langle R_1 \rangle$ serves as the order parameter for this phase transition.    Notice that  the  $\langle R_1 \rangle = 0$ regime does not correspond to a paramagnetic phase since the spherical model will be  frozen into the configuration represented by the leading right eigenvector~\cite{kosterlitz1976spherical}.

  \subsubsection{Limiting case of dense graphs}
We discuss  the  limit of dense graphs by setting $c= n$ and $\rho=0$.   
 Eq.~(\ref{eq:lambda1})  then reduces to 
   \begin{eqnarray}
     \lambda_1 = \left\{\begin{array}{ccc} n \langle J \rangle   && \langle J \rangle  > 0 , \\
     \sqrt{n \langle J^2 \rangle}  &&  \langle J \rangle  \leq 0 ,  \end{array} \right. \label{eq:lambda1IID}
   \end{eqnarray}  
   which is the well-known expression for  the leading eigenvalue $\lambda_1$ of a random matrix with independent and identically distributed matrix elements drawn from a distribution $p_J$, see Refs.~\cite{girko1985circular, bai1997circular, gotze2010circular, tao2010random, bordenave2012around, tao2013outliers}, as well as Refs.~\cite{may1972will, allesina2012stability}.    However, note that the formula~(\ref{eq:lambda1}) holds for graphs with $c \in O_n(1)$ and therefore the correspondence holds  only formally.      Analogously, we obtain in this limit that 
    \begin{eqnarray}
   \frac{\langle R_{\rm 1}\rangle^2}{\langle |R_{\rm 1}|^2\rangle} = \left\{\begin{array}{ccc} 1 &{\rm if}&
   \langle J \rangle >0, \\ 0&{\rm if}& \langle J \rangle \leq
0.\end{array}\right.\nonumber\\ \label{eq:R1x}
\end{eqnarray}

     \subsection{Subleading eigenvalue and  spectral gap}\label{Sec:gap}
The spectral gap is the difference $\lambda_1 - {\rm Re}[\lambda_2]$ between the leading eigenvalue and the real part of the subleading eigenvalue.   
From the  results in Secs.~\ref{outliersub}, \ref{sec:bound}  and \ref{sec:lead}, we readily obtain an expression for the typical value of  the spectral gap when  $c (\rho + 1)>1$,  namely,  
\begin{eqnarray}
\lefteqn{\lambda_1 - {\rm Re}[\lambda_2] }&& 
\nonumber\\ 
 &=& \left\{\begin{array}{ccc} c (\rho + 1) \langle J \rangle - \sqrt{c (\rho + 1) \langle J^2 \rangle} &{\rm if}&  \langle J \rangle > \sqrt{\frac{\langle J^2 \rangle}{c(\rho + 1)}}, \\  0 &{\rm if}& \langle J \rangle \leq    \sqrt{\frac{\langle J^2 \rangle}{c (\rho + 1)}}.\end{array}\right. \nonumber\\ \label{eq:lambdaSG}
\end{eqnarray}     
The expected value of the entries of the right eigenvector associated with the subleading eigenvalue satisfy
 \begin{eqnarray}
   \frac{\langle R_{\rm 2}\rangle^2}{\langle |R_{\rm 2}|^2\rangle} = 0.\label{eq:R2}
\end{eqnarray} 

\subsection{Perron-Frobenius theorem} \label{sec:Perron}
Here we discuss how the theoretical results are related to the   celebrated Perron-Frobenius theorem~\cite{horn2012matrix}, which states that the eigenvalue $\lambda_1$  of  a nonnegative matrix, and the components of its right (left) eigenvector,  are nonnegative numbers.  In other words, the  Perron-Frobenius theorem implies that  $R_{1,j}\geq 0$ for all $j=1,2,\ldots,n$.

Interesting conclusions about the localization of eigenvectors of $\mathbf{A}$ are drawn if we combine the Perron-Frobenius theorem with the result (\ref{eq:R1}). 
If   $c(\rho+1) \leq \langle J^2 \rangle/\langle J \rangle^2$ and $c(\rho+1)>1$, such that $\lambda_1$ is part of  $\partial \sigma_{\rm ac}$, then $\langle R_1 \rangle =0$ and  $\langle R_{1}^2 \rangle = 1$, see Eq.~(\ref{eq:convR2}).   Since according to the Perron-Frobenius theorem $R_1\geq 0$, we obtain that $R_1=0$  holds with probability one.  
  The  two conditions    $\lim_{n\rightarrow \infty} \langle R_1(\mathbf{A}_n)\rangle = 0$ and     $\lim_{n\rightarrow \infty} \langle R^2_1(\mathbf{A}_n)\rangle = 1$ can  be simultaneously valid  provided that a few components of the eigenvector $\vec{R}_1(\bA)$ diverge, such that  $\lim_{n\rightarrow \infty} \langle R^2_1(\mathbf{A}_n)\rangle\neq  \langle \lim_{n\rightarrow \infty} R^2_1(\mathbf{A}_n)\rangle$.     
  
  Hence,  (\ref{eq:R1}) and the Perron-Frobenius theorem imply that for nonnegative matrices  for which the conditions $c(\rho+1) \leq \langle J^2 \rangle/\langle J \rangle^2$ and $c(\rho+1)>1$ are fulfilled,  the right eigenvector $\vec{R}_1$ associated with the leading eigenvalue is localized on a few nodes.

\section{Stability of complex systems on random and directed graphs}\label{sec:app}  
We apply the results from the previous section to a linear stability analysis of dynamical systems defined on random directed graphs.  

Let $\vec{x}^\dagger(t) = (x_1(t), \dots,x_n(t))$ be the state vector of a large dynamical system of interest, and let 
\begin{equation}
\partial_t \vec{x}(t) = \vec{f}[\vec{x}(t)] \label{eq:nonl}
\end{equation}
be a set of nonlinearly coupled differential equations that describe the dynamics of the system of interest. 

We consider a fixed point or stationary state $\vec{x}^\ast$ and study the dynamics described by Eq.~(\ref{eq:nonl}) in the vicinity of  $\vec{x}^\ast$.    A  stationary state is a  vector that satisfies
\begin{equation}
f[\vec{x}^\ast] = 0.
\end{equation} 
Note that a nonlinear system may contain several stationary states~\cite{fyodorov2016nonlinear}, but here we are only interested in the dynamics of $\vec{x}(t)$ in the vicinity of one given stationary state.
According to the Hartman-Grobner theorem  \cite{grobman1959homeomorphism, hartman1960lemma}, the  dynamics described by Eq.~(\ref{eq:nonl})
is in the vicinity of the fixed point  $\vec{x}^\ast$  well approximated by the set of linearly coupled equations  given by  Eq.~(\ref{eq:linx}) with $\mathbf{A}$ the Jacobian of  $f$ and 
\begin{equation}
\vec{y}(t) = \vec{x}(t)-\vec{x}^\ast
\end{equation}
  the deviation vector.

   The  stability of  the   stationary state $\vec{x}^\ast$   is determined by the sign of the real part of the leading eigenvalue  $\lambda_1(\mathbf{A})$.
Indeed, if  the matrix $\mathbf{A}$ is diagonalizable, then  the dynamics of  $\vec{y}^\dagger(t)$ 
is  governed by the   eigenvalues $\lambda_j(\bA)$ and their associated right eigenvectors $\vec{R}_j(\bA) $  and left eigenvectors $\vec{L}_j(\bA)$ \cite{horn1985},  namely,
 \begin{eqnarray}
 \vec{y}^\dagger(t) =  \sum^n_{j=1}  \left( \vec{y}(0) \cdot \vec{R}_j  \right) \: e^{\lambda_j t}   \vec{L}^\dagger_j .\label{jkla}
 \end{eqnarray}    
  In the case when all eigenvalues have negative real parts, then $\lim_{t\rightarrow \infty} \vec{y}^\dagger(t) = 0$, which implies that the stationary state   is stable.  On the other hand, if there exists at least one eigenvalue with a positive real part, then  the stationary state  is unstable.     With a bit more effort, one can show that the stability criterion based on the sign of the real part of the  leading eigenvalue also holds for  systems described by nondiagonalizable matrices, see Appendix~\ref{AppNonDiag}. 
  
   From Eq.~(\ref{jkla}), we also observe that    right and left eigenvectors associated with the leading eigenvalue contain  valuable information about the dynamics of a system in the vicinity of a fixed point.     In particular, the nature of the  mode that destabilizes the system  takes the form of the left eigenvector $\vec{L}_1$.    For instance,  
if the eigenvector $\vec{L}_1$ has a positive mean $\langle L_1\rangle>0$, then the instability  is    reminiscent of a ferromagnetic phase,  whereas if 
$\langle L_1\rangle=0$, then the instability is reminiscent of a spin-glass phase~\cite{mezard2001bethe, mezard2003cavity, franz2001ferromagnet}.

We study here the stability of   large systems 
coupled through random matrices  $\mathbf{A}$ defined on random directed graphs with a prescribed degree distribution $p_{K^{\rm in},K^{\rm out}}$,  as defined in Sec.\ref{eq:modelDef}.    
To this aim, we use the theory from Sec.~\ref{sec:lead} for the leading eigenvalue $\lambda_1$ and the associated values of  $\langle L_1 \rangle$ and $\langle R_1\rangle$ (which in this ensemble are equivalent).

  A first interesting observation  is that for random directed graphs $\lambda_1$ is finite, even in the limit  $n \rightarrow \infty$.  This stands in contrast with  the leading eigenvalue of nondirected random graphs \cite{krivelevich2003largest, chung2004spectra}, which diverges for  increasing $n$.      As a consequence, random directed graphs with a prescribed degree distribution are stable in the limit of large $n$,   which provides an interesting take on the   diversity-stability debate   \cite{mccann2000diversity}.     
The remarkable stability of large dynamical systems defined on directed graphs  follows from their locally tree-like and oriented nature.   Since the local neighborhood of a randomly selected node is  an oriented tree, there exist  no   feedback loops that  can amplify the amplitude of local perturbations.    On the other hand, in nondirected random graphs local perturbations are amplified through   feedback loops  provided by the nondirected links.    As a consequence, dynamical systems on locally tree-like networks with unidirectional interactions are much more stable than dynamical systems defined on networks with bidirectional interactions.

It remains of interest to study how  network architecture affects the stability of large dynamical systems defined on random directed graphs.   Since $\lambda_1$ is a random variable in the limit of infinitely large $n$, we focus first on its typical value $\lambda^\ast$, given by Eq.~(\ref{eq:lambda1}).  Interestingly,  for the  interaction networks defined in Sec.~\ref{eq:theory},  the eigenvalue $\lambda^\ast$  is solely  governed by    three  parameters that characterize the network architecture:  the {\it effective mean degree}  
\begin{equation}
c(1+\rho)
\end{equation}
 that characterizes the effective number of degrees of freedom each node in the network interacts with; the {\it coefficient of variation} 
\begin{equation}
 v_J := \sqrt{\langle J^2 \rangle - \langle J \rangle^2}/\langle J \rangle
 \end{equation}
  that characterizes the fluctuations in the coupling strengths between the constituents of the system; and the {\it effective interaction strength} 
  \begin{equation}
  \alpha := \langle J \rangle/d
  \end{equation}
   that quantifies the relative strength of the interactions with regard to the rate~$d$ of decay.    Hence, the  system stability, characterized by the typical value of the  leading eigenvalue $\lambda_1$, only depends on these three parameters, and thus enjoys a high degree of universality.

      \begin{figure}[t]
\centering
\includegraphics[width=0.5\textwidth]{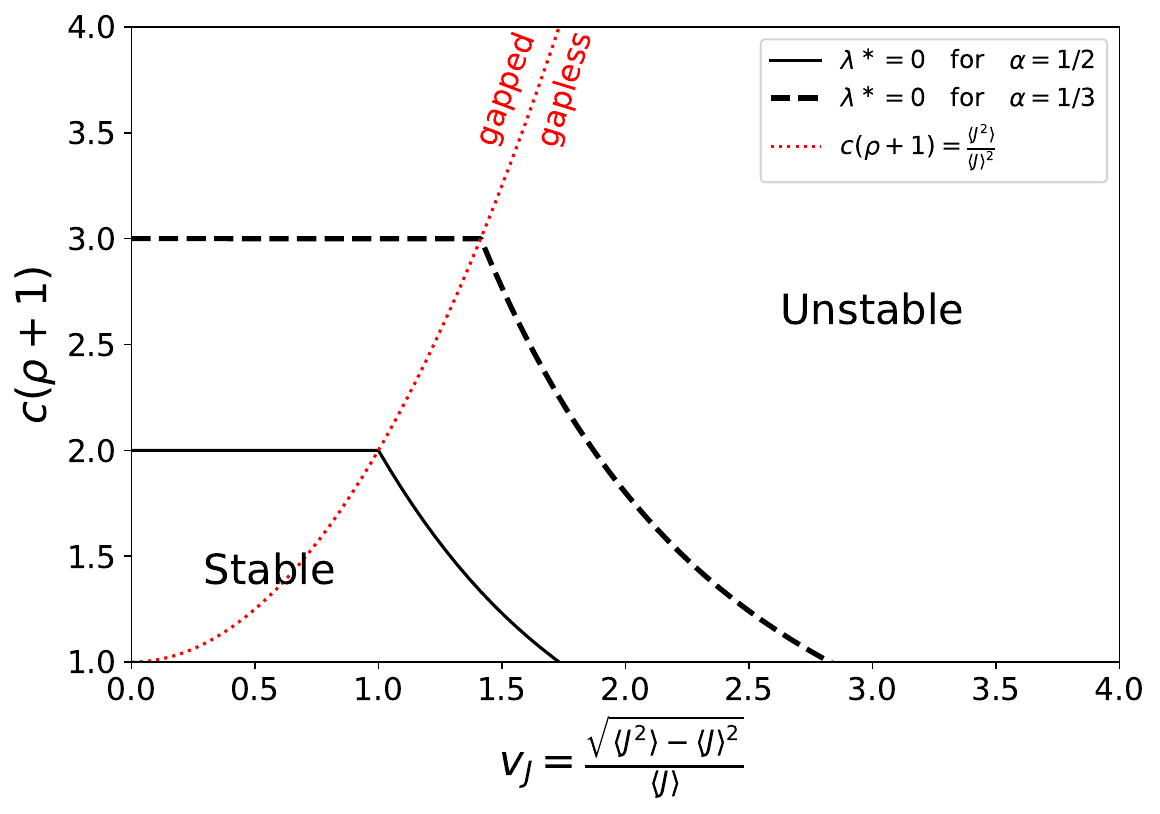}
\caption{{\it Universal phase diagram for the stability of dynamical systems  on random directed graph  with positive $\langle J\rangle$. }       Black solid line  and black dashed line 
separate the unstable phase  at large  effective
connectivity $c(\rho+1)$ from the stable phase at small  connectivity $c(\rho+1)$ for two given values of $\alpha = \langle J\rangle/d$.     The red dotted line separates the gapped phase at small $v_J$ from a gapless phase at high~$v_J$, which can also be considered a   transition line from a ferromagnetic phase (gapped) to a spin-glass phase (gapless).    } \label{fig5}
\end{figure} 

   \begin{figure}[t]
\centering
\includegraphics[width=0.5\textwidth]{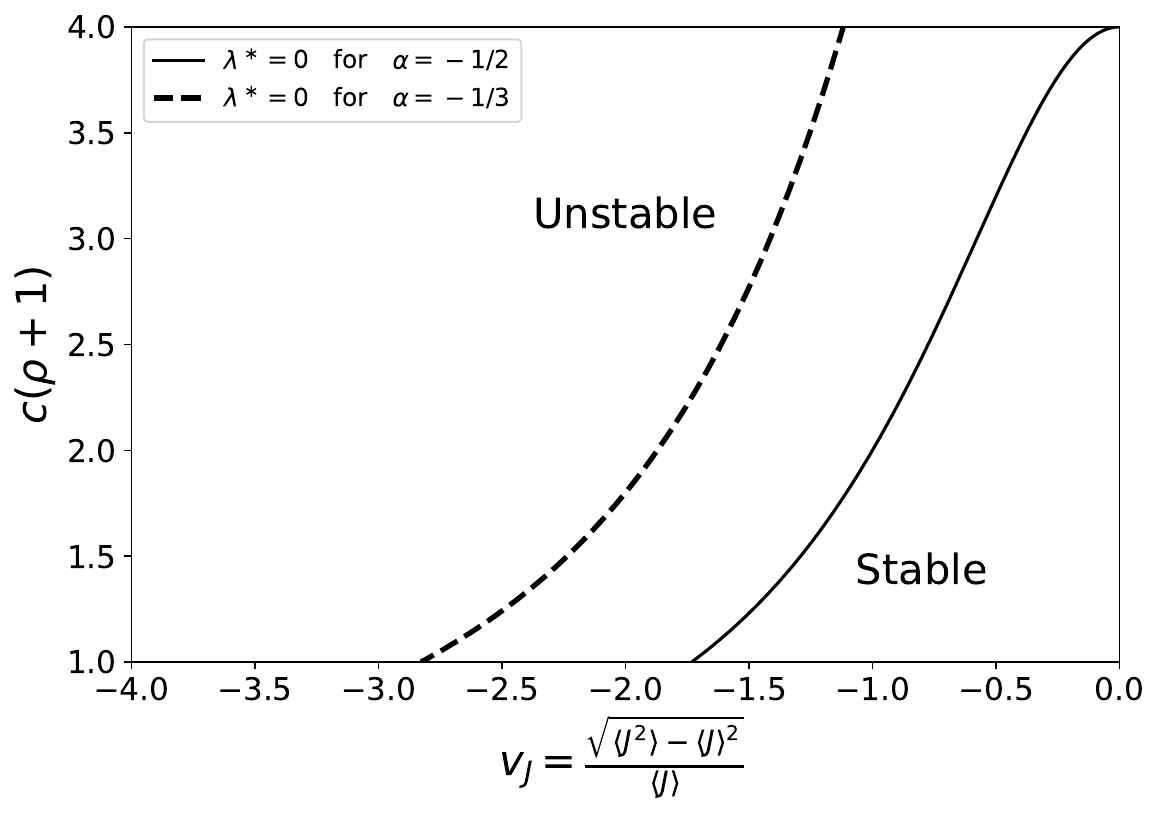}
\caption{{\it Universal phase diagram  diagram for the stability of dynamical systems on random directed graph  with negative $\langle J\rangle$.}    Similar as in figure \ref{fig5}, but now for negative $\alpha$.   In this case there is no gapped (or ferromagnetic) phase. } \label{fig6}
\end{figure}

In order to better understand how the three parameters $c(1+\rho)$, $v_J$, and $\alpha$  govern the stability of dynamical systems on  random directed graphs, we present in Fig.~\ref{fig5}  the   phase diagram of the system  in the $(v_J,c(1+\rho))$ plane, for fixed values of $\alpha\in[0,1]$  and $c(1+\rho)>1$.  The reason we choose these parameter regimes is because for $\alpha >1$ there exist no stable phase and for  $c(1+\rho) <1$ the graph does not have a giant strongly connected component; in the latter regime,  the system falls apart in the sense that it is a union of  a large number of small isolated subsystems, and  thus we are not considering anymore the linear stability of a  large system of interacting degrees of freedom.    

The phase diagram shows the critical connectivity $c_{\ast}$ (black lines)  that separates  the stable phase (${\rm Re}[\lambda^\ast]<0$), for systems at low connectivity $c(1+\rho)$, from the unstable phase (${\rm Re}[\lambda^\ast]>0$), for systems at high connectivity  $c(1+\rho)$.     If $\alpha>0$, then the critical line is determined by 
  \begin{eqnarray}
  c_\ast = \left\{ \begin{array}{ccc}1/\alpha, && v^2_J<1/\alpha-1, \\ 
1/[\alpha^2(v^2_J+1)], && v^2_J\in [1/\alpha-1, v^2_\ast],   \\ 
 \setminus && v^2_J > v^2_\ast, \end{array} \right. \label{eq:critConn}
  \end{eqnarray}
  which provides  the effective connectivity $c_\ast = c(\rho+1)$ for which ${\rm Re}[\lambda^\ast]=0$ as a function of $\alpha$ and $v_J$; in formula (\ref{eq:critConn}) we have used the symbol $v^2_\ast =  \frac{1-\alpha^2}{\alpha^2}$.    Since the critical connectivity is finite for all values of $\alpha$ and $v_J$, it follows that for large  enough  $c(1+\rho)$ any dynamical system is unstable, which is consistent with the results in May's paper~\cite{may1972will} that states that  any large enough fully connected system is unstable.   
 However, as we see from Eq.~(\ref{eq:critConn}) and Fig.~\ref{fig5}, the phase transition  to the stable phase at low connectivities has   three  qualitatively different regimes,  which we discuss in the following paragraphs.

 The critical value $v_\ast$ separates the regime $v_J>v_{\ast}$,  which does not have   stable phase, from the regime $v_J<v_{\ast}$, which has a stable phase at low enough connectivity $c(\rho+1)>1$.   Hence, for small enough fluctuations in the interaction strengths ($v_J<v_{\ast}$) it is  possible to stabilize the system by  rewiring edges in the graph   such
 that the correlation $\rho$ between indegrees and outdegrees decreases.
 Stabilizing the system by rewiring edges is however not possible when $v_J>v_{\ast}$.  

Moreover,  the regime $v_J<v_{\ast}$  consists of two distinct regimes:  a {\it gapped} regime, which appears when the  fluctuations in the interaction strengths   are small ($v^2_J<1/\alpha-1$), and a {\it gapless} regime, which appears when the fluctuations in the  interaction strengths   are large ($v^2_J>1/\alpha-1$).   In Fig.~\ref{fig5}, these two regimes are separated by the  red dotted line.    In the gapped regime the leading eigenvalue is an outlier and   the critical connectivity $c_{\ast}$  is  independent of $v^2_J$.   This implies that fluctuations do not affect the system stability when the leading eigenvalue is an outlier.  On the other hand, in the gapless regime the leading eigenvalue is part of the boundary of the continuous spectrum and the critical connectivity $c_{\ast}$ decreases as $1/v^2_J$.    In this regime, fluctuations in the interaction strengths render the system less stable.  The differences between the gapped and gapless regimes can    be understood in terms of the nature of the destabilizing mode.     In the gapped regime, the mode that destabilizes the system is ferromagnetic, i.e., $\langle L_1\rangle>0$, whereas in the gapless regime, the mode that destabilizes the system is spin-glass-like, i.e., $\langle L_1\rangle=0$.      Hence, increasing the fluctuations $v_J$ for  fixed values of the mean strength  $\alpha$ does not affect the ferromagnetic mode, which gives an intuitive understanding why the location of the outlier is independent of $v_J$.

We can quantify the overall stability of systems coupled through random matrices (\ref{eq:model}) in terms
  of a single parameter $a_{\rm stab}$, defined as the area in figure \ref{fig5} where the system is stable and $c(1+\rho) > 1$.
  The quantity $a_{\rm stab}$ is given by
  \begin{eqnarray}
    a_{\rm stab} &=& \frac{1}{\alpha} \sqrt{\frac{ 1- \alpha}{\alpha} } \left( 1 - \sqrt{\alpha (1 + \alpha) }   \right) \nonumber \\
    &+& \frac{1}{\alpha^2} \left[ \tanh^{-1}{\left( \sqrt{\frac{1 - \alpha^2  }{\alpha^2}  }   \right) } - \tanh^{-1}{\left( \frac{1 - \alpha}{ \alpha} \right) } \right]. \nonumber
  \end{eqnarray}
  The area $a_{\rm stab}$ is a monotonic decreasing function of $\alpha$, which approaches $a_{\rm stab} \rightarrow 0$ as $\alpha \rightarrow 1$ and
  $a_{\rm stab} \rightarrow \infty$ as $\alpha \rightarrow 0$. Thus, the increase of the average interaction strength between the elements
  of a network system, in the sense that $\langle J \rangle$ approaches $d$, makes the system less stable.

In Fig.~\ref{fig6} we present the  phase diagram for $\alpha<0$ or equivalently $\langle J\rangle<0$.     Since in this case the outlier is negative,  the critical connectivity is
 \begin{eqnarray}
  c_\ast = \left\{ \begin{array}{ccc}
1/[\alpha^2(v^2_J+1)], && v^2_J\in [0, v^2_\ast],   \\ 
 \setminus && v^2_J > v^2_\ast \end{array} \right. \label{eq:critConn1}
  \end{eqnarray}
 Note that for small values of $v^2_J$ the system is more stable in the case of negative $\langle J \rangle$  since then there exists no outlier that renders the system less stable. 
  
  Finally, we discuss how the phase diagrams, given by Figs.~\ref{fig5} and \ref{fig6}, are  modified by the presence of small cycles in the network.    As illustrated in Fig.~\ref{fig:IllustrationSketch} and discussed in Sec.~\ref{sec:lead}, there is a finite, albeit small, probability $\nu$ that the leading eigenvalue $\lambda_1$ is larger than $\lambda^{\ast}$.     This happens when a random directed graph contains a cycle that generates a strong enough feedback loop.   As a consequence, one should interpret the phase diagrams Figs.~\ref{fig5} and \ref{fig6} as describing the typical behaviour of dynamical systems defined on random directed graphs in the limit $n\rightarrow \infty$.    There is however a small nonzero  probability that a random directed graph contains a cycle that  destabilizes the system through the feedback loop that it generates.

 \section{Numerical examples on matrices of finite size} \label{sec:examples}
  In this section, we  compare  theoretical results for infinitely large matrices with direct diagonalization results on  matrices of  finite size~$n\sim O(10^3)$.  Such numerical experiments  reveal   the magnitude of  finite size effects,  which are important for applications because  real-world systems are finite. Moreover, this comparison allows us to better  understand the potential  limitations of the theory.

 Since a nonzero $d$ results in   a constant shift of  all eigenvalues by $-d$, i.e.~$\lambda_j\rightarrow  \lambda_j - d$, we set  in all examples   
\begin{equation}
A_{jj} = d=0 , \quad  \forall j\in[n].
\end{equation}     

The numerical experiments are designed as follows. First, we use    the algorithm  presented in  Appendix~\ref{sec:finite} to  sample a matrix  from a random-matrix ensemble  of the type given by  Eq.~(\ref{eq:model}).   Second, we use the subroutine {\it gsl\_eigen\_nonsymmv} from the GNU Scientific Library to compute  the $n$ eigenvalues  of the sampled matrix and  the entries of their  right eigenvectors.   
Third, in order to test the  theory in  Sec.~\ref{eq:theory}, we compute for each matrix sample  $\mathbf{A}$
the  leading eigenvalue $\lambda_1(\mathbf{A})$, the real part  of the subleading eigenvalue $\lambda_2(\mathbf{A})$,  and the 
observable 
\begin{eqnarray}
\mathcal{R}_1(\mathbf{A}) = \frac{\sum^{n}_{j=1} R_{1,j}(\mathbf{A}) }{\sqrt{\sum^{n}_{j=1}|R_{1,j}(\mathbf{A})|^2 }}  ,  \label{eq:mathcalR}
\end{eqnarray}  
which quantifies the mean value of the components of the right eigenvector  associated with $\lambda_1(\mathbf{A})$.   Before we  compute  $\mathcal{R}_1(\mathbf{A})$ with the above equation,  we rotate  all the  elements  $R_{1,j}(\mathbf{A})$ by a constant phase   $e^{i \theta}$, such that the empirical mean $\sum^n_{j=1}R_{1,j}(\mathbf{A})$ is a positive real number, in accordance with our conventions in Eq.~(\ref{eq:convR}).  
Lastly, we compute the   mean values $\overline{\lambda}_1$, $\overline{\lambda}_2$, and
$\overline{\mathcal{R}_1}$  of  the sampled populations, together with the standard deviations for each quantity.    Empirical mean values for, say $\overline{\lambda}_1$, are  compared with either  the theoretical ensemble averages $\langle \lambda_1 \rangle$ or with the typical value of $\lambda_1$ provided by the deterministic outlier $\lambda_{\rm isol}$ or the by the boundary $|\lambda_{\rm b}|$ of the continuous part of the spectrum.     Note that we use the notation $\langle \lambda_1 \rangle$ for theoretical ensemble averages, while $\overline{\lambda_1}$ is used for  empirical mean values over the sampled populations, which forms an estimate of $\langle \lambda_1 \rangle$.

The present section is organized into three subsections.
 In  Sec.~\ref{subSec:rhoN},  we consider     adjacency matrices of directed random graphs with negative  degree correlations ($\rho<0$) and unweighted links ($J_{ij}=1$).    For this ensemble, we have derived in Sec.~\ref{sec:unweightedGraphs} exact results for the statistics of the leading eigenvalue $\lambda_1$ in the limit $n\rightarrow \infty$.   Hence, we expect  a good correspondence between theory and experiment in all parameter regimes.  Deviations between theory and experiment will  be due to finite size effects and finite sampling statistics only.   
 
 In   Sec.~\ref{subSec:rhoP},  we consider  the  adjacency matrices of directed random graphs with positive  degree correlations ($\rho>0$) and weighted links.   For this ensemble, we have derived in Sec.~\ref{sec:weighted} exact results for the typical value of $\lambda_1$ in the regime  $c(\rho+1)>1$  Hence, we  expect  in this regime a good correspondence between theory and experiment, and  deviations between theory and experiment will  be due to finite size effects,  finite sampling statistics, and because of deviations between the mean and typical value of $\lambda_1$.   
 
In Sec.~\ref{subSec:powerLaw}, we apply the theoretical results of Sec.~\ref{eq:theory}  to  adjacency matrices of  random  directed graphs with power-law degree distributions, which have diverging moments.   Since the graphs are unweighted, the theory of Sec.~\ref{sec:unweightedGraphs}  applies.     However,  since for power-law random graphs the tails of the degree distributions decay very slowly, we expect  to observe deviations between theory and experiment, and in Sec.~\ref{subSec:powerLaw} we test  the limitations of the theory for power-law random graphs.

    Lastly, in Sec.~\ref{sec:pr}, we test the predictions given by Eqs.~(\ref{eq:pRb}) and (\ref{eq:pRIsol})  for the number of zero-valued entries in the right eigenvector  $\vec{R}_1$. 

\subsection{Adjacency matrices of  unweighted and directed random graphs with negative degree correlations}   \label{subSec:rhoN} 
      
\begin{figure*}[t]
\centering
 \hspace{-0.5cm}
{\includegraphics[width=0.4\textwidth]{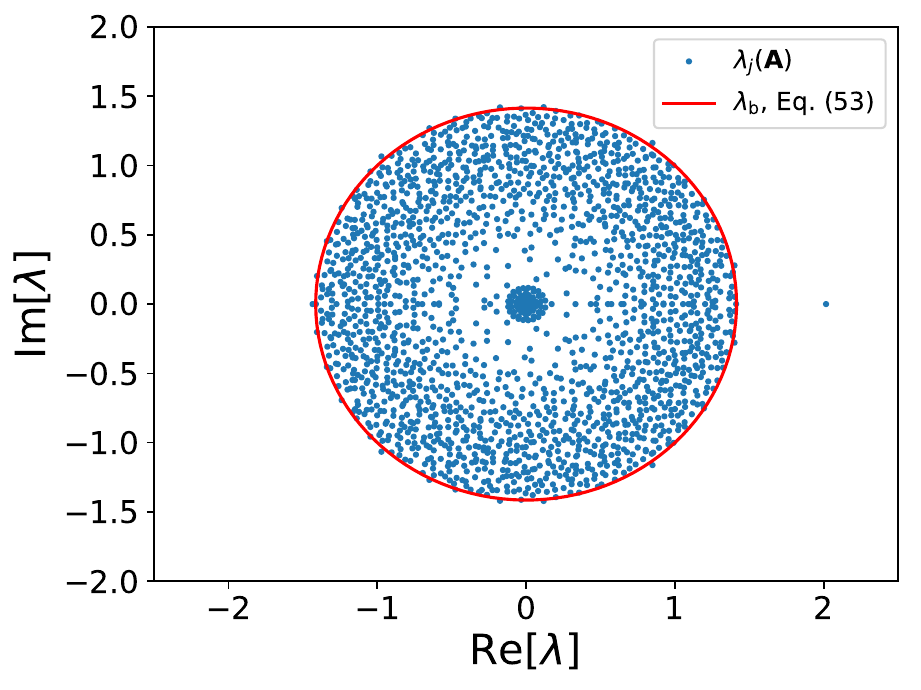}\label{fig:adjneg1}   \put(-25,32){\Large (a)}} 
{\includegraphics[width=0.4\textwidth]{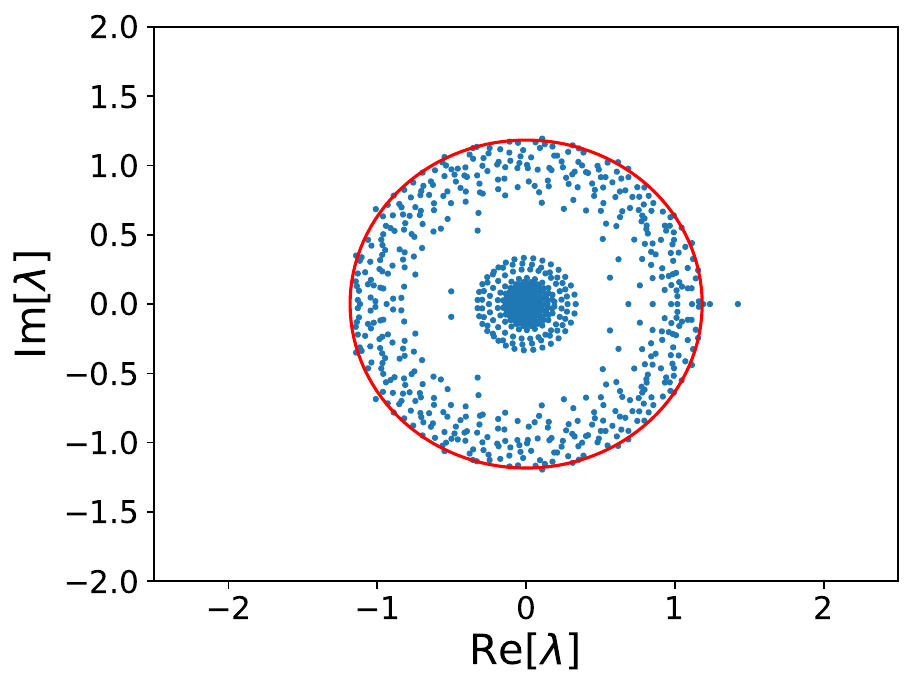}\label{fig:adjneg2}    \put(-25,32){\Large (b)}}
{\includegraphics[width=0.4\textwidth]{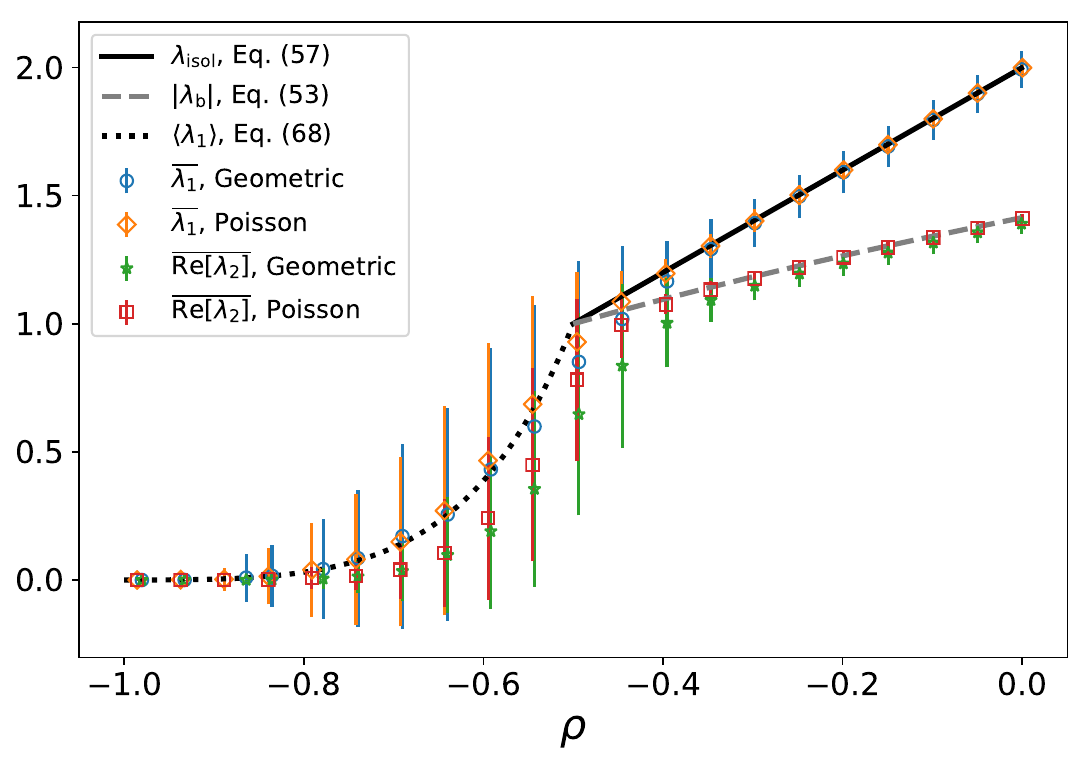}\label{fig:adjneg3}\put(-25,30){\Large (c)}} 
{\includegraphics[width=0.4\textwidth]{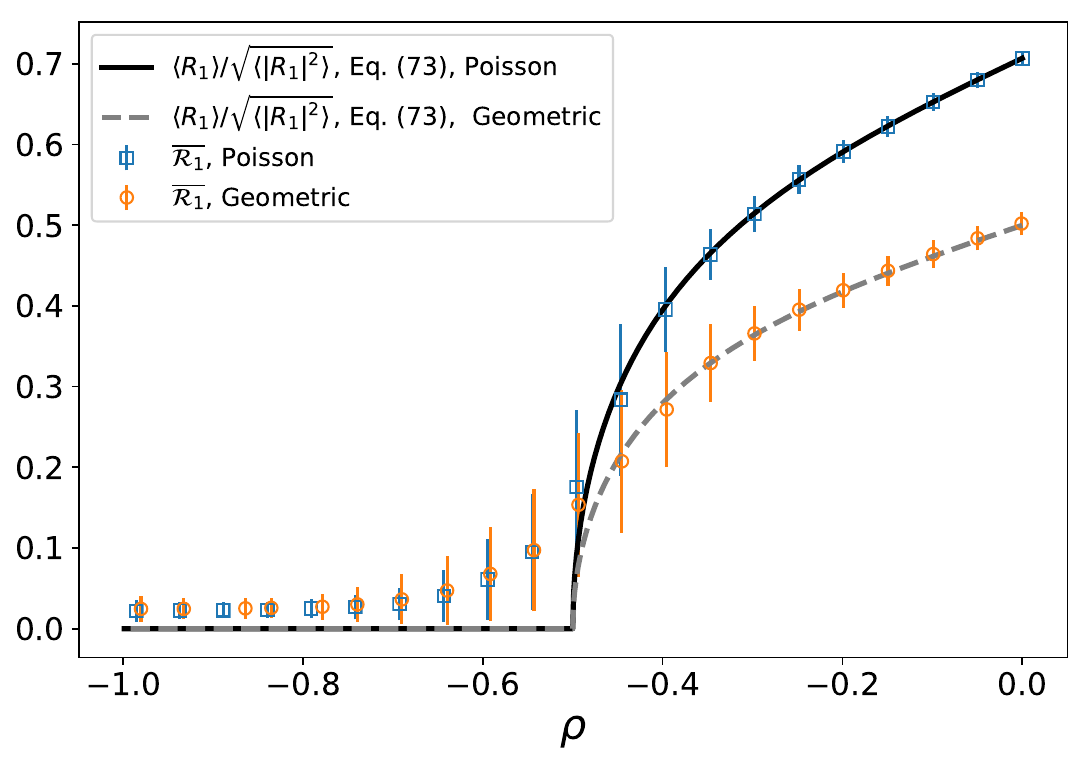}\label{fig:adjneg4} \put(-25,30){\Large (d)}} 
\caption{{\it  Effect of negative $\rho$ on the spectral properties of  the adjacency matrices of random directed graphs.}      Spectral properties for   the adjacency matrices of Poissonian [see Eq.~(\ref{eq:neg1})]  or geometric [see Eq.~(\ref{eq:neg2})]  random directed graphs    with a mean degree $c=2$  and negative $\rho$ are presented. 
  Direct diagonalization results for  matrices of size $n=4000$ (markers) are compared  with the theoretical results for infinitely large matrices  (lines) derived in  Sec.~\ref{eq:theory}.      Panels (a) and (b): eigenvalues $\lambda_j(\mathbf{A})$  of the adjacency matrices of two  Poissonian random graphs with $\rho=0$ [Panel (a)] and $\rho=-0.3$ [Panel (b)], respectively, are plotted and compared with the theoretical boundary $\lambda_b$ for the spectrum given by Eq.~(\ref{eq:boundary}).    Panel (c): Mean values of the leading eigenvalue  $\overline{\lambda}_1$  and real part of the subleading eigenvalue  $\overline{{\rm Re}[\lambda_2]}$ are plotted as a function of $\rho$ and compared with theoretical results $\lambda_{\rm isol} = 2(\rho+1)$  and  $|\lambda_{\rm b}| = \sqrt{2  (\rho+1)}$ if $\rho>-0.5$ and $\langle \lambda_1 \rangle  = 1- [1-c (\rho +1)] e^{c (\rho +1)}  $ if $\rho<-0.5$.
Panel (d):   Mean value $\overline{\mathcal{R}_1}$ for the entries of the right eigenvector associated with the leading eigenvalue are plotted as a function of $\rho$ and compared with the theoretical results  $\frac{\langle R_1\rangle}{\sqrt{\langle |R_1|^2\rangle}} = \sqrt{ \frac{1 + 2\rho }{ 2+\rho   -2\rho^2   } } $ and   $\frac{\langle R_1 \rangle }{\sqrt{\langle R^2_1 \rangle}} = \sqrt{\frac{1+2\rho }{   2 (2  +\rho - 2\rho^2)} }$ for the Poissonian and  geometric ensemble, respectively, when $\rho\geq-0.5$, and with $\frac{\langle R_1\rangle}{\sqrt{\langle |R_1|^2\rangle}} = 0$ when $\rho<-0.5$.    In   Panels (c) and (d),  direct diagonalization results are  the sample means    over $1000$  matrix realizations  and error bars represent the sample standard deviations.    
    } \label{fig1}
\end{figure*}  

 We consider the adjacency matrices of  Poissonian random graphs --- also called  Erd\H{o}s-R\'{e}nyi random graphs --- and 
 geometric random graphs with negative degree correlation coefficient $\rho\in[-1,0]$ and with constant weights $J_{ij}=1$.
 
 For {\it Poissonian} random graphs, the prescribed degree distribution is given by
\begin{eqnarray}
\lefteqn{p_{K^{\rm in},K^{\rm out}}(k,\ell) = (1+\rho) \: p_{\rm p}(k;c)  p_{\rm p}(\ell;c)} \quad   \quad&&
\nonumber\\ 
&& -  \frac{\rho}{2} \left[ \delta_{k,0}\:  p_{\rm p}(\ell;2c) +   \delta_{\ell,0} \: p_{\rm p}(k;2c)    \right], \label{eq:neg1}
\end{eqnarray}
where $k,\ell\in\left\{0,1,\ldots,n-1\right\}$ and where   
\begin{eqnarray}
p_{\rm p}(k;c) = \frac{1}{\mathcal{N}_{\rm p}}\frac{c^{k}}{k!}, \label{eq:pP}
\end{eqnarray}
with   $\mathcal{N}_{\rm p} = \sum^{n-1}_{k=0}c^k/k!$  the normalization constant.   For $n\rightarrow \infty$,   $p_{\rm p}(k;c)$ is the Poisson distribution  with mean degree $c$ and  $\mathcal{N}_{\rm p} =e^c$.   
For {\it geometric} random graphs, the prescribed degree distribution is given by
\begin{eqnarray}
\lefteqn{p_{K^{\rm in},K^{\rm out}}(k,\ell) = (1+\rho) \: p_{\rm g}(k;c)  p_{\rm g}(\ell;c)} &&
\nonumber\\ 
&& -  \frac{\rho}{2} \left[ \delta_{k,0} \: p_{\rm g}(\ell;2c) + \delta_{\ell,0} \: p_{\rm g}(k;2c)   \right], \label{eq:neg2}
\end{eqnarray}
where $k,\ell\in\left\{0,1,\ldots,n-1\right\}$ and where 
\begin{eqnarray}
p_{\rm g}(k;c) =  \frac{1}{\mathcal{N}_{\rm g}}\left(\frac{c}{1+c}\right)^k,  \label{eq:expP}
\end{eqnarray}  
with $\mathcal{N}_{\rm g} = \sum^{n-1}_{k=0}\left(\frac{c}{1+c}\right)^k$  the normalization constant.
For  $n\rightarrow \infty$,  $p_{\rm g}(k;c) $ is the geometric distribution with mean degree $c$ and $\mathcal{N}_{\rm g} = c+1$.

Throughout this subsection, we consider unweighted  graphs for which  $J_{jk}=1$ for all $j \neq k$, and thus
\begin{eqnarray}
p_J(x) = \delta(x-1).
\end{eqnarray}

  In Fig.~\ref{fig1}, we shows how the   degree correlation coefficient $\rho$ affects the spectral properties of  adjacency matrices of directed random graphs with mean degree $c=2$.   We compare the theoretical results given by Eqs.~(\ref{eq:outlier}), (\ref{eq:boundary}), (\ref{eq:R}), (\ref{eq:meanlambda1}), (\ref{eq:R1}) and (\ref{eq:lambdaSG}) with direct diagonalization results. 
       
  In the Panels (a) and (b) of Fig.~\ref{fig1}, we provide a global picture of the spectra of adjacency matrices of Poissonian random graphs by  comparing the spectra of matrices  with $\rho=0$  and    $\rho=-0.3$.     We observe how negative degree correlations contract the spectrum: for $\rho=-0.3$ the  leading eigenvalue is smaller  than for $\rho=0$, and the spectrum concentrates around the origin when $\rho$ is more negative.

  In Panel (c) of Fig.~\ref{fig1},  we present a more detailed  analysis of the behaviour of the leading eigenvalue $\lambda_1$ and  the subleading eigenvalue $\lambda_2$ as a function of $\rho$.    As discussed in Sec.~\ref{eq:theory},  for $c(\rho+1)>1$, the leading  and subleading eigenvalues are self-averaging and given by   $\lambda_1 = \lambda_{\rm isol}$ and ${\rm Re}[\lambda_2] = |\lambda_{\rm b}|$, respectively.    These findings are  well corroborated by the numerical results   in Fig.~\ref{fig1}(c).    We observe that $\lambda_1 = \lambda_2$ at the critical percolation threshold $\rho= -1+1/c= -0.5$, as predicted by the theory.     For  $c(\rho+1)<1$, there does not exist a giant strongly connected component, see Sec.~\ref{sec:sizeConn}, and therefore the leading eigenvalue is either $0$ or $1$, depending on whether the graph contains an oriented ring or not.   In this regime, the mean value $\langle \lambda_1\rangle$ is given by Eq.~(\ref{eq:meanlambda1}) and its variance is given by Eq.~(\ref{eq:Var}), both findings which are well corroborated by numerical results in Fig.~\ref{fig1}(c).      
       
  In Fig.~\ref{fig1}(d), we present a systematic study of the first moment $\langle R_1 \rangle$ of the eigenvector $\vec{R}_1$  associated with the leading eigenvalue, which is an outlier for $\rho\geq -0.5$.  The theoretical result  Eq.~(\ref{eq:R}) is well  corroborated by direct diagonalization results for the empirical observable $\overline{\mathcal{R}_1}$, defined in Eq.~(\ref{eq:mathcalR}).      We observe a phase transition from a  phase with $\langle  R_1 \rangle = 0$, for $\rho < -0.5$, to a  phase with $\langle  R_1 \rangle > 0$, for $\rho > -0.5$.  Note that $\langle  R_1 \rangle = 0$ for $\rho < -0.5$ since in this regime there exists no giant SCC, and therefore the right eigenvector is localized on an isolated component with a finite number of nodes.

 Taken together, the results in Fig.~\ref{fig1} illustrate how  the leading eigenvalue of the adjacency matrix of  a random  directed graph increases as a function of $\rho$.    These results imply that  one can reduce $\lambda_1$  significantly reduced through a rewiring procedure that decreases correlations between indegrees and outdegrees.   These results  are in agreement with the phase diagram in Fig.~\ref{fig5}, which shows that     dynamical systems coupled through  graphs with  negative  $\rho$ are more stable than  those coupled through graphs with positive $\rho > 0$.

\subsection{Adjacency matrices of  weighted and directed random graphs with positive degree correlations}  \label{subSec:rhoP}    
We analyze the spectral properties of  the adjacency matrices of  Poissonian and geometric  random graphs with positive $\rho$ and random weights.   

  The {Poissonian ensemble with positive $\rho$ has a prescribed degree distribution
\begin{eqnarray}
p_{K^{\rm in},K^{\rm out}}(k,\ell) = (1-c\rho) p_{\rm p}(k)  p_{\rm p}(\ell)  
 + c\rho \:p_{\rm p}(\ell) \delta_{k, \ell}, \nonumber\\ \label{eq:posAss}
\end{eqnarray}
where $\rho\in[0,1/c]$, and where $p_{\rm p}$ is the truncated  Poisson distribution defined by Eq.~(\ref{eq:pP}).  The geometric ensemble  with positive $\rho$ has  the prescribed degree distribution  
\begin{eqnarray}
\lefteqn{p_{K^{\rm in},K^{\rm out}}(k,\ell)} && \nonumber\\  
&&  = \left(1- \frac{c \rho}{c+1}\right) p_{\rm g}(k)  p_{\rm g}(\ell)  
 +  \frac{c \rho}{c+1} \:p_{\rm g}(\ell) \delta_{k, \ell},  \nonumber\\ \label{eq:posAss2}
\end{eqnarray}
where $\rho\in[0,(c+1)/c]$, and $p_{\rm g}$ is the truncated geometric distribution defined by Eq.~(\ref{eq:expP}).    

The off-diagonal matrix entries $J_{jk}$ are i.i.d.~random variables drawn either from   a Gaussian distribution
\begin{eqnarray}
  p_J(x) = \frac{1}{\sqrt{2\pi v^2}} e^{-\frac{(x-\mu_0)^2}{2v^2}}, \label{eq:gauss}
  \label{gauss}
\end{eqnarray}
 or from a bimodal distribution 
\begin{eqnarray}
  p_J(x) =  b \delta(x-x_0) + (1-b)\delta(x+x_0),
  \label{bim}
\end{eqnarray}
with the parametrization $x_0 = \sqrt{\mu_0^2+v^2}$ and $2b = 1 + \mu_0/x_0$. In each case, the parameters  $\mu_0$ and $v$ denote, respectively, the mean and the standard
deviation of the distribution $p_J(x)$.

\begin{figure*}[t]
\centering
 \hspace{-0.5cm}
{\includegraphics[width=0.4\textwidth]{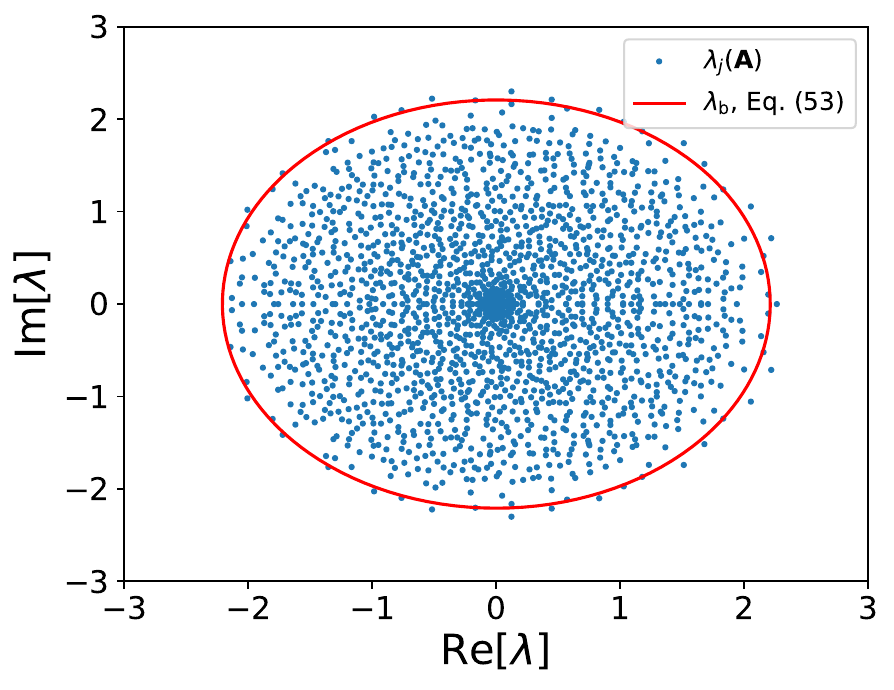}\label{fig:2a} \put(-25,32){\Large (a)}} 
{\includegraphics[width=0.4\textwidth]{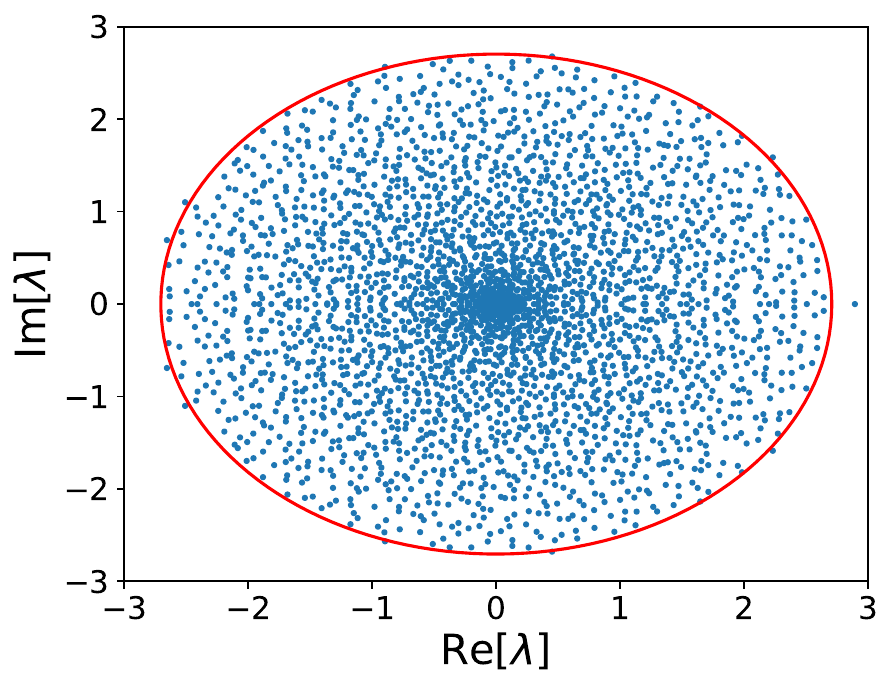}\label{fig:2b} \put(-25,32){\Large (b)}}
{\includegraphics[width=0.4\textwidth]{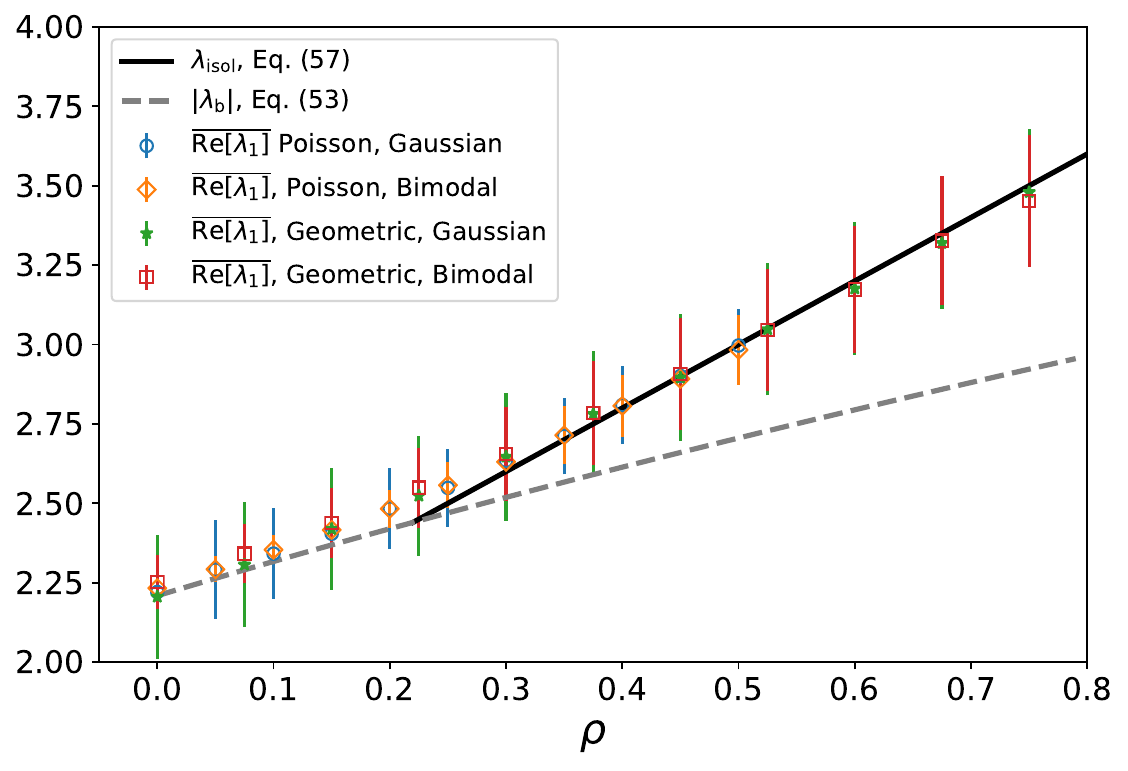}\label{fig:2c} \put(-25.5,29){\Large (c)}} 
{\includegraphics[width=0.4\textwidth]{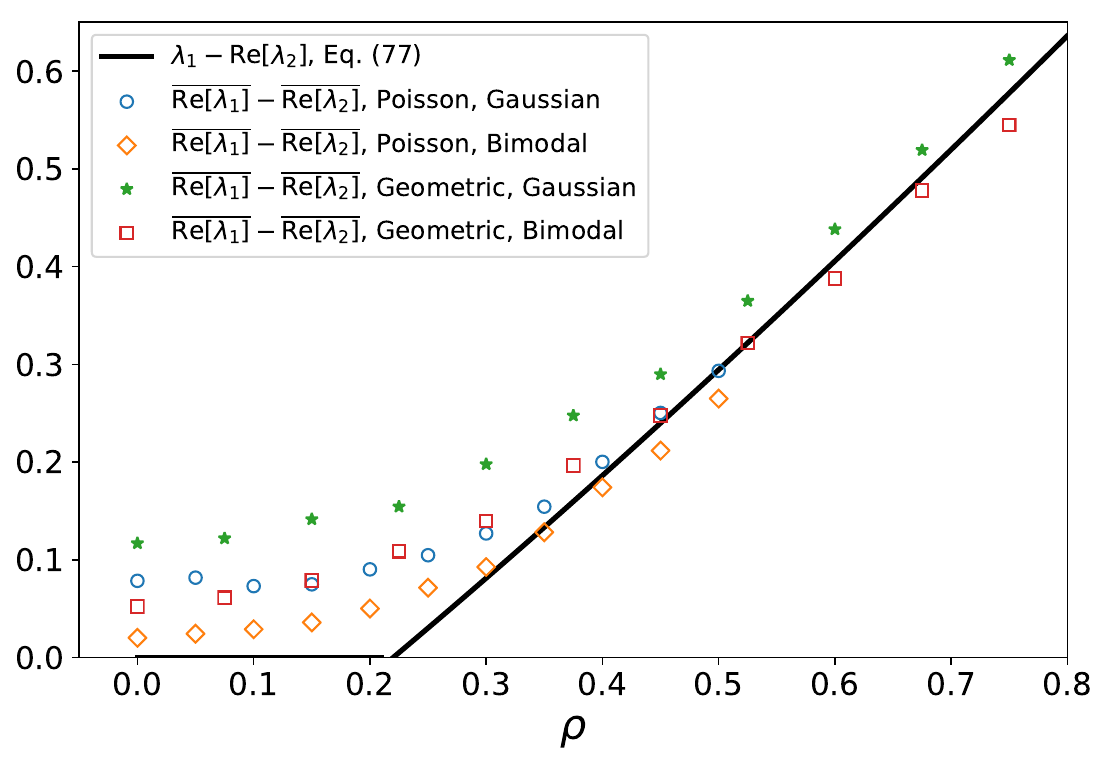}\label{fig:2d} \put(-28,29){\Large (d)}} 
{\includegraphics[width=0.4\textwidth]{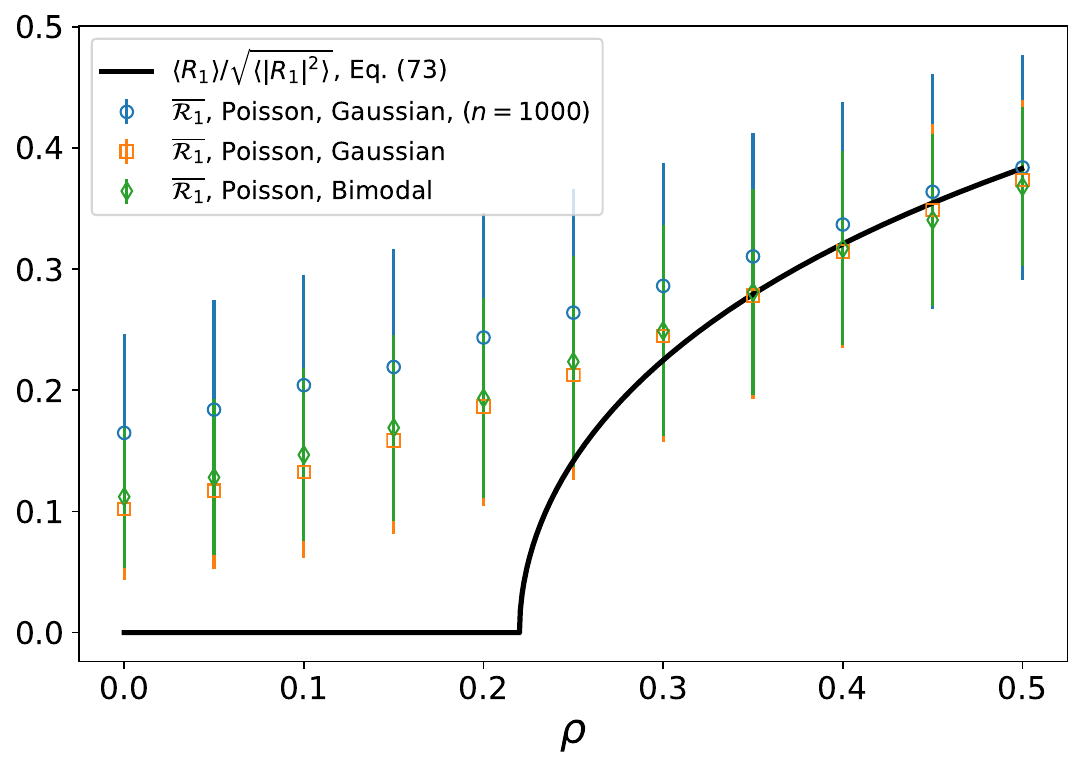}\label{fig:2e} \put(-24,29){\Large (e)}} 
{\includegraphics[width=0.4\textwidth]{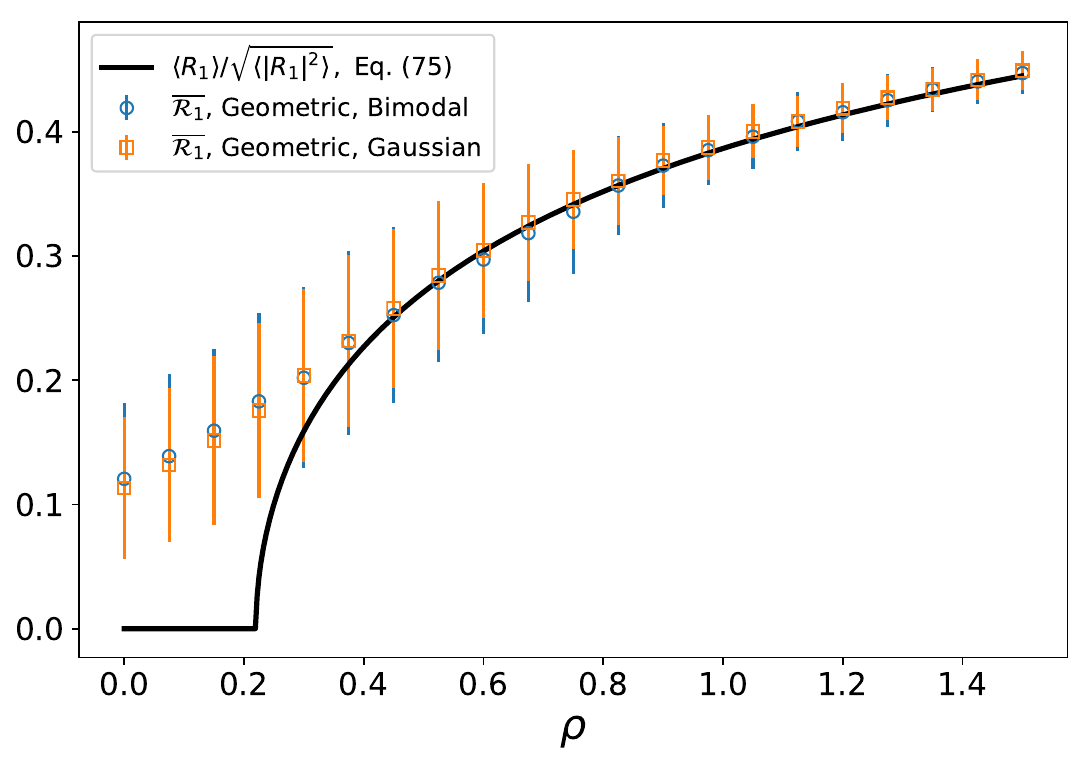}\label{fig:2f} \put(-23,29){\Large (f)}} 
\caption{ {\it Effect of positive $\rho$ on the spectral properties of  adjacency matrices of weighted, random, directed graphs.}     
Spectral properties for   the adjacency matrices of Poissonian [see  Eq.~(\ref{eq:posAss})]  or geometric [see Eq.~(\ref{eq:posAss2})]  random directed graphs    with a mean degree $c=2$ and positive $\rho$  are presented.     The off-diagonal weights are drawn from a Gaussian or a bimodal distribution with mean $\mu_0 = 1$ and    
   standard deviation  $v = 1.2$ (see Eqs.~(\ref{gauss}) and (\ref{bim})).   
   Direct diagonalization results of  matrices of size $n=4000$ (markers) are compared  with  theoretical results for $n \rightarrow \infty$  (lines), presented in Sec.~\ref{eq:theory}.       Panels (a) and (b): eigenvalues $\lambda_j(\mathbf{A})$  of the adjacency matrices of two  Poissonian random graphs with $\rho=0$ [Panel (a)] and $\rho=0.5$ [Panel (b)], respectively, are plotted and compared with the theoretical boundary $\lambda_b$ for the spectrum given by Eq.~(\ref{eq:boundary}).
Panels (c)-(f): the sample means for the leading eigenvalue $\overline{{\rm Re}[\lambda_1]}$,  the spectral gap $\overline{{\rm Re}[\lambda_1]} - \overline{{\rm Re}[\lambda_2]}$ and the first moment of the right eigenvector $\overline{\mathcal{R}_1}$ are plotted as a function of $\rho$ and compared with the theoretical expressions $\lambda_1$,  $\lambda_1 - {\rm Re}[\lambda_2] $ and $\langle R_1\rangle$ derived in Sec.~\ref{eq:theory}.   Sample means are  over  $1000$ matrix realizations  of size $n=4000$ (except for the blue circles in Panel (e), which are for $n=1000$). The error bars denote sample standard deviations. } \label{fig2}
\end{figure*}  
 \begin{figure*}[t]
\centering
 \hspace{-0.5cm}
{\includegraphics[width=0.4\textwidth]{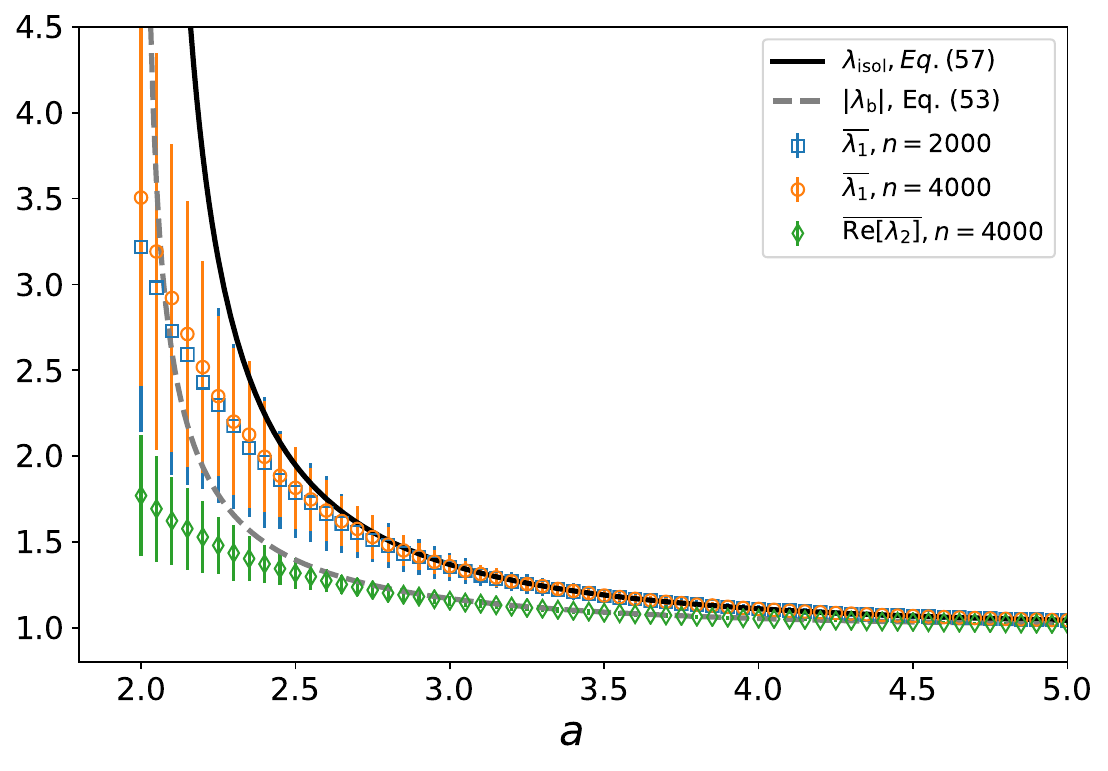}\label{fig:4a} \put(-26,37){\Large (a)}} 
{\includegraphics[width=0.4\textwidth]{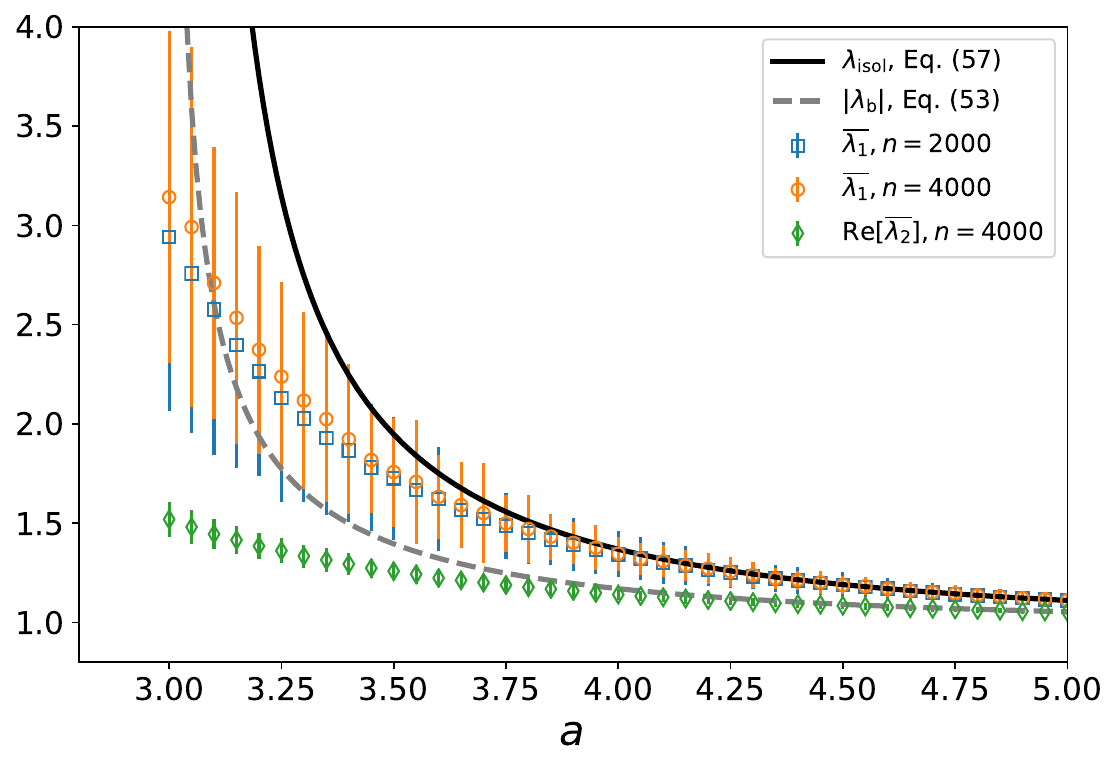}\label{fig:4b} \put(-28,37){\Large (b)}}
{\includegraphics[width=0.4\textwidth]{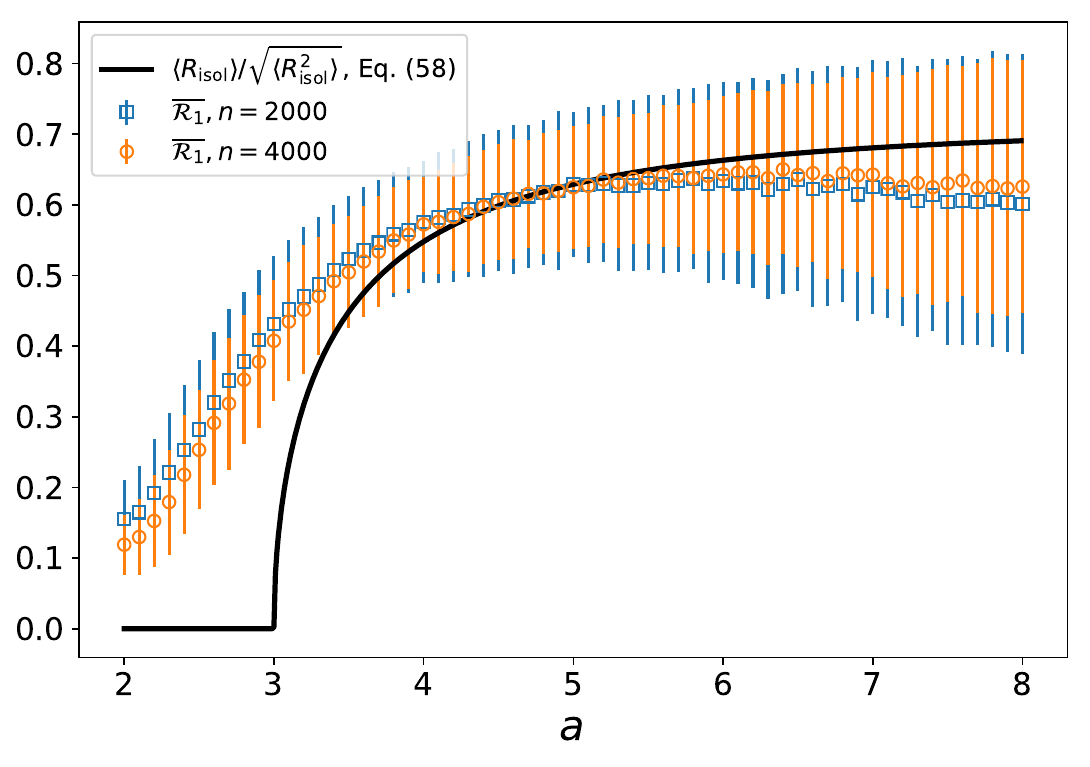}\label{fig:4c} \put(-25,32){\Large (c)}} 
{\includegraphics[width=0.4\textwidth]{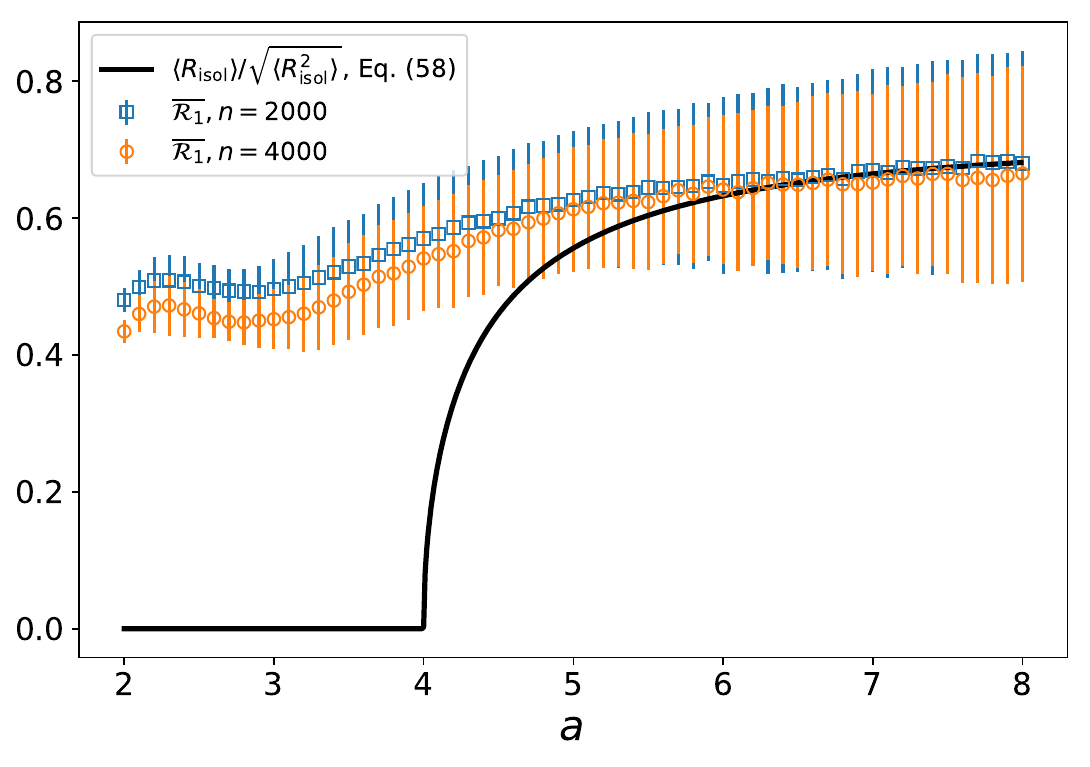}\label{fig:4d} \put(-25,32){\Large (d)}} 
\caption{ 
{\it Leading and subleading eigenvalues for adjacency matrices of power-law random graphs with prescribed degree distributions.} 
Panels (a) and (b):     Results for the leading eigenvalue $\overline{\lambda}_1$ and the real part of the subleading eigenvalue $\overline{{\rm Re}[\lambda_2]}$  are presented as a function of the exponent $a$ of power-law random graphs with degree distributions given by either Eq.~(\ref{eq:ensemblePow1}) [Panel (a)], for which $\rho=0$, or by Eq.~(\ref{eq:ensemblePow2}) [Panel (b)], for which $\rho>0$.    Direct diagonalization results (markers) in Panel (a) and (b) are compared with  the theoretical results (lines) given by $|\lambda_{\rm b}|^2 = \lambda_{\rm isol} = \zeta(a-1)/\zeta(a)$ and  $|\lambda_{\rm b}|^2 = \lambda_{\rm isol} = \zeta(a-2)/\zeta(a-1)$, respectively.  Panels (c) and (d):  
Results for the mean values $\overline{\mathcal{R}_1}$ and $\overline{\mathcal{R}_2}$  of the entries of right eigenvectors associated with the leading and subleading eigenvalue, respectively, are presented as a function of the exponent $a$ for power-law random graphs with degree distributions given by either Eq.~(\ref{eq:ensemblePow1}) [Panel (c)], for which $\rho=0$, or by Eq.~(\ref{eq:ensemblePow2}) [Panel (d)], for which $\rho>0$.      In Panel(c), direct diagonalization results are compared with $\frac{\langle R_{\rm isol} \rangle}{\sqrt{\langle R^2_{\rm isol} \rangle}}= \sqrt{\frac{\zeta(a-1)[\zeta(a-1)-\zeta(a)]}{\zeta(a)[\zeta(a-2) - \zeta(a-1)]}}$ if $a> 3$, and with $\frac{\langle R_{\rm isol} \rangle}{\sqrt{\langle R^2_{\rm isol} \rangle}}= 0$ if $a<3$.   In Panel (d), direct diagonalization results are compared with    $\frac{\langle R_{\rm isol} \rangle}{\sqrt{\langle R^2_{\rm isol} \rangle}} = \sqrt{\frac{ \zeta(a-1)\zeta(a-2)  [\zeta(a-2)- \zeta(a-1)]}{ \zeta(a) \zeta(a-1)\zeta(a-3) -  \zeta^2(a-1) \zeta(a-2)  +[\zeta(a-1)-\zeta(a)]   \zeta^2(a-2)  }}$, if $a>4$, and with $\frac{\langle R_{\rm isol} \rangle}{\sqrt{\langle R^2_{\rm isol} \rangle}}= 0$ if $a<4$.    
In all  four panels weights are set equal to $J_{jk} = 1$ and  markers are  sample means over either $2000$ or $1000$  matrices of size $n=2000$  or $n=4000$, respectively.
   Error bars denote  standard deviations over the population of different matrix realizations.      } \label{fig4}
\end{figure*} 

In Fig.~\ref{fig2}, we  analyze how positive values of $\rho$ affect the spectral properties of adjacency matrices of randomly weighted directed   graphs. We compare the spectral properties for different values of $\rho$ and fixed  parameters  $c=2$,  $\mu_0=1$, and $v = 1.2$. We
compare theoretical results from Sec.~\ref{eq:theory} (lines) with  direct diagonalization results for matrices of size $n=4000$ (markers).   
  
In Panels (a) and (b) of Fig.~\ref{fig2}, we provide a global picture of  the spectra  of Poissonian random graphs by comparing the spectrum  of a graph without degree correlations ($\rho=0$, Panel (a)) with the spectrum of a graph with positive degree correlations ($\rho = 0.5$, Panel (b)). In the latter case, the correlations  are perfect in the sense that $K^{\rm in}_j = K^{\rm out}_j$  for each node $j$.   The direct diagonalization results corroborate well  the formula Eq.~\ref{eq:boundary} for the boundary of the continuous part of the spectrum. We also observe that the leading eigenvalue $\lambda_1(\mathbf{A})$ increases as a function of $\rho$, that  $\lambda_1(\mathbf{A})$ is located at the boundary $\partial \sigma_{\rm ac}$ for $\rho=0$ [Panel (a)],  and that   $\lambda_1(\mathbf{A})$ is an outlier for $\rho=0.5$ [Panel (b)].
  
   In the Panels (c) and (d) of Fig.~\ref{fig2}, we provide a more detailed view of  the eigenvalues $\lambda_1$ and $\lambda_2$ as a function of $\rho$.      We  observe that both $\overline{\lambda_1}$ and $\overline{\lambda_2}$ are monotonically increasing functions of $\rho$, and that there is a continuous transition from a gapless phase for $\rho< \langle J^2\rangle/(c\langle J\rangle^2)  -1 \approx 0.22$ to a gapped phase for $\rho> \langle J^2\rangle/(c\langle J\rangle^2)  -1$.   We observe that the values of  $\overline{\lambda_1}$ and $\overline{\lambda_2}$ are universal, in the sense that they depend  on the  distributions $p_J$ and $p_{K^{\rm in},K^{\rm out}}$ only  through the parameters $c$, $\rho$, $\langle J\rangle$ and $\langle J^2\rangle$.  Theoretical results are  well corroborated with direct diagonalization results, although finite size effects  are more significant for the spectral gap.

   Lastly,  Panels (e) and (f) of Fig.~\ref{fig2}  compare the theoretical result $\langle R_1\rangle$ of Sec.~\ref{eq:theory} with the sampled average $\overline{\mathcal{R}_1}$ of the quantity $\mathcal{R}_1$, as defined in Eq.~(\ref{eq:mathcalR}).      In the gapless phase, we have $\langle R_1 \rangle = 0$, while in the gapped phase we obtain $\langle R_1 \rangle >  0$, which is   reminiscent of a continuous phase transition between a spin-glass phase and  a ferromagnetic phase.    
   We observe that in the gapped phase  direct diagonalization results are in very good agreement with the theoretical expressions for infinitely large matrices,  whereas in the gapless phase there are significant deviations between theory and direct diagonalization results.    These deviations are due to  finite size effects, which are significant because of our convention to normalize the eigenvectors with  Eq.~(\ref{eq:convR}).
   In spite of that, we observe that direct diagonalization results slowly converge to the theoretical values as the matrix size $n$ increases.

Overall, we  conclude that the theoretical results for the typical values of  $ \lambda_1$, $ \lambda_2$, and $\langle R_1 \rangle$, presented in Sec.~\ref{eq:theory}, describe well the numerically  estimates for their ensemble average.     This is because in the regime   $c(\rho+1)>1$ it is unlikely that a stochastic outlier eigenvalue exists.     
 
 \subsection{Adjacency matrices of random graphs with  power-law degree distributions}  \label{subSec:powerLaw} 
 In this subsection, we  analyze the spectral properties of  the adjacency matrices of  random directed graphs with power-law degree distributions.   
Power-law random graphs are  interesting from a practical point of view, since degree distributions of real-world systems  often have  tails that are fitted well by power-law distributions~\cite{amaral2000classes, RevModPhys.74.47, dunne2002food, clauset2009power}.   For example, the 
 World Wide Web is a directed graph with a power-law degree distribution of the form $p_{K^{\rm in},K^{\rm out}}(k,\ell)\sim  k^{-2.1}\ell^{-2.7}$  \cite{broder2000graph}.   Since  power-law  degree distributions have diverging moments, these ensembles exhibit strong finite size effects and large sample-to-sample fluctuations, and it is thus interesting to test the possible limitations of the theory in Sec.~\ref{eq:theory}  for power-law random graphs. 
   
 We consider two classes  of  power-law random graphs, namely, an ensemble  without correlations between indegrees and outdegrees ($\rho=0$), and an ensemble  with perfect  degree correlations, where  $K^{\rm in}_j = K^{\rm out}_j$ for all nodes $j$  ($\rho > 0$).  The ensemble without degree correlations has a  prescribed degree distribution  
  \begin{eqnarray}
 p_{\rm deg}\left(k,\ell \right) =\frac{k^{-a}\ell^{-a} }{\mathcal{N}^2_{\rm pow}},  \label{eq:ensemblePow1}
 \end{eqnarray} 
with $k,\ell\in[n-1]$ and with  $\mathcal{N}_{\rm pow} = \sum^{n-1}_{k=1}k^{-a}$, while the ensemble with perfect degree correlations has the prescribed degree distribution  
   \begin{eqnarray}
 p_{\rm deg}\left(k,\ell\right) = \frac{k^{-a}}{\mathcal{M}_{\rm pow}}  \delta_{k,\ell}, \label{eq:ensemblePow2}
 \end{eqnarray} 
   with $k,\ell\in[(n-1)/2]$ and with  $\mathcal{M}_{\rm pow} = \sum^{(n-1)/2}_{k=1}k^{-a}$ the normalization constant. The parameter $a$ controls how fast the degree distribution decays
   for large degrees.

 \begin{figure*}[t]
\centering
 \hspace{-0.5cm}
{\includegraphics[width=0.4\textwidth]{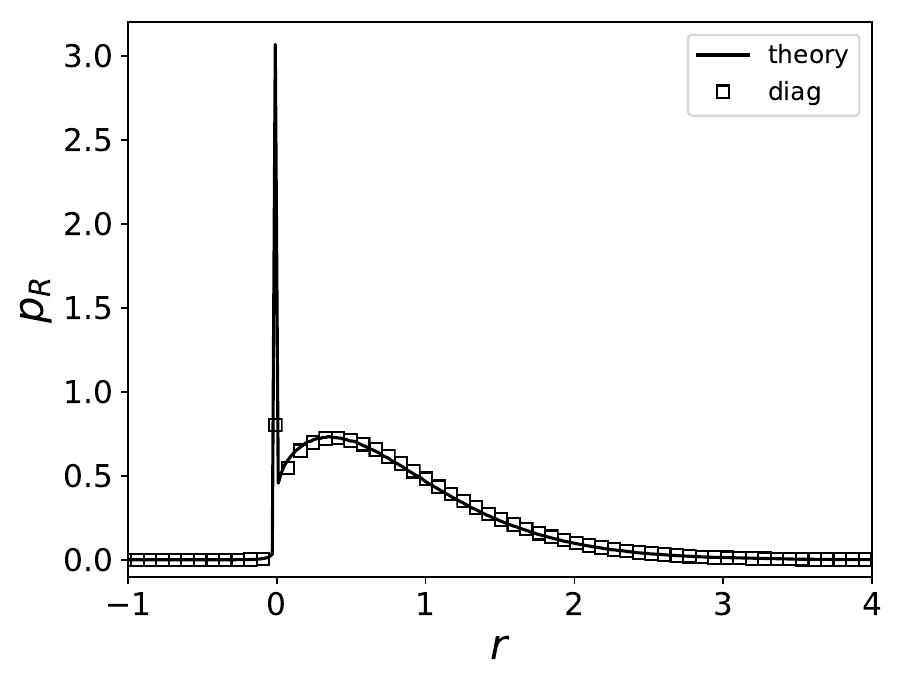} \put(-25,36){\Large (a)} \label{fig:eigv}} 
{\includegraphics[width=0.4\textwidth]{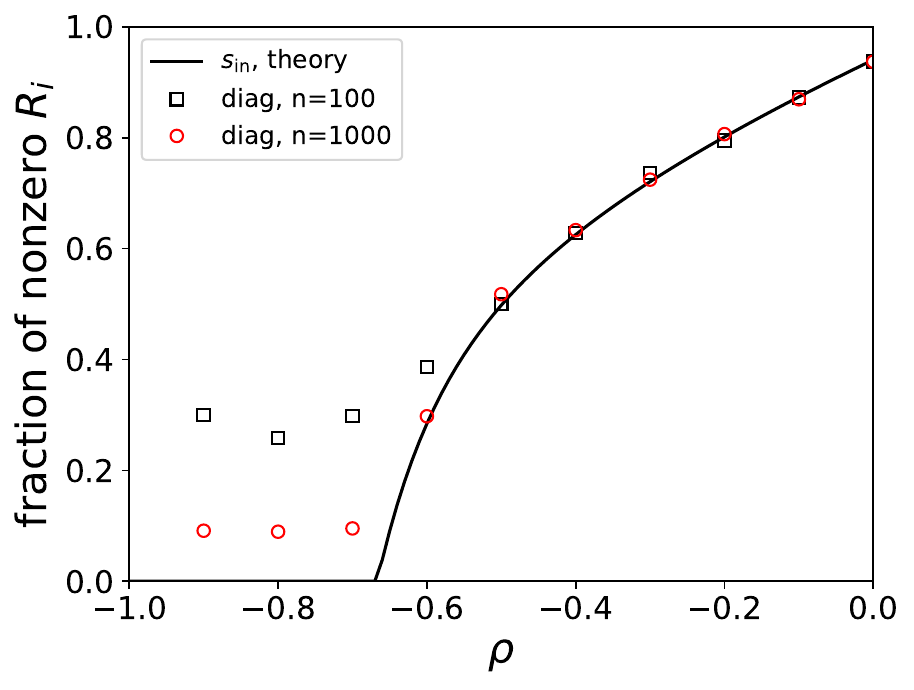} \put(-29,36){\Large (b)} \label{fig:sout}} 
\caption{ {\it Properties of the distribution $p_{R_1}$.}   Results are for the adjacency matrices of random directed graphs with a Poisson degree distribution  given by Eq.~(\ref{eq:neg1}) and with mean $c=3$.     Edges are weighted by random couplings $J_{ij}$ drawn from a Gaussian distribution [see Eq.~(\ref{eq:gauss})] with  mean $\mu_0 = 1$ and variance $v^2 = 0.2$.     Panel (a):    Theoretical results for $p_R$ solving Eqs.~(\ref{eq:pRecR}-\ref{eq:qRecR}) with $\lambda=\lambda_{\rm isol}$   (solid lines) are compared with a histogram of the entries of $\vec{R}_1$ obtained from  direct diagonalizing $2e+4$ matrices   of size $n=1000$ (markers).   The degree correlation coefficient $\rho=0$.   Panel (b):  Fraction of nonzero entries of  the leading right eigenvector $\vec{R}_1$  as a function of the degree-correlation coefficient $\rho$.   Direct diagonalization results  for matrices of size $n=100$ and $n=1000$ (markers are sample averages over $100$ and $20$ samples, respectively) are compared with  theoretical results for $s_{\rm in}$ (solid line)  obtained from solving Eqs.~(\ref{eq:sout}) and (\ref{eq:bb}).   In the numerical experiments, we have used the criterion $|R_i|^2<1e-20$    to identify a zero-valued entry. 
  } \label{figeigv}
\end{figure*} 

We discuss the values of the parameters $c$ and $\rho$ in the limit $n\rightarrow \infty$.  
If $a>2$,  then the mean degree is given by 
   \begin{eqnarray}
   c = \zeta(a-1)/\zeta(a),
   \end{eqnarray}
   with $\zeta(x)$  the Riemann zeta function.   Also, if $a>2$, then the ensemble of Eq.~(\ref{eq:ensemblePow1}) has a degree-correlation coefficient
       \begin{eqnarray}
    \rho = 0,
       \end{eqnarray}
while  if $a>3$, then the ensemble of Eq.~(\ref{eq:ensemblePow2}) has a degree-correlation coefficient
       \begin{eqnarray}
\rho = \frac{\zeta(a-2)\zeta(a)}{\zeta^2(a-1)} -1.
       \end{eqnarray}  
       Note that $c(\rho+1)>1$, and therefore the power-law graphs we consider have a giant SCC.  
 
 We consider unweighted power-law random graphs with $J_{jk}=1$ for all $j,k \in [n]$.

We now resort to direct diagonalization in order to gain a better understanding of the statistics of the  leading eigenvalue 
of power-law random graphs.    
In  Panel (a) of Fig.~\ref{fig4}, we plot  the sample mean $\overline{\lambda_1}$ of the  leading eigenvalue $\lambda_{1}(\mathbf{A})$ and  the sample mean $\overline{{\rm Re}[\lambda_2]}$ of the real part of the subleading eigenvalue   $\lambda_{2}(\mathbf{A})$ as a function of $a$ in the ensemble defined by Eq.~(\ref{eq:ensemblePow1}) for which $\rho=0$.      We observe that for   $a\gtrsim 3$  the theoretical expressions Eqs.~(\ref{eq:outlier}) and (\ref{eq:boundary}) for $\lambda_{\rm isol}$ and $|\lambda_{\rm b}|$,  respectively, are in very good agreement with direct diagonalization results for the leading and subleading eigenvalue. In the regime $a \lesssim 3$, we  observe significant deviations between theory and numerical experiments.      Such deviations are expected, since $c\rightarrow \infty$  for $a\rightarrow 2^+$, and therefore the theoretical expressions for $\lambda_{\rm isol}$ and $|\lambda_{\rm b}|$  diverge for $a\rightarrow 2^+$.   Analogously, in Panel (b) of Fig.~\ref{fig4},   we present results  for $\overline{\lambda_1}$ and $\overline{{\rm Re}[\lambda_2]}$ as a function of $a$ for the  ensemble defined by Eq.~(\ref{eq:ensemblePow2}) for which $\rho > 0$.    In this case, the theory works well when $a\gtrsim 4$, whereas for $a \lesssim 4$ we   observe significant deviations between theory and numerical experiments.    This is because for $a\rightarrow 3^+$ the degree correlation coefficient $\rho$ diverges, and therefore the theoretical expressions for $\lambda_{\rm isol}$ and $|\lambda_{\rm b}|$ also diverge. Overall, these results show that the  Eqs.~(\ref{eq:outlier}) and (\ref{eq:boundary}) work remarkably well for power-law random graphs.

Lastly, in Panels (c) and (d) of Fig.~\ref{fig4}, we plot the empirical mean $\overline{\mathcal{R}_1}$  as a function of $a$ and compare results from the direct diagonalization of  randomly sampled matrices   with 
 the theoretical expression for $\langle R_{\rm isol}\rangle$ given by  Eq.~(\ref{eq:R}).      We observe  a reasonable good agreement between theoretical results and numerical experiments, considering that power-law random graphs exhibit significant finite-size effects and fluctuations.    Interestingly,  the normalized mean  $\langle R_{\rm isol}\rangle/\sqrt{\langle R^2_{\rm isol}\rangle}$  vanishes at $a = 3$ and $a = 4$ for  ensembles with degree distributions (\ref{eq:ensemblePow1}) and  (\ref{eq:ensemblePow2}), respectively.
Since the Perron-Frobenius theorem applies to this ensemble, this is a transition from a delocalized phase ($\langle R_{\rm isol}\rangle/\sqrt{\langle R^2_{\rm isol}\rangle}>0$) to a localized phase ($\langle R_{\rm isol}\rangle/\sqrt{\langle R^2_{\rm isol}\rangle} = 0$), as argued in Sec.~\ref{sec:Perron}.    In other words, the leading eigenvector is {\it localized} when the exponent $a$ that characterizes the decay of the power-law degree distribution is small enough.

\subsection{Distribution $p_R$  for the entries of the right eigenvectors} \label{sec:pr} 
So far, we have studied the mean value of the distribution $p_R$.    Differently, in   this section we analyze properties of the full distribution $p_R$. 

Equations  (\ref{eq:pRb}) and (\ref{eq:pRIsol}) state that the distribution $p_R$ contains a delta peak at the origin with weigth  $1-s_{\rm in}$, where  $s_{\rm in}$ is the relative size of the IN component.   In other words, the number of nonzero entries in a right eigenvector is equal to the size of the IN component.  
Figure~\ref{figeigv} tests this prediction for the adjacency matrix of a directed random graph with a Poissonian degree distribution given by Eq.~(\ref{eq:neg1}) with $c=3$.

In Panel (a) of  Fig.~\ref{figeigv}, we compare theoretical predictions for $p_R$, obtained by solving the recursive distributional Eqs.~(\ref{eq:pRecR}-\ref{eq:qRecR}) at $\lambda=\lambda_{\rm isol}$ through a population dynamics algorithm \cite{abou1973selfconsistent, mezard2001bethe, metz2010localization, neri2016eigenvalue}, with a histogram of the entries of the right eigenvector associated with the leading eigenvalue $\lambda_1$, obtained  through direct diagonalization results.    We have set $\rho=0$ and the couplings $J_{ij}$ are drawn from a Gaussian distribution.   In Fig.~\ref{figeigv}, we observe an excellent agreement between theory and numerical experiments and we also observe  a delta peak at the origin, which is clearly discernible in both theory and numerical experiments.   

In order to  quantify the  weight of the delta peak at the origin, we plot in Panel (b) of Fig.~\ref{figeigv} the fraction of  entries  $R_i$ that are not equal to zero.   We compare direct diagonalization results for  right eigenvectors associated with the leading eigenvalue $\lambda_1$ with the  theoretical expression  $s_{\rm in}$ obtained by solving Eqs.~(\ref{eq:sout}) and (\ref{eq:bb}).     We find again an excellent agreement between theory and numerical experiments, confirming that the   number of nonzero elements in $\vec{R}$ equals the size $s_{\rm in}$ of the IN component

 \section{Extensions of the theory}\label{sec:ext}  
 Here we extend the theory from Sec.~\ref{sec:rec}  to the case of  random matrices with diagonal disorder and graphs that  contain nondirected links.          
  \subsection{Random matrices with diagonal disorder}
We consider  random matrices of the form
\begin{eqnarray}
 \bA =  - \mathbf{D} +  \bJ\circ\bC, \label{eq:modelxx}
 \end{eqnarray} 
where  $\bJ$ and $\bC$ are defined in exactly the same way as in Eq.~(\ref{eq:model}), but where  $\mathbf{D}$ is now  a diagonal matrix with   entries  $[\mathbf{D}]_{jj} = D_j$ that are i.i.d.~random variables drawn from a probability distribution $p_D(x)$ with $x\in\mathbb{R}^+$.    Note that   $p_D$ has a support on the positive real axis since otherwise ${\rm Re}[\lambda_1]>0$ and the dynamical system described by $\mathbf{A}$ will not be stable.      In the special case when 
 \begin{eqnarray}
 p_D(x) = \delta(x-d), \label{eq:delta}
 \end{eqnarray}
 we recover  the model given by Eq.~(\ref{eq:model}).

The theory of Sec.~\ref{sec:rec} applies to the model  given by Eq.~(\ref{eq:modelxx}) after  some minor modifications.
As shown in Appendix~\ref{sec:der2}, for the present model the distribution $p_R$  solves the recursion relation
 \begin{eqnarray}
 \lefteqn{p_R(r) = \sum^{\infty}_{k=0} \sum^{\infty}_{\ell=0}p_{K^{\rm in}, K^{\rm out}}(k,\ell)    \int {\rm d}y \:p_D(y) }&& 
 \nonumber\\ 
&& \int \prod^{\ell}_{j=1}{\rm d}^2 r_j  q_R(r_j) \int \prod^{\ell}_{j=1}{\rm d} x_j  p_J(x_j) \delta \left[r - \frac{\sum^{\ell}_{j=1}x_j r_j }{\lambda+y}\right], \nonumber\\ \label{eq:pRecRD}
 \end{eqnarray} 
 and $q_R$ solves  the recursion relation
  \begin{eqnarray}
 \lefteqn{q_R(r) = \sum^{\infty}_{k=0} \sum^{\infty}_{\ell=0}p_{K^{\rm in}, K^{\rm out}}(k,\ell)     \frac{k}{c}  \int {\rm d}y \:p_D(y)}&& 
 \nonumber\\ 
&& \int \prod^{\ell}_{j=1}{\rm d}^2 r_j  q_R(r_j) \int \prod^{\ell}_{j=1}{\rm d} x_j  p_J(x_j) \delta \left[r - \frac{\sum^{\ell}_{j=1}x_j r_j }{\lambda+y}\right].\label{eq:qRecRD} \nonumber\\
 \end{eqnarray}  
If  $D$ is deterministic, then Eq.~(\ref{eq:delta}) holds and we recover the recursion Eqs.~(\ref{eq:pRecR}) and (\ref{eq:qRecR}).
 
 In Appendix~\ref{sec:der2}, we derive the values of $\lambda$ for which the recursion  Eqs.~(\ref{eq:pRecRD}) and (\ref{eq:qRecRD}  admit normalizable solutions.  In this way, we obtain that the deterministic outliers of  $\mathbf{A}$ solve the equation 
  \begin{eqnarray}
 c(\rho+1)\langle J \rangle\Big \langle \frac{1}{\lambda_{\rm isol}+D}\Big\rangle  = 1, \label{eq:lambdabD}
 \end{eqnarray} 
  and the eigenvalues $\lambda_{\rm b}\in\mathbb{C}$ at the boundary of the continuous part of the spectrum solve  \begin{eqnarray}
 c(\rho+1)\langle J^2 \rangle\Big \langle \frac{1}{|\lambda_{\rm b}+D|^2}\Big\rangle  = 1. \label{eq:IsolD}
 \end{eqnarray}     
Using Eqs.~(\ref{eq:lambdabD}) and (\ref{eq:IsolD}),  it is possible to derive phase diagrams for  the stability  of dynamical systems with disorder in the decay rates $D_j$, similar to  those presented in Figs.~\ref{fig5} and \ref{fig6}.   We  leave this open for future studies.

      \subsection{Nondirected graphs with random couplings}
We consider   random matrices of the form
\begin{eqnarray}
 \bA_n =  -d\:\mathbf{1}_n +  \tilde{\bJ}_n\circ\tilde{\bC}_n , \label{eq:modelxxxxx}
 \end{eqnarray} 
 where  $\tilde{\bC}_n$ is  the adjacency matrix of an nondirected random graph with a prescribed degree distribution $p_{\rm deg}(k)$, and where  $\tilde{\mathbf{J}}_n$ is   a  random matrix with zero entries on the diagonal and with  offdiagonal  pairs $(\tilde{J}_{jk},\tilde{J}_{kj})$ that are  i.i.d.~random variables  with a distribution $p_{\tilde{J}_1, \tilde{J}_2}(x,y)$.  We assume that  $p_{\tilde{J}_1, \tilde{J}_2}$ satisfies  the symmetry property 
 \begin{equation}
 p_{\tilde{J}_1, \tilde{J}_2}(x,y) = p_{\tilde{J}_1, \tilde{J}_2}(y,x).
 \end{equation}
 The  random matrix model defined by Eq.~(\ref{eq:modelxxxxx})  is locally tree-like, but it is in general  not  locally oriented.
 Nevertheless, locally oriented ensembles can be recovered in the limiting case 
      \begin{equation}
     p_{\tilde{J}_1, \tilde{J}_2}(x,y) = \frac{1}{2}p_{J}(x)\delta(y) + \frac{1}{2}p_{J}(y)\delta(x). 
     \end{equation}
     In this case,   the model given by Eq.~(\ref{eq:modelxxxxx}) is the adjacency matrix of a directed random graph with  a joint degree distribution 
 \begin{eqnarray}
 \lefteqn{ p_{K^{\rm in}, K^{\rm out}}(k, \ell)  }&&  \nonumber \\
  && = \sum^{\infty}_{m=0}  \frac{p_{\rm deg}(m)}{2^m} \sum^{m}_{n=0}\frac{m!}{n!(m-n)!} \delta_{k, m}\delta_{\ell, m-k},  
 \end{eqnarray}
 which is a special case of the model defined by  Eq.~(\ref{eq:model}).      On the other hand, if  
       \begin{equation}
       p_{\tilde{J}_1, \tilde{J}_2}(x,y) = \delta(x-y)p_J(x), \label{eq:symrand}
     \end{equation} 
then Eq.~(\ref{eq:modelxxxxx}) defines symmetric random matrices.

In Appendix~\ref{sec:der2}, we derive a set of recursion relations for $p_R$.     We obtain that the  distribution $p_R$ is the marginal 
\begin{eqnarray} 
p_{R}(r) = \int {\rm d}^2 g   \: p_{G,R}(g,r)  
  \end{eqnarray}   
 of the joint distribution $p_{G,R}(g,r)$ that solves the recursion relation
   \begin{eqnarray} 
\lefteqn{p_{G,R}(g,r)   =   \sum^{\infty}_{k=0}p_{{\rm deg}}(k)  } && 
\nonumber \\  
&& \times \int \prod^{k}_{\ell=1} {\rm d}^2g_{\ell}{\rm d}^2r_{\ell} \: q_{R, G}(r_{\ell},g_{\ell})  \int \prod^{k}_{\ell=1} {\rm d}x_{\ell} \: {\rm d}y_{\ell}  \:p_{J_1, J_2}(x_{\ell},y_{\ell}) \nonumber \\ 
\nonumber\\ 
&& \times \delta\left(r  + \frac{\sum^{k}_{\ell=1}x_\ell r_{\ell}}{-\lambda - d - \sum^{k}_{\ell=1}x_\ell g_{\ell} y_\ell}\right) \nonumber\\
&& \times  \delta\left(g -  \frac{1}{-\lambda - d -\sum^{k}_{\ell=1}x_\ell g_{\ell} y_\ell  }\right),\label{eq:distriO1}
 \end{eqnarray}  
and $q_{G,R}$  solves the equation
 \begin{eqnarray} 
\lefteqn{q_{G,R}(g,r)   =   \sum^{\infty}_{k=0} \frac{k\: p_{{\rm deg}}(k)}{c}} &&  
\nonumber \\  
&& \times \int \prod^{k-1}_{\ell=1} {\rm d}^2g_{\ell}{\rm d}^2r_{\ell} \: q_{R, G}(r_{\ell},g_{\ell})  \int \prod^{k-1}_{\ell=1} {\rm d}x_{\ell} \: {\rm d}y_{\ell}  \:p_{J_1, J_2}(x_{\ell},y_{\ell}) 
\nonumber\\ 
&& \times \delta\left(r  +  \frac{\sum^{k-1}_{\ell=1}x_\ell r_{\ell}}{-\lambda - d - \sum^{k-1}_{\ell=1}x_\ell g_{\ell} y_\ell }\right) \nonumber\\
&& \times  \delta\left(g -  \frac{1}{-\lambda -d- \sum^{k-1}_{\ell=1}x_\ell g_{\ell} y_\ell  }\right).
\nonumber\\  \label{eq:distriO2}
 \end{eqnarray}   
Note that in the special case of symmetric random matrices [i.e., when Eq.~(\ref{eq:symrand}) holds], Eqs.~(\ref{eq:distriO1}-\ref{eq:distriO2}) are equivalent to those derived in Refs.~\cite{kabashima2010cavity, kabashima2012first, takahashi2014fat, susca2019top}.

The outliers $\lambda_{\rm isol}$ and the boundary $\lambda_b$ of the continuous part of the spectrum are found as values of $\lambda$ for which the relations (\ref{eq:distriO1}-\ref{eq:distriO2}) admit normalizable solutions.  In the present case, we do not know how to derive compact analytical expressions for $\lambda_{\rm isol}$ and  $\lambda_b$.   However, Eqs.~(\ref{eq:distriO1}-\ref{eq:distriO2}) can  be solved numerically  with a population dynamics algorithm, as described in Refs.~\cite{abou1973selfconsistent, mezard2001bethe, metz2010localization}, and consequently a stability phase diagram as in Figs.~\ref{fig5} and \ref{fig6} can be derived. We leave such a study open for future work.

    \section{Discussion}\label{sec:discu} 
 Random matrices  appear in  linear stability analyses of large dynamical  systems.   So far, most studies have considered dynamical systems for which the system constituents interact  with either  a  number of degrees of freedom that increases with system size, see e.g.~Refs.~\cite{may1972will,sear2003instabilities, rajan2006eigenvalue,  allesina2012stability, grilli2016modularity, ahmadian2015properties, PhysRevLett.114.088101, PhysRevE.93.022302, kuczala2016eigenvalue, gibbs2018effect, PhysRevE.100.032307, haas2019subpopulations}, or interact through a one-dimensional chain, see e.g.~Refs~\cite{hatano1996localization,amir2016non, zhang2019eigenvalue}.  However, real-world systems often consist of components interacting through large  networks,  see e.g.~Refs.~\cite{barrat2008dynamical, newman2010networks, barthelemy2018spatial, dorogovtsev2013evolution, barabasi2016network}.   Therefore, an  interesting  question  is    how  network topology affects system stability.

 In this paper, we have  analysed the linear stability of large dynamical systems defined on  random  directed  graphs  with a prescribed degree distribution  $p_{K^{\rm in},K^{\rm out}}$, which serve as a model for, among others, the World Wide Web~\cite{broder2000graph, pastor2007evolution} and neural networks~\cite{brunel2000dynamics, amari2003handbook, sporns2010networks}.    We have shown that 
  dynamical systems defined on  random  directed  graphs   are more stable than systems defined on nondirected graphs.    Indeed, we have shown that for random directed graphs the leading eigenvalue  is with probability one finite in the limit of infinitely large $n$.     This results brings an interesting perspective in the diversity-stability debate~\cite{mccann2000diversity}.    Dynamical systems defined on dense matrices or nondirected graphs are unstable when $n$ is large enough: in the former because $\lambda_1$  is of the order $O(\sqrt{n})$ \cite{may1972will}, while in the latter because $\lambda_1$  is of the order $O(\sqrt{k_{\rm max}})$~\cite{krivelevich2003largest, chung2004spectra, susca2019top}.    Hence,  a large and complex systems will be in general unstable \cite{gardner1970connectance, may1972will}.   However,  if the system is defined on a random directed graph, then  it can be infinitely large and stable since $\lambda_1$ converges to a finite limit for large $n$.  The stabilising nature of  random directed graphs is a consequence of their locally tree-like and oriented structure, which implies that there exist no feedback loops that amplify local perturbations.

A second surprising result is that the stability of dynamical systems defined on directed random graphs exhibits  a universal character, in the sense that it   is governed by only three network parameters:  the effective mean degree  $c(\rho+1)$, the coefficient of variation $v_J = \sqrt{\langle J^2\rangle -\langle J\rangle^2}/\langle J\rangle$, and the ratio $\alpha = \langle J \rangle/d$ between the mean interaction strength and the decay rate.   This result follows from the analytical expression, given by Eq.~(\ref{eq:lambda1}), for the typical value of the  leading eigenvalue of the adjacency matrix that encodes the  network of interactions  between the  system constituents.        From the analytical expression for the typical value of the leading eigenvalue, we obtain the  universal  phase diagrams of Figs.~\ref{fig5} and \ref{fig6}.  
   
 Analyzing these phase diagrams, we obtain the following interesting conclusions on how  network topology affects system stability.  First,  negative correlations between   indegrees and outdegrees stabilize large dynamical systems, whereas the   mean coupling strength $\alpha$  and the coupling fluctuations $v_J$  render dynamical systems less stable.    Second, when the fluctuations $v_J$ of the coupling strengths are small enough, then  the stability is controlled by an outlier and  is independent of $v_J$.   On the other hand, when $v_J$ is large enough, then the leading eigenvalue is determined by the boundary of the continuous part of the spectrum and the system stability decreases as a function of $v_J$.    Moreover, in the first scenario, the unstable mode is ferromagnetic ($\langle R \rangle>0$) whereas in the second scenario it is spin-glass-like ($\langle R\rangle=0$).   Lastly,   systems with coupling fluctuations  $v_J$ larger than  the critical value $v_\ast = \sqrt{\frac{1-\alpha^2}{\alpha^2}}$ do not contain a stable phase, no matter how large the negative correlations between  indegrees and outdegrees are.

The universal phase diagrams of Figs.~\ref{fig5} and \ref{fig6}  have been derived with a mathematical method, akin to the cavity method in statistical physics, which computes the typical value of the  leading eigenvalue of  random directed graphs that have a giant SCC in the limit of $n\rightarrow \infty$.    The cavity method computes the typical value of $\lambda_1$ by neglecting contributions of cycles of finite length.    However, if the graph  contains disorder in the weights $J_{ij}$, then the leading eigenvalue is not a self-averaging quantity and there exists a finite (albeit) small probability that the leading eigenvalue comes from a cycle of finite length that is part of the graph, as sketched in Fig.~\ref{fig:IllustrationSketch}.       Hence, short cycles can destabilize  large dynamical systems defined on random directed graphs when they induce strong enough  feedback loops.  

The derived theoretical results for the spectra of  large sparse non-Hermitian random matrices may also be useful for applications other than the linear stability analysis of large dynamical systems described by differential equations.        For example, the theory is also useful to study the stability of dynamical systems in discrete time \cite{hastings1982may},  which are relevant for the study of systemic  risk in networks of banks connected through financial contracts~\cite{bardoscia2017pathways, PhysRevE.100.032307}.    For discrete-time systems, the stability is controlled  by the spectral radius $r(\mathbf{A}) = {\rm max}\left\{|\lambda_1|, |\lambda_2|,\ldots, |\lambda_n| \right\}$.
Another example of an application is the analysis of  spectral algorithms that use the    right or left eigenvector associated with the (sub)leading eigenvalue to obtain information about a system, e.g.,   spectral clustering algorithms \cite{pentney2005spectral, krzakala2013spectral},  centrality measures based on  eigenvectors \cite{bonacich2001eigenvector, langville2011google, ermann2015google}, or algorithms for the low-rank matrix estimation problem~\cite{lesieur2017constrained, bordenave2020detection}. Detectability thresholds of spectral algorithms often depend on the location of the leading and subleading eigenvalue  \cite{krzakala2013spectral, bordenave2015non,zdeborova2016statistical,  kawamoto2018algorithmic, bordenave2020detection}.   A fourth example of an application is the analysis of stochastic processes with  spectra of Laplacian or Markov matrices \cite{mohar1991laplacian, staring2003random, kuhn2015spectra, agliari2017exact,margiotta2019glassy}:  the stationary state of a Markov process is the right (or left) eigenvector associated with the leading  eigenvalue of a Markov matrix \cite{margiotta2019glassy},  the relaxation time is provided by the spectral gap \cite{levin2017markov, monthus2009eigenvalue, agliari2013immune},  and the  cumulant generating function of a time-additive observable can be expressed in terms of the leading eigenvalues of a  sequence of Markov matrices \cite{donsker1975asymptotic, donsker1975asymptotic2, donsker1976asymptotic3, donsker1983asymptotic4, de2016rare}.   A fifth application is the study of nonHermitian  quantum mechanics on random graphs \cite{hatano1996localization, feinberg1999non, goetschy2011non}, which is currently an active research field.    Lastly, we remark that the subleading eigenvalue, and its associated right (left) eigenvector,  provide not only information about the asymptotic stability of large dynamical systems, but also about their response to random perturbations as shown in Ref.~\cite{PhysRevResearch.2.023333}.   Taken together, we conclude that the  spectral theory presented in this paper can be used in  various contexts.    
   
    The  theoretical results obtained in this paper are conjectures about the spectral properties of directed random matrices.  
Reference \cite{bordenave2020detection}  provides a mathematical proof for Eqs.~(\ref{eq:meanlambda1}) and (\ref{eq:lambdaSG}) for the leading and subleading eigenvalues in the special case of   directed Erd\H{o}s-R\'{e}nyi graphs with $J_{ij}=1$.    To our knowledge, there exist no proofs of Eqs.~(\ref{eq:boundary}) and (\ref{eq:outlier}) for the deterministic outlier eigenvalue $\lambda_{\rm isol}$ and the boundary of the continuous part of the spectrum $\lambda_{\rm b}$ for  graph ensembles different than  directed Erd\H{o}s-R\'{e}nyi graphs.  Also, we are note aware of proofs for  Eqs.~(\ref{eq:Rb1}), (\ref{eq:Rb2}) and (\ref{eq:R}) on the mean value of the distribution of right eigenvector elements, the  recursion relations   for $p_R$ given by  Eqs.~(\ref{eq:pRecR}) and (\ref{eq:qRecR}), the algebraic multiplicity of the trivial $-d$-eigenvalue given by Eq.~(\ref{eq:muOrientedxx}), and  Eqs.~(\ref{eq:pRb}) and (\ref{eq:pRIsol})  for the number of zero entries of right eigenvectors.   The results in the present paper are thus interesting conjectures about the spectral properties of sparse non-Hermitian random matrices.

In the present paper,     we have focused on systems that are locally tree-like and oriented.  For future work, it would be  interesting to understand  how network topology affects the linear stability of  non-oriented systems \cite{neri2012spectra}  and  systems that contain small cycles or motifs \cite{metz2011spectra, bolle2013spectra, newman2019spectra, aceituno2019universal}.      Based on the results in the present paper, we would expect that those systems are in general less stable than locally tree-like and oriented systems.

  \appendix
  
  \section{Stability criterion for a linear dynamical system described by the Eqs.~(\ref{eq:linx})}\label{AppNonDiag} 
 We call a linear dynamical system, described by the Eqs.~(\ref{eq:linx}), {\it stable} if  
\begin{eqnarray}
\lim_{t\rightarrow \infty}\vec{y}(t) = 0 \label{eq:stable}
\end{eqnarray}
 for all initial states $\vec{y}(0)$.

  In this appendix,   we show that a linear dynamical system is stable if and only if all the eigenvalues of $\mathbf{A}$ have negative real parts.     To this aim, we express the matrix  $\mathbf{A}$ in its canonical form, which contains as few as possible nonzero offdiagonal matrix entries.     For a diagonalizable matrix, the canonical form is diagonal, whereas for a nondiagonalizable matrix, the canonical form is a  Jordan matrix~\cite{horn1985}.
     
     \subsection{Diagonalizable matrices}
    If $\mathbf{A}$ is a diagonalizable matrix of size $n$, then there exists  a nonsingular matrix $\mathbf{S}$ such that  \cite{horn1985}
     \begin{eqnarray}
\mathbf{A} = \mathbf{S} \Delta \mathbf{S}^{-1}, \label{eq:ACanD}
\end{eqnarray} 
where $\Delta$ is a diagonal matrix with diagonal elements $[\Delta]_{jj} = \lambda_j(\mathbf{A})$.  As a consequence, the $j$-th column of  $\mathbf{S}$ is a right eigenvector $\vec{R}_j$ associated with the eigenvalue $\lambda_j$, and the $j$-th row of $\mathbf{S}^{-1}$ is a left eigenvector $\vec{L}^\dagger_j$ associated with the eigenvalue $\lambda_j$.    Since $\mathbf{S}$ is a nonsingular matrix, both right eigenvectors and left eigenvectors form a set of $n$ independent vectors that span $\mathbb{C}^n$, and because $\mathbf{S}^{-1}\mathbf{S} = \mathbf{1}_n$  right and left eigenvectors form a biorthonormal system, 
\begin{eqnarray}
\vec{L}_j\cdot \vec{R}_k = \delta_{j,k}. \label{eq:biortho}
\end{eqnarray}
As a consequence, the matrix $\mathbf{A}$ can be written as 
\begin{eqnarray}
\mathbf{A} = \sum^n_{j=1} \lambda_j \vec{R}_j \vec{L}^\dagger_j.    \label{eq:eigdA}
\end{eqnarray}

We can  decompose $\vec{y}^\dagger(t)$ into the basis of left eigenvectors $\vec{L}^\dagger_j$, such that  
\begin{eqnarray}
\vec{y}^\dagger(t) = \sum^n_{j=1}  c_j(t)  \vec{L}^\dagger_j. \label{eq:daggerSol}
\end{eqnarray} 
The coefficients
\begin{eqnarray}
c_j(t) = \vec{y}(t) \cdot \vec{R}_j  
\end{eqnarray}
follow from the biorthonormality condition (\ref{eq:biortho}) of left and right eigenvectors.

Substituting  the canonical form of $\mathbf{A}$, given by Eq.~(\ref{eq:eigdA}), into Eq.~(\ref{eq:linx}), and using that the decomposition (\ref{eq:daggerSol}) is unique, we obtain the $n$ independent and  linear differential equations 
\begin{eqnarray}
\partial_t c_j(t)    =   \lambda_j  \:  c_j(t) .\label{eq:firstOrder}
\end{eqnarray}   
Finally, solving Eqs.~(\ref{eq:firstOrder}) we obtain 
\begin{eqnarray}
c_j(t) = e^{\lambda_j t} \: c_j(0)\label{eq:ytSol}
\end{eqnarray} 
and the expression  Eq.~(\ref{jkla}) for $\vec{y}^\dagger(t) $ after substituting (\ref{eq:ytSol}) into (\ref{eq:daggerSol}).

The expression (\ref{jkla}) for $\vec{y}^\dagger(t) $ implies that  a system described by a diagonalizable matrix is stable, if and only if, the real part of all eigenvalues is negative.

      \subsection{Nondiagonalizable matrices} 
      A matrix $\mathbf{A}$ is nondiagonalizable  if there does not exist a nonsingular matrix  $\mathbf{S}$ for which  relation (\ref{eq:ACanD}) holds with $\Delta$ being a diagonal matrix.  Nondiagonalizable matrices contain at least one eigenvalue with a  geometric multiplicity that is smaller than its algebraic multiplicity.    The algebraic multiplicity of an eigenvalue $\lambda$  is the multiplicity of $\lambda$ as the root of the polynomial equation ${\rm det}[\mathbf{A}-\lambda \mathbf{1}_n] = 0$.  The geometric multiplicity an eigenvalue is the number of linearly independent right eigenvectors associated with this eigenvalue.

      If   $\mathbf{A}$ is a nondiagonalizable matrix of size $n$, then there exists a nonsingular matrix $\mathbf{S}$
 such that \cite{horn1985}
      \begin{eqnarray}
\mathbf{A} = \mathbf{S} \mathbf{H} \mathbf{S}^{-1} \label{eq:ACan}
\end{eqnarray} 
where  $\mathbf{H}$ is a Jordan matrix.   The Jordan matrix has the form 
\begin{eqnarray}
\mathbf{H} = \left[\begin{array}{cccc} J_{n_1}(\lambda_{\ell_1}) & & &  \text{\huge0}  \\  &J_{n_2}(\lambda_{\ell_2}) & &   \\ & &\ddots& \\ \text{\huge0} & &  & J_{n_m}(\lambda_{\ell_m})  \end{array}\right],
\end{eqnarray}
     where 
     \begin{eqnarray}
     \ell_{\alpha} =  1+ \sum^{\alpha-1}_{\beta=1}n_{\beta}, \quad {\rm with}\quad   \alpha\in [m] , \label{eq:ellj}
     \end{eqnarray}
     and 
\begin{eqnarray}
J_{n}(\lambda) =  \left[\begin{array}{ccccc} \lambda & 1 &0 & \ldots&0 \\  0&  \lambda  & 1 & &     \\  \vdots& &\ddots & \ddots& \\ &  & &  &  1 \\ 0 & & &  & \lambda
 \end{array}\right]
\end{eqnarray}
is a Jordan block of size $n$.    

The number of Jordan blocks associated with an eigenvalue $\lambda$ equals the  geometric multiplicity of the eigenvalue $\lambda$.   Hence, the number of independent right eigenvectors of the matrix $\mathbf{A}$ is equal to the number $m$ of Jordan blocks in the matrix $\mathbf{H}$.

The columns of $\mathbf{S}$ are the generalized right eigenvectors $\vec{\mathcal{R}}_j$ of the matrix $\mathbf{A}$.   Since the matrix  $\mathbf{S}$ is nonsingular, the generalized right eigenvectors form a set of $n$ independent vectors that span $\mathbb{C}^n$.     
Analogously, the rows of $\mathbf{S}^{-1} $ are the generalized left eigenvectors  $\vec{\mathcal{L}}^\dagger_j$ of $\mathbf{A}$.   Also,  generalized left and  right eigenvectors form a biorthonormal system, 
\begin{eqnarray}
\vec{\mathcal{L}}_j\cdot \vec{\mathcal{R}}_k = \delta_{j,k},  \quad {\rm with}\quad j,k\in [n], \label{eq:biortho2} 
\end{eqnarray}
because $\mathbf{S}\mathbf{S}^{-1} = \mathbf{1}_n$.
The $m$ right eigenvectors of $\mathbf{A}$ are 
\begin{eqnarray}
\vec{R}_{\alpha} = \vec{\mathcal{R}}_{\ell_{\alpha}},\quad {\rm with} \quad  \alpha\in [m],
\end{eqnarray} 
 with the $\ell_\alpha$ as defined in Eq.~(\ref{eq:ellj}).
Analogously, the $m$ left eigenvectors are 
\begin{eqnarray}
\vec{L}_\alpha = \vec{\mathcal{L}}_{\ell_\alpha + n_\alpha-1},\quad {\rm with} \quad  \alpha\in [m].
\end{eqnarray}  

Since the generalized right and generalized left eigenvectors form a biorthonormal system, the matrix $\mathbf{A}$ can be expressed in the form 
\begin{eqnarray}
\mathbf{A} &=& \sum^m_{\alpha=1} \sum^{n_{\alpha}-2}_{\kappa=0} \vec{\mathcal{R}}_{\ell_{\alpha} + \kappa} \left(\lambda_{\ell_{\alpha}}\vec{\mathcal{L}}^\dagger_{\ell_{\alpha} + \kappa} + \vec{\mathcal{L}}^\dagger_{\ell_{\alpha} + \kappa+1}\right)  \nonumber \\  
&& +  \sum^m_{\alpha=1}\lambda_{\ell_{\alpha}} \vec{\mathcal{R}}_{\ell_{\alpha} + n_{\alpha}-1}    \vec{\mathcal{L}}_{\ell_{\alpha} + n_{\alpha}-1}  .  \label{eq:can2}
\end{eqnarray}

Since $\mathbf{S}$ is nonsingular, the  $\vec{\mathcal{L}}^\dagger_j$ form a basis of $\mathbb{C}^n$, and therefore we can  decompose $\vec{y}^\dagger(t)$ into this basis, namely,  
\begin{eqnarray}
\vec{y}^\dagger(t) = \sum^n_{j=1}  c_j(t)  \vec{\mathcal{L}}^\dagger_j. \label{eq:daggerSol2}
\end{eqnarray} 
The coefficients
\begin{eqnarray}
c_j(t) = \vec{y}(t) \cdot \vec{\mathcal{R}}_j  , \quad j\in[n]
\end{eqnarray}
follow from the biorthonormality relation (\ref{eq:biortho2}) of generalized left and  generalized right eigenvectors.

Substituting  the canonical form of $\mathbf{A}$, given by Eq.~(\ref{eq:can2}), into Eq.~(\ref{eq:linx}), and using that the decomposition (\ref{eq:daggerSol2}) is unique, we obtain a set of  $m$ independent   linearly coupled  differential equations corresponding with each of the  $m$ Jordan blocks of the matrix $\mathbf{A}$.       For the $\alpha$-th   Jordan block of size $n_{\alpha}$, we obtain the  differential equation  
\begin{eqnarray}
 \partial_t c_{\ell_{\alpha}+n_{\alpha}-1}(t) &=& \lambda_{\alpha } c_{\ell_{\alpha}+n_{\alpha}-1}(t)\label{eq:firstOrderxxx}
\end{eqnarray}     
together with the coupled differential equations
\begin{eqnarray}
\partial_t c_{\ell_{\alpha}+\kappa}(t)      &=&      \lambda_{\ell_{\alpha}}   c_{\ell_{\alpha}+\kappa}(t)  +   c_{\ell_{\alpha}+\kappa+1}(t)    \label{eq:firstOrderx} 
\end{eqnarray}     
for $\kappa=0,1,\ldots, n_{\alpha}-2$.
The coupled Eqs.~(\ref{eq:firstOrderx})  represent a feedforward loop~\cite{ahmadian2015properties}.   
Solving Eq.~(\ref{eq:firstOrderxxx}),  we obtain that
\begin{eqnarray}
c_{\ell_{\alpha}+n_{\alpha}-1}(t)   =  e^{\lambda_{\alpha}t} c_{\ell_{\alpha}+n_{\alpha}-1}(0) ,
\end{eqnarray}    
and solving the Eqs.~(\ref{eq:firstOrderx}), we obtain that 
\begin{eqnarray}
c_{\ell_{\alpha}+\kappa}(t)  & =&  e^{\lambda_{\alpha} t} p^{(\alpha)}_{\kappa}(t) ,
\end{eqnarray}     
where 
\begin{eqnarray}
 p^{(\alpha)}_{\kappa}(t)  = \sum^{n_{\alpha}-1-\kappa}_{\beta=0} \frac{t^{\beta}}{\beta!}  c_{\ell_{\alpha}+\kappa+\beta}(0) 
\end{eqnarray}
is a polynomial of degree $n_{\alpha}-1-\kappa$  and where 
 $\kappa=0,1,\ldots, n_{\alpha}-1$.   Substituting the explicit solution of the  coefficients $c_{j}(t)$   in  Eq.~(\ref{eq:daggerSol2}), we find that 
\begin{eqnarray}
\vec{y}^\dagger(t) = \sum^m_{\alpha=1}  e^{\lambda_{\alpha}t} \sum^{n_{\alpha}-1}_{\kappa=0}p^{(\alpha)}_{\kappa}(t)  \vec{\mathcal{L}}^\dagger_{\ell_\alpha + \kappa}. \label{eq:daggerSol3}
\end{eqnarray}

For large $t\rightarrow \infty$, the dynamics of  $\vec{y}^\dagger(t)$ is dominated by the  eigenvalue with the largest real part, say $\lambda_1$, such that
\begin{eqnarray}
\vec{y}^\dagger(t) = O(e^{{\rm Re}[\lambda_1]t}t^{n_1-1}) .
\end{eqnarray}  
Hence, the dynamical system is stable if  ${\rm Re}[\lambda_1]<0$   and it is unstable if ${\rm Re}[\lambda_1]>0$.      

This proves that $\vec{y}(t)$ is stable if and only if  all eigenvalues of $\mathbf{A}$ have negative real parts and it is unstable if and only if there exists at least one eigenvalue with a positive real part.

\section{Directed graphs with a  prescribed degree distribution $p_{K^{\rm in},K^{\rm out}}$} \label{sec:finite}    
In this appendix, we define   the random, directed graphs with a  prescribed degree distribution $p_{K^{\rm in},K^{\rm out}}$ that we use throughout this paper.   Subsequently, we detail the algorithm we use to sample  graphs from this ensemble.

\subsection{Definition} \label{sec:defRandomG}
A random graph  $\mathcal{G}$ of size $n$ is   a random set $E\subset [n]\times  [n]$ of directed links. 

In the present paper, we consider random graphs with a  given prescribed degree distribution $p_{K^{\rm in}, K^{\rm out}}$.  In this ensemble, graphs are drawn  with  probability  
\begin{eqnarray}
{\rm Prob}(E=e)  = \frac{p_{\left\{K^{\rm in}_j, K^{\rm out}_j\right\}_{j\in [n]}}\left(\left\{k^{\rm in}_j, k^{\rm out}_j\right\}_{j\in [n]}\right)}{n\left(\left\{k^{\rm in}_j, k^{\rm out}_j\right\}_{j\in [n]}\right)} \nonumber\\ \label{eq:unif}
\end{eqnarray} 
where  $p_{\left\{K^{\rm in}_j, K^{\rm out}_j\right\}_{j\in [n]}}$  is the probability distribution of a degree sequence and where $n\left(\left\{k^{\rm in}_j, k^{\rm out}_j\right\}_{j\in [n]}\right)$ is the number of   graphs with a  degree sequence $\left\{k^{\rm in}_j, k^{\rm out}_j\right\}_{j\in [n]}$.    The probability distribution of a degree sequence is proportional to
\begin{eqnarray}
\lefteqn{p_{\left\{K^{\rm in}_j, K^{\rm out}_j\right\}_{j\in [1,n]}} }&& 
\nonumber\\ 
&\sim& 
 \delta_{\sum^{n}_{j=1}k^{\rm in}_j, \sum^{n}_{j=1}k^{\rm out}_j} \prod^n_{j=1}p_{K^{\rm in}, K^{\rm out}}(k^{\rm in}_j,k^{\rm out}_j).
\end{eqnarray}    
This model is called the uniform    model  \cite{dembo2010gibbs}.   It is the configuration model \cite{bollobas2001random, newman2010networks, dorogovtsev2013evolution} conditioned on the event that there are no self-links and multiple edges.
However, since in the configuration model self-links and multiple edges are rare, the results in this paper apply both to the configurational model and the uniform model (the local neighborhood of a randomly selected node is for both models  the same in the limit  $n\rightarrow \infty$).

\subsection{Algorithm}
We detail the algorithm we use in this paper to sample graphs from the ensemble defined in~Sec.~\ref{sec:defRandomG}.     We consider the specific case of a distribution of the form
\begin{eqnarray}
 p_{K^{\rm in}, K^{\rm out}}(k,\ell) =  q \:p_{\rm deg}(k)  p_{\rm deg} (\ell)  
 + (1-q) p_{\rm deg}(k)  \delta_{k, \ell}. \nonumber\\
 \end{eqnarray} 
The algorithm we have used to generate random graphs from this degree distribution consists  of the following steps: 
\begin{enumerate}
\item We generate a sequence of $n$ i.i.d.~variables  $k^{\rm in}_j$    from the distribution $p_{K^{\rm in}}$;
\item We generate a sample of $n$ i.i.d.~Bernoulli random variables $x_j \in \left\{
0,1\right\}$, which take the value  $x_j=1$ with probability $q$ and $x_j=0$ with probability $1-q$;
\item If $x_j = 0$, then we set $k^{\rm out}_j = k^{\rm in}_j$;
\item We generate a random permutation $\zeta$ on the set of indices $j\in [1,n]$ for which  $x_j=1$;
\item  If $x_j=1$, then we set $k^{\rm out}_j = k^{\rm in}_{\zeta(j)}$;
\item To each $j$ we associate $k^{\rm in}_j$ insockets and $k^{\rm out}_j$ outsockets;
\item We randomly connect pairs of insockets with outsockets by starting  with the node with the highest  total degree $k^{\rm in}_j+k^{\rm out}_j$ and connecting its sockets to   $k^{\rm in}_j$ randomly selected outsockets and  $k^{\rm out}_j$  randomly selected insockets.  Two connected sockets create a directed edge. 
\item   We do not allow for self-links and we do not allow for multiple edges.   Sometimes step seven in the algorithm fails because connecting two sockets would  create either a self-link or a multiple edge.  In this case, we the algorithm restarts  step seven.  
\item We repeat step seven  until the algorithm has found a proper set of edges that defines an oriented simple random graph.  
\end{enumerate}
This algorithm works very well for most of the degree distributions discussed in this paper, except for  power-law random graphs with a small  exponent $a$, see Section~\ref{subSec:powerLaw}.  This is because for power-law random graphs it can be difficult to avoid multiple edges or self-links.    Generating  graphs with a power-law degree distribution with a small exponent $a$ requires  more sophisticated algorithms, such as, algorithms using Markov chains~\cite{coolen2009constrained, annibale2009tailored}.  Alternatively, one could consider the configurational model instead of the uniform model and allow for self-links and multiple edges.      One should however bare in mind that for power-law random graphs with small exponent $a$ the configuration model and the uniform model may not be equivalent anymore because finite size effects will be significant.   

\section{Oriented rings in random and directed graphs}\label{OrientedRing}
An oriented ring graph of size $\ell$ is a subgraph of  size $\ell$ that has an  adjacency matrix of the form \begin{eqnarray}
A_{ij} = \left\{\begin{array} {ccc} J_i  \delta_{j,i+1} && i\neq \ell, \\  J_\ell \delta_{j,1} &&i=\ell, \end{array}\right.
\end{eqnarray}
where $J_i\in\mathbb{R}$.

   Oriented ring graphs may contribute stochastic outliers to the spectra of random graphs with a prescribed degree distribution $p_{K^{\rm in},K^{\rm out}}$.   Here, we first derive  explicit expressions for the eigenvalues of an isolated oriented ring graph, and then count the number of oriented ring graphs in random and directed graphs.

\subsection{Eigenvalues of an oriented ring graph}
The eigenvalues $\lambda_j$ of an oriented ring graph are located on the circle centred at the origin with radius 
\begin{eqnarray}
\gamma = \left(\prod^{\ell}_{j=1}|J_j|\right)^{1/\ell}
\end{eqnarray}
and are given by 
\begin{eqnarray}
\lambda_j = \gamma \:{\rm sign}\left(\prod^{\ell}_{j=1} J_j \right)\: e^{{\rm i}\frac{2\pi}{\ell}(j-1) }, \quad j\in[\ell].
\end{eqnarray}  

Notice that for simplicity we have used that $A_{ii}=0$.   If $A_{ii}=-d$, then the eigenvalues are located on the circle with radius $\gamma$ centred at $-d$.

\subsection{Number of oriented ring graphs in a random and directed graph}
We count the  average number $\langle N(\ell)\rangle$ of  oriented ring graphs of  length~$\ell$ located in a random and directed graph with a prescribed degree distribution. 

Before considering the general case, we count the average number of oriented rings in the directed Erd\"{o}s-R\'{e}nyi ensemble, viz.~\cite{bianconi2005loops},
\begin{eqnarray}
\langle N(\ell)\rangle = \frac{1}{\ell}\left(\frac{c}{n}\right)^\ell n(n-1)\ldots (n-\ell+1),
\end{eqnarray}
which is the probability of drawing $\ell$ edges multiplied by the total number of ordered sequences of $\ell$  indices.      
In the limit of large $n$, 
\begin{eqnarray}
\langle N(\ell)\rangle = \frac{c^\ell}{\ell}.   \label{eq:cyclesER}
\end{eqnarray} 
The expected   number of cycles of finite length is given by 
 \begin{eqnarray}
\langle N \rangle =  \sum^{\infty}_{\ell=2}\langle N(\ell)\rangle =  -\ln(1-c) - c .
\end{eqnarray}

Consider now a random and directed graph with a prescribed degree distribution $p_{K^{\rm in},K^{\rm out}}$.   The distribution of outdegrees obtained by  following a link in a directed graph is given by
\begin{eqnarray}
\frac{p_{K^{\rm in},K^{\rm out}}(k,\ell)k}{c}.  
\end{eqnarray}
  Hence, the average number of oriented rings of length $\ell$ is given by 
\begin{eqnarray}
\langle N(\ell)\rangle =  \frac{1}{\ell}\frac{\langle K^{\rm out} K^{\rm in} \rangle^{\ell}}{(cn)^{\ell}}   n(n-1)\ldots (n-\ell+1)\nonumber\\
\end{eqnarray} 
and in the limit of large $n$ 
\begin{eqnarray}
\langle N(\ell)\rangle = \frac{1}{\ell}\frac{\langle K^{\rm out} K^{\rm in} \rangle^{\ell}}{c^{\ell}}    = \frac{1}{\ell}[c (\rho +1)]^{\ell}. \label{eq:cyclesr}
\end{eqnarray}  
 If $\rho=0$, then Eq.~(\ref{eq:cyclesr}) is equivalent to Eq.~(\ref{eq:cyclesER}).

 The total expected number of cycles of finite length is given by 
 \begin{eqnarray}
\langle N \rangle =  \sum^{\infty}_{\ell=2}\langle N(\ell)\rangle  =  -\ln[1-c (\rho +1)] - c(\rho+1) \nonumber.
\end{eqnarray} 
 
 The distribution of $N$, the number of oriented cycles of finite length, is a Poisson distribution with mean $\langle N\rangle$.    
 
 The probability $p_+$ to have at least one cycle of length larger than $2$ is given by 
 \begin{eqnarray}
 p_+ =1- e^{-\langle N \rangle } =1- (1-c (\rho +1))e^{c(\rho+1)}. \nonumber
 \end{eqnarray}
Note that $p_+\rightarrow 0$ for $c(\rho+1)\rightarrow 0$ and $p_+\rightarrow 1$ for $c(\rho+1)\rightarrow 1$.   Hence, at the percolation transition of the  SCC  there exists with probability one at least one cycle of finite length.

 \section{The  algebraic  multiplicity  of the $-d$-eigenvalue in  random and directed graphs}\label{App:specConn} 
We show that the  spectral distribution $\mu(z)$  of  the adjacency matrix $\mathbf{A}$  of random and directed graphs, as defined in Eq.~(\ref{eq:model}), takes the form 
\begin{eqnarray}
\mu(z) = (1-s_{\rm sc})\delta(z+d) +s_{\rm sc}\: \tilde{\mu}(z), \label{eq:muOriented}
\end{eqnarray}
where $s_{\rm sc}$ is the relative size of the giant strongly connected component, see Sec.~\ref{sec:concomp}, and where $ \tilde{\mu}(\lambda)$ is the normalized spectral distribution associated with the giant strongly connected component.     Since $d$ only contributes a trivial shift  $\lambda_{j}\rightarrow \lambda_j-d$ to all eigenvalues,  we can set $d=0$ without loss of generality.

In order to demonstrate Eq.~(\ref{eq:muOriented}), we use the spectral theory for sparse non-Hermitian random matrices from Refs.~\cite{rogers2009cavity, metz2018spectra}.     We focus on the case  for which $J_{jk}=1$, but the derivation can readily be extended to the $J_{jk}\neq 1$ case.    As shown in those references, the spectral distribution $\mu(z)$ of  matrices of the form given by Eq.~(\ref{eq:model}) can be expressed as
\begin{eqnarray}
\mu(z) =  \frac{1}{\pi} \lim_{\eta\rightarrow 0} \partial_{z^\ast}\int {\rm d}^2\mathsf{g}\: p_{\mathsf{G}}(\mathsf{g})\:[\mathsf{g}]_{21}, \label{eq:genmu}
\end{eqnarray}
where  $\partial_{z^\ast} = (\partial_x + {\rm i}\partial_y)/2 $ and where $\mathsf{g}$ is a $2\times 2$ square matrix with complex-valued entries.     The distribution $p_{\mathsf{G}}$ solves the recursive distributional equation 
\begin{eqnarray}
\lefteqn{p_{\mathsf{G}}(\mathsf{g}) =   \sum^{\infty}_{k=0} \sum^{\infty}_{\ell=0}p_{K^{\rm in}, K^{\rm out}}(k,\ell)    }&& 
 \nonumber\\ 
&& \times \int \prod^{k}_{j=1}{\rm d}^2 \mathsf{g}_j  q_{\rm in}(\mathsf{g}_j) \int \prod^{\ell}_{j=1}{\rm d}^2 \mathsf{h}_j  q_{\rm out}(\mathsf{h}_j) 
\nonumber\\  && 
\times \delta \left[\mathsf{g} - \frac{1}{\mathsf{z} - {\rm i}\eta \mathbf{1}_2 - \sigma_- \sum^{k}_{j=1}\mathsf{g}_j \sigma_+ -  \sigma_+ \sum^{\ell}_{j=1} \mathsf{h}_j\sigma_-}\right], \nonumber\\ \label{eq:padf}
\end{eqnarray}  
where 
\begin{eqnarray}
\mathsf{z}  = \left(\begin{array}{cc} 0&z \\ z^\ast&0 \end{array}\right) , \quad \mathbf{1}_2  = \left(\begin{array}{cc} 1&0 \\ 0&1 \end{array}\right) 
\end{eqnarray}
and 
\begin{eqnarray}
\sigma_-  = \left(\begin{array}{cc} 0&0 \\ 1&0 \end{array}\right) , \quad \sigma_+= \left(\begin{array}{cc} 0&1 \\ 0&0 \end{array}\right) .
\end{eqnarray}
The distributions $q_{{\rm out}}$ and $q_{\rm in}$ solve the recursive distributional equations
\begin{eqnarray}
\lefteqn{q_{{\rm out}}(\mathsf{g}) =   \sum^{\infty}_{k=0} \sum^{\infty}_{\ell=0} \frac{k \:p_{K^{\rm in}, K^{\rm out}}(k,\ell) }{c}   }&& 
 \nonumber\\ 
&& \times \int \prod^{k-1}_{j=1}{\rm d}^2 \mathsf{g}_j  q_{\rm in}(\mathsf{g}_j) \int \prod^{\ell}_{j=1}{\rm d}^2 \mathsf{h}_j  q_{\rm out}(\mathsf{h}_j) 
\nonumber\\  && 
\times \delta \left[\mathsf{g} - \frac{1}{\mathsf{z} - {\rm i}\eta \mathbf{1}_2 - \sigma_- \sum^{k-1}_{j=1}\mathsf{g}_j \sigma_+ -  \sigma_+ \sum^{\ell}_{j=1} \mathsf{h}_j\sigma_-}\right]  \label{eq:rel1out}\nonumber\\
\end{eqnarray}  
and 
\begin{eqnarray}
\lefteqn{q_{{\rm in}}(\mathsf{h}) =   \sum^{\infty}_{k=0} \sum^{\infty}_{\ell=0} \frac{\ell \: p_{K^{\rm in}, K^{\rm out}}(k,\ell) }{c}   }&& 
 \nonumber\\ 
&& \times \int \prod^{k}_{j=1}{\rm d}^2 \mathsf{g}_j  q_{\rm in}(\mathsf{g}_j) \int \prod^{\ell-1}_{j=1}{\rm d}^2 \mathsf{h}_j  q_{\rm out}(\mathsf{h}_j) 
\nonumber\\  && 
\times \delta \left[\mathsf{h} - \frac{1}{\mathsf{z} - {\rm i}\eta \mathbf{1}_2 - \sigma_- \sum^{k}_{j=1}\mathsf{g}_j \sigma_+ -  \sigma_+ \sum^{\ell-1}_{j=1} \mathsf{h}_j\sigma_-}\right], \nonumber\\  \label{eq:rel2out}
\end{eqnarray} 
respectively.

In order to derive the result   (\ref{eq:muOriented}), we use the ansatz
\begin{eqnarray}
\lefteqn{q_{\rm out}(\mathsf{g}) }&& 
\nonumber\\ 
&& =  b  \int {\rm d}^2x  \: \hat{q}_{\rm out}(x)   \delta\left[\mathsf{g} -   \left(\begin{array}{cc}x & 1/z^\ast\\ 1/z& 0 \end{array}\right)\right] \nonumber \\ 
&& + (1-b)\tilde{q}_{\rm out} (\mathsf{g})\label{eq:qoutA}
\end{eqnarray}    
and 
\begin{eqnarray}
\lefteqn{q_{\rm in}(\mathsf{g}) }&& 
\nonumber\\ 
&& =  a   \int {\rm d}^2x \: \hat{q}_{\rm in}(x)   \delta\left[\mathsf{g} -   \left(\begin{array}{cc}0 & 1/z^\ast\\ 1/z& x \end{array}\right)\right] \nonumber \\ 
&& + (1-a)\tilde{q}_{\rm int} (\mathsf{g}), \label{eq:qoutB}
\end{eqnarray}    
where $a,b\in[0,1]$, and $\hat{q}_{\rm out}(x)$, $\tilde{q}_{\rm out} (\mathsf{g})$, $\hat{q}_{\rm in}(x)$,  and $\tilde{q}_{\rm in} (\mathsf{g})$ are normalized distributions.  

Using the ansatz (\ref{eq:qoutA})-(\ref{eq:qoutB}) in the relations (\ref{eq:rel1out}-\ref{eq:rel2out}), we obtain that $a$ and $b$ solve the self-consistent equations (\ref{eq:a}) and (\ref{eq:bb}), and  
the distributions  $\hat{q}_{\rm out}(x)$, $\tilde{q}_{\rm out} (\mathsf{g})$, $\hat{q}_{\rm in}(x)$,  and $\tilde{q}_{\rm in} (\mathsf{g})$  solve  a set of recursive distributional equations,  whose precise form will not matter.

Using the ansatz  (\ref{eq:qoutA})-(\ref{eq:qoutB})  in Eq.~(\ref{eq:padf}),  we obtain 
\begin{eqnarray}
\lefteqn{p_{\mathsf{G}}(\mathsf{g})}&& 
\nonumber\\
 && = (1-s_{\rm wc} + s_{\rm t}) \delta\left(\mathsf{g} - \mathsf{z}^{-1}\right)  
\nonumber\\ 
&&  + (s_{\rm wc}-s_{\rm out}-s_{\rm t}) \int {\rm d}x  \: \hat{p}_{\rm in}(x)   \delta\left[\mathsf{g} -   \left(\begin{array}{cc}x  & 1/z^\ast\\ 1/z& 0 \end{array}\right)\right] 
\nonumber\\ 
&& 
 +  (s_{\rm wc}-s_{\rm in}-s_{\rm t})    \int {\rm d}x  \: \hat{p}_{\rm out}(x)   \delta\left[\mathsf{g} -   \left(\begin{array}{cc}0 & 1/z^\ast\\ 1/z& x \end{array}\right)\right]
 \nonumber \\ 
 &&+(s_{\rm in} + s_{\rm out}+s_{\rm t}-s_{\rm wc}) \tilde{p}(\mathsf{g})  ,  
\label{eq:pLast}\end{eqnarray} 
where $s_{\rm in}$, $s_{\rm out}$, $s_{\rm wc}$ and $s_{\rm t}$ denote the relative sizes of the giant incomponent, outcomponent,  weakly connected component, and  tendrils, respectively (see Sec.~\ref{Sec:ConComp} or Refs.~\cite{broder2000graph, dorogovtsev2001giant, timar2017mapping}).   The distributions $\hat{p}_{\rm in}(x) $, $\hat{p}_{\rm out}(x)$ and $ \tilde{p}(\mathsf{g})  $ solve  a set of recursive distributional equations that  we  have omitted because their  precise form does  not matter here.

Eqs.~(\ref{eq:genmu}) and (\ref{eq:pLast}), together with  the formulae
\begin{eqnarray}
\frac{1}{\pi}\partial_{z^\ast} \frac{1}{z} = \delta(z)
  \end{eqnarray}
and  
\begin{eqnarray}
s_{\rm sc } = s_{\rm in} + s_{\rm out}+s_{\rm t}-s_{\rm wc},
    \end{eqnarray}
imply the final result (\ref{eq:muOriented}), which we aimed to prove in this appendix.

\section{Derivation of  the recursive distributional equations for $p_R$  in random and directed graphs with a prescribed degree distribution $p_{K^{\rm in},K^{\rm out}}$} \label{sec:der} 
In this appendix, we derive the  recursive distributional equations  (\ref{eq:pRecR}-\ref{eq:qRecR}) for the distribution $p_R$ of entries of right eigenvectors in random  and directed graphs with a prescribed degree distribution $p_{K^{\rm in},K^{\rm out}}$.    The relations (\ref{eq:pRecR}-\ref{eq:qRecR}) apply to the eigenvalues $\lambda_{\rm b}$ located at the boundary of the continuous part $\sigma_{\rm ac}$ of the spectrum and to deterministic eigenvalue outliers $\lambda_{\rm isol}$.  

The derivations we present are based on the theory of Ref.~\cite{neri2016eigenvalue} that relies on the cavity method~\cite{mezard2003cavity, mezard2001bethe, rogers2008cavity,  bordenave2010resolvent, metz2018spectra} (also known as the objective method    in probability theory \cite{aldous2004objective, bordenave2010resolvent}  and belief propagation in computer science \cite{bickson2008gaussian, weiss2000correctness, yedidia2003understanding}).   

The theory of   Ref.~\cite{neri2016eigenvalue}  builds on  two properties of random and directed graphs with a prescribed degree distribution, namely, that these random graphs are {\it locally tree-like} and  {\it oriented}.    In addition, it uses that eigenvalues $\lambda_{\rm b}$ and $\lambda_{\rm isol}$ are {\it stable}, i.e., insensitive to small matrix perturbations.          

In a first subsection, we clarify what we mean by a matrix  being locally tree-like and oriented, and in the second and third subsections we derive the recursive distributional Eqs.~(\ref{eq:pRecR}-\ref{eq:qRecR}).

\subsection{Locally tree-like  and oriented random matrices} \label{sec:locor}
 We say that
an nondirected graph  is a tree if it is connected and it does not contain cycles, see Ref.~\cite{bollobas2013modern}, and we say that a matrix is oriented if  $A_{ij}A_{ji} = 0$ for all pairs $i$ and $j$.   In the following, we extend these global definitions  to local definitions that apply to sequences of random matrices $\mathbf{A}_n$, with  $n\in\mathbb{N}$.   

First we define the concept of local tree-likeness.  
Let $\mathbf{A}_n$, with $n\in\mathbb{N}$, be a sequence of random matrices  and  let $\mathbf{C}_n$ be their associated adjacency matrices, i.e., $C_{kj}=1$ when $A_{kj}\neq 0$ and $C_{kj}=0$ when $A_{kj}=0$.      We also consider 
the associated  symmetrized adjacency matrices  
$\tilde{\mathbf{C}}_n$ with entries  $\tilde{C}_{jk} = {\rm max}\left\{C_{jk},  C_{kj}\right\}$, which are the adjacency matrices of  nondirected simple  graphs.     We define the nondirected $\ell$-neighborhood of a node $i$ as the  subgraph of  $\tilde{\mathbf{C}}_n$  formed by the nodes that are separated no more than  a distance  $\ell$ from $i$.      We say that the sequence of random matrices $\mathbf{A}_n$ is locally tree-like if, for each $\ell \in \mathbb{N}$, the nondirected $\ell$-neighborhood of a uniformly and randomly selected node in $\tilde{\mathbf{C}}_n$   is in the limit  $n\rightarrow \infty$ with probability one a tree, see Ref.~\cite{bordenave2010resolvent}.

Second, we define local orientedness of a sequence of random matrices $\mathbf{A}_n$.    We say that the sequence of random matrices $\mathbf{A}_n$ is locally oriented if, for each $\ell \in \mathbb{N}$, the principal submatrix of $\mathbf{A}_n$  formed by the nodes in the nondirected $\ell$-neighborhood of a uniformly and randomly selected node $i$ is in the limit $n\rightarrow \infty$ with probability one  oriented.  

\subsection{Recursion relations for $p_R$ in locally tree-like and oriented matrices } 

Let $\lambda$ be an  eigenvalue of  the   matrix $\bA$ and let $\vec{R}$ be the right eigenvector associated with $\lambda$. Equation (\ref{eq:eigvDef}) implies that
\begin{eqnarray}
R_j &=& \frac{1}{\lambda+A_{jj}} \sum^{n}_{k=1; k\neq j}A_{jk}R_k , 
\end{eqnarray}  
for all $j \in [n]$.   Using  Eq.~(\ref{eq:model}) and the graph definitions in Sec.~\ref{sec:modeldef}, we obtain 
\begin{eqnarray}
R_j &=& \frac{1}{\lambda+d} \sum_{k\in \partial^{\rm out}_j}J_{jk} R_k .  \label{eq:rel} 
\end{eqnarray}
In general, the random variables  $R_k$ are correlated with the entries $J_{jk}$ and the degree $K^{\rm out}_j$, and therefore, Eq.~(\ref{eq:rel}) is not useful to derive a selfconsistent distributional equation.  However, if $\mathbf{A}$ is a locally  tree-like and  oriented  matrix,  then the $R_k$ are statistically independent from the $J_{jk}$ and $K^{\rm out}_j$.

The statistical independence between $R_k$ and $A_{jk}$ can be understood from a recursive argument.
Let $\bA^{(j)}$ be the principal submatrix obtained from $\mathbf{A}$  by deleting its $j$-th column and row, and let $\vec{R}^{(j)}$ be the right eigenvector of  $\bA^{(j)}$ associated with the  same eigenvalue $\lambda$; hence, $\lambda$ is an eigenvalue of both $\mathbf{A}$ and $\bA^{(j)}$.  Then, for a locally   tree-like and oriented matrix it holds, in the limit $n\rightarrow \infty$,  that 
\begin{eqnarray}
R_k =   R^{(j)}_k , \label{eq:relx}
\end{eqnarray}  
for all $k\in[n]$ and  $j\in \partial^{\rm in}_k$, where  $R^{(j)}_k$ is the $k$-th element of the right eigenvector $\vec{R}^{(j)}$.  For a detailed derivation of Eq.~(\ref{eq:relx}), we refer to the next Appendix~\ref{sec:derivrkrjk}.   Importantly, the derivation of Eq.~(\ref{eq:relx}) in Appendix~\ref{sec:derivrkrjk} relies on the assumption that $\lambda$ is either a deterministic eigenvalue outlier $\lambda_{\rm isol}$ or an eigenvalue $\lambda_{\rm b}$ located at the boundary of the continuous part of the spectrum.

The Eqs.~(\ref{eq:rel}) and (\ref{eq:relx}) imply that 
\begin{eqnarray}
R^{(j)}_k &=&\frac{1}{\lambda+d} \sum_{\ell\in \partial^{\rm out}_k}J_{k\ell} R^{(k)}_\ell, \label{eq:relxxx}
\end{eqnarray}     
for all $k\in[n]$ and  $j\in \partial^{\rm in}_k$.
Since we are interested in the statistics of $R$, we will also use the relation 
\begin{eqnarray}
R_j &=&\frac{1}{\lambda+d} \sum_{k\in \partial^{\rm out}_j}J_{jk} R^{(j)}_k  ,\label{eq:relxx}
\end{eqnarray}   
which  also follows from    Eqs.~(\ref{eq:rel}) and (\ref{eq:relx}).

In the remaining part of this appendix, we use the Eqs.~(\ref{eq:relxxx})  and (\ref{eq:relxx})  to derive the recursion relations  (\ref{eq:pRecR}-\ref{eq:qRecR}).   We define the distributions of right eigenvector entries $R_j$, 
\begin{eqnarray}
 p_R(r|\bA) = \frac{1}{n}\sum^n_{j=1} \delta(r-R_j)  \label{eq:uniform}
\end{eqnarray}
and the distribution of entries $R^{(j)}_k$, 
\begin{eqnarray}
q_{R}(r|\bA) = \frac{1}{c\:n}\sum_{k=1}^n \sum_{j \in \partial^{\rm in}_{k}} \delta(r-R_k^{(j)} ),  \label{eq:out}
\end{eqnarray}
where $c$ is the mean outdegree. The distribution $p_R(r|\bA)$ is obtained by selecting uniformly at random a node $j$ and asking what is the corresponding eigenvector entry $R_j$, whereas  the distribution $q_{R}(r|\bA)$  is obtained  by selecting uniformly at  random an edge $j\rightarrow k$  and asking what is the  eigenvector entry $R_k^{(j)}$.        Since  the model defined in Sec.~\ref{sec:modeldef}  is locally tree-like, all random variables on the right hand side of  Eqs.~(\ref{eq:relxxx}) and   (\ref{eq:relxx}) are independent.  In addition, using that all nodes are statistically equivalent,  we obtain the   recursion relations   (\ref{eq:pRecR}-\ref{eq:qRecR}), which we were meant to derive.

\section{Recursion relations for the entries $R_j$ of right eigenvectors}\label{sec:derivrkrjk}  
We derive a set of recursion relations for the entries $R_j$ of the right eigenvectors associated with deterministic outliers $\lambda_{\rm isol}$ or eigenvalues located at the boundary of the continuous part of the spectrum $\sigma_{\rm ac}$ in locally tree-like random matrices in the limit of infinitely large $n$.     In the special case of locally tree-like and oriented matrices, see Appendix~\ref{sec:locor} for a definition, we show that  Eq.~(\ref{eq:relx})  holds.    

In order to clearly show how the assumptions  of locally tree-likeness and locally orientedness enter in the theory, we first derive a set of recursion relations in the entries $R_j$ of a general matrix $\mathbf{A}$.    As we will demonstrate, the relations for general matrices are not closed and are thus not useful.   In order to close these equations, we make the assumption that $\mathbf{A}$ is locally tree-like.   Lastly, we show how the recursion relations simplify when $\mathbf{A}$ is also locally tree-like and oriented.  

\subsection{General matrices}
 The derivations we present rely on a recursive implementation of the Schur formula to the resolvent of $\mathbf{A}$.

The resolvent of $\mathbf{A}$ is defined by 
\begin{eqnarray}
\mathbf{G}_{\bA}(z) = \frac{1}{\bA - z \:\mathbf{1}_n},  \quad z \in \mathbb{C}\setminus \left\{\lambda_1, \lambda_2, \ldots, \lambda_n\right\},  \label{eq:Gx}\nonumber\\
\end{eqnarray}
with $\mathbf{1}_n$ the identity matrix of size $n$.      In the limit of $n\rightarrow \infty$, the resolvent $\mathbf{G}_{\bA}(z)$ only exists for values $z\notin \sigma_{\rm ac}$.

 Let $\lambda$ be a nondegenerate eigenvalue, and let $\vec{R}$ and $\vec{L}$  be a corresponding left and right eigenvector.     We assume  that  there exists a path in the complex plane that reaches $\lambda$ and along which  $\mathbf{G}_{\bA}(\lambda-\eta)$ exists.           In addition, we assume that   
 $\lambda$ is a {\it stable} eigenvalue of $\bA$, i.e., we assume that  if $\lambda$ is  an eigenvalue of $\mathbf{A}$,  then $\lambda$ is also an eigenvalue of the principal submatrix $\bA^{(j)}$, which we obtain from $\bA$ by deleting the $j$-th row and column.   Hence, $\lambda$ is either a deterministic eigenvalue outlier or is located at the boundary of the continuous part of the spectrum $\sigma_{\rm ac}$.

Since there exists a path that reaches  $\lambda$ and along which  $\mathbf{G}_{\bA}(\lambda-\eta)$ exists, we can write
\begin{eqnarray}
\lim_{\eta \rightarrow 0} \eta \mathbf{G}_{\bA}(\lambda-\eta)= \vec{R}\: \vec{L}^\dagger + O(|\eta|). \label{eq:c10x}
\end{eqnarray}  
Note that relation (\ref{eq:c10x}) also holds when the matrix $\bA$ is not diagonalizable since we can decompose $\mathbf{G}_{\bA}(\lambda)$ in a biorthonormal system of generalized left and right eigenvectors, analogous to the decomposition of $\mathbf{A}$ in Eq.~(\ref{eq:can2}).    Eq.~(\ref{eq:c10x}) implies that the components $R_{j}$ of $\vec{R}$ are  given by
\begin{eqnarray}
R_{j} =  \vec{e}_j\cdot \vec{R}  =   \lim_{\eta \rightarrow 0} \eta \frac{\sum^n_{\ell=1}G_{j\ell}(\lambda - \eta )}{\vec{L} \cdot \vec{1}},  \label{eq:r1x}
\end{eqnarray}
where $G_{j\ell}(\lambda - \eta ) = \left[\mathbf{G}_{\mathbf{A}}(\lambda)\right]_{j\ell}$, $\vec{1}$ is the column vector with all components equal to one, and $\vec{e}_j$ is the column vector with all components equal to zero, except for the $j$-th component, which is equal to one.  

To compute the    elements $G_{j\ell}(\lambda)$ of the resolvent matrix, we use the Schur formula, which is a common tool in random matrix theory (see for instance
section 2.4.3 in Ref.~\cite{tao2012topics} and also Refs.~\cite{ bai2008methodologies, bordenave2010resolvent, bordenave2012around}). Let
\begin{equation} 
 \left(\begin{array}{cc}\mathbf{a} & \mathbf{b} \\ \mathbf{c} & \mathbf{d} \end{array}\right)
 \end{equation}
 be a block matrix, then 
  \begin{equation}
    \mathbf{s}_\mathbf{a}  := \mathbf{d}-\mathbf{c}\mathbf{a}^{-1}\mathbf{b}
 \end{equation} 
is the Schur complement of block $\mathbf{a}$, and   
 \begin{equation}
\mathbf{s}_\mathbf{d}   := \mathbf{a}-\mathbf{b}\mathbf{d}^{-1}\mathbf{c}
 \end{equation} 
is the Schur complement of block $\mathbf{d}$.  
If $\mathbf{a}$ and its Schur-complement  $\mathbf{s}_{\mathbf{a}}$ are invertible matrices,  then the following block inversion formula holds 
 \begin{eqnarray}
 \left(\begin{array}{cc}\mathbf{a} & \mathbf{b} \\ \mathbf{c} & \mathbf{d} \end{array}\right)^{-1} =  \left(\begin{array}{cc} \mathbf{s}^{-1}_\mathbf{d} & -\mathbf{s}^{-1}_\mathbf{d} \:\mathbf{b}\mathbf{d}^{-1} \\ -\mathbf{d}^{-1}\mathbf{c}\: \mathbf{s}^{-1}_\mathbf{d}&\mathbf{s}^{-1}_\mathbf{a}\end{array}\right), \label{SchurInversion}
 \end{eqnarray} 
 which we  call  the {\it Schur  formula}.

We use the Schur formula to derive recursion relations for the elements of the resolvent $\mathbf{G}_{\mathbf{A}}$ and eventually  Eqs.~(\ref{eq:relxxx}) and (\ref{eq:relxx}).    
Applying the Schur formula to the off-diagonal elements $G_{j\ell} $  of the resolvent, we obtain
\begin{eqnarray}
G_{j\ell}  &=&   -G_{jj}\sum^n_{k=1; (k\neq j)}A_{jk}G^{(j)}_{k\ell}  ,
\end{eqnarray}
where
\begin{eqnarray}
G^{(j)}_{k\ell}= \left[\mathbf{G}_{\mathbf{A}^{(j)}}\right]_{k\ell},
\end{eqnarray}
and  where
\begin{eqnarray}
 \mathbf{G}_{\mathbf{A}^{(j)}} = (\mathbf{A}^{(j)}_{n-1}- \lambda \mathbf{1}_{n-1})^{-1}.
\end{eqnarray}
Summing over the index $\ell$, we obtain
\begin{eqnarray}
\sum^n_{\ell=1}G_{j \ell} = G_{jj}\left(1 -  \sum^n_{k=1; (k\neq j)}A_{jk}\sum^n_{\ell=1;(\ell\neq j)} G^{(j)}_{k\ell}\right).  \nonumber\\ 
\end{eqnarray}   
Finally, using Eq.~(\ref{eq:r1x}), we find 
\begin{eqnarray}
R_{j} &=&  \lim_{\eta\rightarrow 0}\eta \frac{G_{jj}}{\vec{L} \cdot \vec{1}} 
\nonumber\\ 
 &&
  -G_{jj}    \sum^n_{k=1; (k\neq j)}A_{jk}  \lim_{\eta\rightarrow 0}\eta \frac{\sum^n_{\ell=1;(\ell\neq j)} G^{(j)}_{k\ell}}{ \vec{L} \cdot \vec{1}} + O(|\eta|).  \nonumber\\
\end{eqnarray}   
The first term is an order $O(1/n)$ smaller than  the second term,and we  identify 
\begin{eqnarray}
R^{(j)}_{k} =  \lim_{\eta\rightarrow 0}\eta \frac{\sum^n_{\ell=1;(\ell\neq j)} G^{(j)}_{k\ell}}{   \vec{L}^{(j)} \cdot \vec{1} } ,
\end{eqnarray}   
where we have used  that $ \vec{L} \cdot \vec{1}  \approx  \vec{L}^{(j)} \cdot \vec{1}$ for large enough~$n$.   Hence, we obtain the relation
\begin{eqnarray}
R_{ j} &=&  -G_{jj}(\lambda-\eta)   \sum^n_{k=1;(k\neq j)}A_{jk}R^{(j)}_{k}. \label{eq:recursionfinal1A} 
\end{eqnarray}

We can repeat the above line of reasoning to obtain a recursion relation for the entries $R^{(j)}_k$ of the right eigenvector $\vec{R}^{(j)}$ associated with the eigenvalue $\lambda$ of $\mathbf{A}^{(j)}$.    We obtain then  that  
\begin{eqnarray}
R^{(j)}_{k} &=&  -G^{(j)}_{kk}(\lambda-\eta)   \sum^n_{\ell =1;\ell \neq j,k}A_{k\ell }R^{(j),(k)}_{\ell}, \label{eq:recursionfinal1x}
\end{eqnarray} 
where $R^{(j),(k)}_\ell$ is the $\ell$-th entry of the right eigenvector $\vec{R}^{(j),(k)}$  associated with the eigenvalue $\lambda$ of the principal submatrix $\mathbf{A}^{(j),(k)}$.     The principal submatrix $\mathbf{A}^{(j),(k)}$ is obtained from $\mathbf{A}$ by removing both the $j$-th and $k$-th rows and columns.    

In order to obtain an expression for the diagonal elements $G_{kk}$ and $G^{(j)}_{kk}$ that appear in the recursion Eqs.~(\ref{eq:recursionfinal1A}) and (\ref{eq:recursionfinal1x}), we use again the Schur formula Eq.~(\ref{SchurInversion}).   We obtain that 
\begin{eqnarray}
G_{jj}(z) = \frac{1}{-z + A_{jj}  -  \sum^n_{k,k'=1 (k,k'\neq j)} A_{jk} G^{(j)}_{k k'}(z) A_{kj} } , \nonumber\\ \label{eq:gjj}
\end{eqnarray}
and 
\begin{eqnarray}
G^{(j)}_{kk}(z)= \frac{1}{-z + A_{kk}  -  \sum^n_{\ell,\ell'=1 (\ell,\ell'\neq k,j)} A_{j\ell } G^{(j,k)}_{\ell \ell'}(z) A_{\ell' j} } ,  \nonumber\\\label{eq:gjjz}
\end{eqnarray}
for all $z\notin \sigma_{\rm ac}$.   

It is insightful to rewrite these equations using the notation 
\begin{equation}
A_{jk} = -D_{j}\delta_{j,k} + (1-\delta_{j,k}) J_{jk}C_{jk} 
\end{equation}
 where $C_{jk}\in \left\{0,1\right\}$ is the adjacency matrix denoting whether  $A_{kj}\neq 0$  ($C_{kj}=1$) or  $A_{kj}=0$ ($C_{kj}=0$).  
 
The Eqs.~(\ref{eq:recursionfinal1A}) and (\ref{eq:recursionfinal1x})  read then 
 \begin{eqnarray}
R_{j} &=&  -G_{jj}(\lambda-\eta)   \sum_{k\in \partial^{\rm out}_j}J_{jk }R^{(j)}_{k}, \label{eq:recursionfinal1B}
\end{eqnarray}
and 
\begin{eqnarray}
R^{(j)}_{k} &=&  -G^{(j)}_{kk}(\lambda-\eta)   \sum_{\ell\in \partial^{\rm out}_k\setminus \left\{j\right\}}J_{k\ell }R^{(j),(k)}_{\ell}, \label{eq:recursionfinal1BS}
\end{eqnarray} 
where $j\in \partial^{\rm in}_k$, and  Eqs.~(\ref{eq:gjj}) and (\ref{eq:gjjz}) read
\begin{eqnarray}
G_{jj} = \frac{1}{-z- D_j  -  \sum_{k\in \partial^{\rm out}_j,k'\in\partial^{\rm in}_j} J_{jk} G^{(j)}_{k k'} J_{k'j} } \nonumber\\ \label{eq:gjjB}
\end{eqnarray}
and 
\begin{eqnarray}
G^{(j)}_{kk}= \frac{1}{-z - D_k -  \sum_{\ell\in \partial^{\rm out}_{k} \setminus \left\{j\right\}}  \sum_{\ell'\in \partial^{\rm in}_{k} \setminus \left\{j\right\}} J_{k\ell } G^{(j), (k)}_{\ell \ell'}J_{\ell' k} } .  \nonumber\\ \label{eq:gjjzB}
\end{eqnarray}

The recursion Eqs.~(\ref{eq:recursionfinal1B}-\ref{eq:gjjzB})  hold for general matrices $\mathbf{A}$.  However,   they do not form a   closed set of equations and are thus not useful.    In order to close this set of  recursion relations, we make the assumption that the  matrices $\mathbf{A}$  are locally tree-like.

\subsection{Locally tree-like matrices}\label{sec:locTreeRanA} 
We show how the set of  recursion Eqs.~(\ref{eq:recursionfinal1B}-\ref{eq:gjjzB}) simplify for random matrices  $\mathbf{A}$ that are   locally tree-like.

For matrices that are locally tree-like,  it holds  that 
\begin{eqnarray}
R^{(j),(k)}_{\ell} = R^{(k)}_{\ell},
\end{eqnarray} 
for all $\ell\in[n]$, $k \in \partial^{\rm in}_\ell$ and  $j\in \partial^{\rm in}_k$.   This is because $\ell$ and $j$ belong to disjoint trees of the forest represented by the matrix $\mathbf{A}^{(k)}$.      As a consequence, Eqs.~(\ref{eq:recursionfinal1BS}) simplify into
\begin{eqnarray}
R^{(j)}_{k} &=&  -G^{(j)}_{kk}(\lambda-\eta)   \sum_{\ell\in \partial^{\rm out}_k\setminus \left\{j\right\}}J_{k\ell }R^{(k)}_{\ell}, \label{eq:recursionfinal1xxA} 
\end{eqnarray}
for all $k\in[n]$ and $j\in \partial^{\rm in}_k$.

The resolvent Eqs.~(\ref{eq:gjjB}-\ref{eq:gjjzB}) also simplify because for locally tree-like graphs it holds that 
 \begin{eqnarray}
 G^{(j)}_{k k'}  = 0 \label{eq:g0}
\end{eqnarray} 
for all $k\in \partial^{\rm out}_j$ and $k\in \partial^{\rm in}_j$, and 
 \begin{eqnarray}
 G^{(j), (k)}_{\ell \ell}  =  G^{(k)}_{\ell \ell} \label{eq:g1}
\end{eqnarray}
for all $\ell\in[n]$, $k \in \partial_\ell$ and  $j\in \partial_k\setminus \left\{\ell\right\}$.

The relations (\ref{eq:g0}) and (\ref{eq:g1}) follow from the fact that for values  $z\notin \sigma_{\rm ac}$, we can develop the series expansion
\begin{eqnarray}
\mathbf{G}_{\mathbf{A}_n}(z) =  -\frac{1}{z}\sum^{\infty}_{m=0} \frac{\mathbf{A}^{m}}{z^m}, 
\end{eqnarray}
and hence 
\begin{eqnarray}
G_{jk}(z) := \left[\mathbf{G}_{\mathbf{A}_n}(z)\right]_{jk} =  -\frac{1}{z}\sum^{\infty}_{m=0} \frac{\left[\mathbf{A}^{m}\right]_{jk}}{z^m}.  \label{eq:series} 
\end{eqnarray}  
 
Eq.~(\ref{eq:g0})  follows now from  Eq.~(\ref{eq:series}) and the fact that for locally tree-like matrices it holds that 
\begin{eqnarray}
\left[\left(\mathbf{A}^{(j)}\right)^{m}\right]_{kk'} = 0
\end{eqnarray}
for all $k\in \partial^{\rm out}_j$, $k'\in \partial^{\rm in}_j$, and $m\in \mathbb{N}$, since $k\in \partial^{\rm out}_j$ and $k'\in \partial^{\rm in}_j$     belong to disjoint trees of the forest represented by the adjacency matrix $\mathbf{A}^{(j)}$, and hence there exists no path of finite length that connects $k$ to $k'$.   

Eq.~(\ref{eq:g1}) on the other hand, follows from Eq.~(\ref{eq:series})  and the fact that 
\begin{eqnarray}
\left[\left(\mathbf{A}^{(j), (k)}\right)^{m}\right]_{\ell \ell} = \left[\left(\mathbf{A}^{(k)}\right)^{m}\right]_{\ell \ell} 
\end{eqnarray}
for all $\ell\in[n]$, $k \in \partial_\ell$,  $j\in \partial_k\setminus \left\{\ell\right\}$ and  $m\in \mathbb{N}$, since $j$ and $\ell$ belong to disconnected trees of the forest represented by $\mathbf{A}^{(k)}$.

Using Eqs.~(\ref{eq:g0}-\ref{eq:g1}) in Eqs.~(\ref{eq:gjjB}-\ref{eq:gjjzB}), we obtain \cite{neri2016eigenvalue, metz2018spectra}
\begin{eqnarray}
G_{jj} = \frac{1}{-z-D_j  -  \sum_{k\in \partial^{\rm out}_j} J_{jk} G^{(j)}_{k k} J_{kj} }  \nonumber\\ \label{eq:gjjB2}
\end{eqnarray}
and 
\begin{eqnarray}
G^{(j)}_{kk}= \frac{1}{-z - D_k -  \sum_{\ell\in \partial^{\rm out}_{k} \setminus \left\{j\right\}}  J_{k\ell } G^{(k)}_{\ell \ell}J_{\ell k} } .  \nonumber\\ \label{eq:gjjzB2}
\end{eqnarray}
Note that for symmetric random matrices, Eqs.~(\ref{eq:gjjB2}-\ref{eq:gjjzB2}) are equivalent to the recursion relations for the resolvent  derived in~Refs.~\cite{abou1973selfconsistent, rogers2009cavity, metz2010localization}.

The Eqs.~(\ref{eq:recursionfinal1B}), (\ref{eq:recursionfinal1xxA}), (\ref{eq:gjjB2}) and (\ref{eq:gjjzB2}) form  a  closed set of recursion relations.      They can either be solved on a given graph instance or we can solve  these equations in a distributional sense for a random graph ensemble.  In either case, we obtain information about the statistics of~$R_j$.

\subsection{Locally tree-like and oriented matrices} 
For random matrices that are locally tree-like and oriented, we can use in Eqs.~(\ref{eq:gjjB2}) and (\ref{eq:gjjzB2}) that 
\begin{eqnarray}
J_{jk} J_{kj} = 0.
\end{eqnarray}
As a consequence, we obtain the explicit expressions 
\begin{eqnarray}
G_{jj} = \frac{1}{-z- D_j  } , \label{eq:gd}
\end{eqnarray}
and 
\begin{eqnarray}
G^{(j)}_{kk}= \frac{1}{-z - D_k}.  \label{eq:gd2}
\end{eqnarray} 
Substituting Eqs.~(\ref{eq:gd}) and (\ref{eq:gd2}) into  Eqs.~(\ref{eq:recursionfinal1B}) and (\ref{eq:recursionfinal1xxA}), we obtain 
 \begin{eqnarray}
R_{j} &=&  \frac{1}{\lambda+D_j}  \sum_{k\in \partial^{\rm out}_j}J_{jk }R^{(j)}_{k}, \label{eq:recursionfinal1Bxxx}
\end{eqnarray}
and 
 \begin{eqnarray}
R^{(j)}_{k} &=&  \frac{1}{\lambda+D_k} \sum_{\ell\in \partial^{\rm out}_k\setminus \left\{j\right\}}J_{k\ell }R^{(k)}_{\ell} .\label{eq:recursionfinal2Bxxx}
\end{eqnarray}

From Eqs.~(\ref{eq:recursionfinal1Bxxx}) and (\ref{eq:recursionfinal2Bxxx}), we observe that 
\begin{eqnarray}
R_k = R^{(j)}_k
\end{eqnarray}
for all $k\in[n]$ and  $j\in \partial^{\rm in}_k$, since then $j\notin \partial^{\rm out}_k$ and thus the right-hand-side of the Eqs.~(\ref{eq:recursionfinal1Bxxx}) and (\ref{eq:recursionfinal2Bxxx}) are identical.     

This concludes the derivation of Eq.~(\ref{eq:relx}), which we were meant to show. 

 \section{Normalizable solutions to the recursive distributional Eqs.~(\ref{eq:pRecR}) and (\ref{eq:qRecR}) for $p_R$} \label{sec:SolRec}
In this  appendix,  we derive analytical results for $\lambda_b\in\partial\sigma_{\rm ac}$, and  $\lambda_{\rm isol}$ by  identifying values of $\lambda$ for which the Eqs.~(\ref{eq:pRecR}) and (\ref{eq:qRecR})  admit a normalizable solution.

   Since Eqs.~(\ref{eq:pRecR}) and (\ref{eq:qRecR})  are linear distributional equations, we can derive a set of fixed-point equations for the lower-order moments of $R$ and $L$.   In order to distinguish averages with respect to $p_R$ with those with respect to $q_R$, we introduce the notation  
   \begin{eqnarray}
  \langle f(R) \rangle = \int {\rm d}^2 r \: p_R(r) f(r)  
          \end{eqnarray}
          and 
             \begin{eqnarray}
     \langle f(R) \rangle_q = \int {\rm d}^2 r \: q_R(r) f(r) ,
             \end{eqnarray}
             where $f$ is an arbitrary function.
From Eq.~(\ref{eq:qRecR}), we obtain that 
 \begin{eqnarray}
 \langle R \rangle_q &=& \frac{\langle K^{\rm in}K^{\rm out}\rangle}{c(\lambda+d)} \langle J \rangle \langle R\rangle_q ,\label{eq:o}\\ 
       \langle R^2 \rangle_q &=& \frac{\langle K^{\rm in}K^{\rm out}\rangle}{c(\lambda+d)^2} \langle J^2 \rangle \langle R^2\rangle_q  \nonumber\\
      &&   +  \frac{\langle K^{\rm in}K^{\rm out}(K^{\rm out}-1)\rangle}{c (\lambda+d)^2}  \langle J \rangle^2 \langle R\rangle^2_q,   \label{eq:aa}\\ 
    \langle |R|^2 \rangle_q &=& \frac{\langle K^{\rm in}K^{\rm out}\rangle}{c|\lambda+d|^2}\langle |J|^2  \rangle \langle |R|^2\rangle_q  \nonumber\\  && +  \frac{\langle K^{\rm in}K^{\rm out}(K^{\rm out}-1)\rangle}{c |\lambda+d|^2}  |\langle J \rangle|^2 |\langle R\rangle_q |^2,\label{eq:b}
 \end{eqnarray}   
 and from  Eq.~(\ref{eq:pRecR}) we obtain 
  \begin{eqnarray}
 \langle R \rangle &=& \frac{c}{\lambda+d} \langle J \rangle \langle R\rangle_q, \label{eq:ox}\\ 
       \langle R^2 \rangle &=& \frac{c}{(\lambda+d)^2} \langle J^2 \rangle \langle R^2\rangle_q  \nonumber\\
      &&   +  \frac{\langle \left(K^{\rm out}\right)^2\rangle -c}{ (\lambda+d)^2}  \langle J \rangle^2 \langle R\rangle^2_q,  \\ 
    \langle |R|^2 \rangle &=& \frac{c}{|\lambda+d|^2}\langle |J|^2  \rangle \langle |R|^2\rangle_q  \nonumber\\  && +  \frac{\langle \left(K^{\rm out}\right)^2\rangle -c}{ |\lambda+d|^2}  |\langle J \rangle|^2 |\langle R\rangle_q |^2.\label{eq:bx}
 \end{eqnarray}  
 
  The Eqs.~(\ref{eq:o}-\ref{eq:bx})  admit three kind of solutions.    The first type of  solution is obtained when  $\langle R \rangle_q\neq 0$.  We denote this
  solution by $\lambda = \lambda_{\rm isol}$ and $R = R_{\rm isol}$ since it identifies the outliers of the random matrix ensemble.   In this case,  Eq.~(\ref{eq:o}) implies that     
\begin{eqnarray}
 \frac{\langle K^{\rm in}K^{\rm out}\rangle}{c(\lambda_{\rm isol}+d)} \langle J \rangle = 1,
\end{eqnarray} 
which gives the result Eq.~(\ref{eq:outlier}) for the outlier eigenvalue. Since $\lambda_{\rm isol}\in\mathbb{R}$, it holds that $R_{\rm isol}\in \mathbb{R}$.
Consequently, we obtain Eq.~(\ref{eq:R}) for $\langle R_{\rm isol} \rangle$ by solving Eqs.~(\ref{eq:o}-\ref{eq:bx}) at $\lambda = \lambda_{\rm isol}$.     

The second type of solution is obtained when  $\langle R \rangle_q =0$ and $\lambda\notin \mathbb{R}$.   We denote this solution as  $\lambda = \lambda_{\rm b}$ and $R = R_{\rm b}$.       Solving    Eq.~(\ref{eq:b}), we obtain the relation 
\begin{eqnarray}
\frac{\langle K^{\rm in}K^{\rm out}\rangle}{c|\lambda_{\rm b}+d|^2}\langle |J|^2  \rangle = 1, 
\end{eqnarray} 
which leads to Eq.~(\ref{eq:boundary}), if we use the degree correlation coefficient $\rho$ as defined in~(\ref{eq:assort}). In this case, $R_b$ is a complex random variable and its first two moments are zero.

The third type of solution is obtained when $\langle R \rangle_q = 0$ and  $\lambda\in \mathbb{R}$, and we denote this solution  also
by $\lambda = \lambda_{\rm b}$ and $R = R_{\rm b}$.    Solving  Eq.~(\ref{eq:o}), we obtain 
\begin{eqnarray}
\frac{\langle K^{\rm in}K^{\rm out}\rangle}{c(\lambda_{\rm b}+d)^2}\langle |J|^2  \rangle = 1 .
\end{eqnarray}  
For this solution, we have that $\langle R_{\rm b} \rangle = 0$, but the value of  $\langle R^2_{\rm b} \rangle \neq 0$ depends on   the normalization of $R_{\rm b}$.

 \section{Fraction of zero-valued entries in right eigenvectors of directed random graphs} \label{Sec:ZeroValuedR}
 We analyze how the topology of  random and directed graphs, in the sense of connected components as discussed in Sec.~\ref{Sec:ConComp} and illustrated in Fig.~\ref{fig:bowtie}, affects the   distribution $p_R(r)$.   
 In particular, we show that for the right eigenvectors of directed random graphs it holds that 
\begin{eqnarray}
p_{R} = (1-s_{\rm in})\delta(r) + s_{\rm in}\:\tilde{p}_{R}(r), \label{eq:pRDelta}
\end{eqnarray}
where $s_{\rm in}$ is the relative size of the IN component.    

Using the 
ansatz 
\begin{eqnarray}
q_R(r) &=& b \delta(r) + (1-b)\tilde{q}_{R}(r). \label{eq:ansatz}
\end{eqnarray}
in Eq.~(\ref{eq:qRecR}), we obtain that $b$ solves Eq.~(\ref{eq:bb}) and 
  \begin{eqnarray}
 \lefteqn{\tilde{q}_R(r) = \sum^{\infty}_{k=0} \sum^{\infty}_{\ell=0}p_{K^{\rm in}, K^{\rm out}}(k,\ell)     \frac{k}{c} }&& 
 \nonumber\\ 
  &&    \quad\quad\quad\times \sum^{\ell}_{m=1} b^{\ell-m} \left(\begin{array}{c} \ell\\ m \end{array}\right) \int \prod^{m}_{j=1}{\rm d}^2 r_j  \tilde{q}_R(r_j) 
 \nonumber\\ 
&& \quad\quad\quad\times \int \prod^{m}_{j=1}{\rm d} x_j  p_J(x_j) \delta \left[r - \frac{\sum^{m}_{j=1}x_j r_j }{\lambda+d}\right].\label{eq:TqRec} \nonumber\\
 \end{eqnarray}  
 Furthermore, using   (\ref{eq:ansatz}) in  (\ref{eq:pRecR}), we obtain Eq.~(\ref{eq:pRDelta}).

Note that analogously, for the distribution $p_L$  of entries of left eigenvectors, it holds that 
 \begin{eqnarray}
 p_L(l) = (1-  s_{\rm out}) \delta(l) +  s_{\rm out} \tilde{p}_{L}(l), 
 \end{eqnarray}
  with   $s_{\rm out}$ the relative size of the OUT component  of the underlying graph.

\section{ Mathematical derivations for  random matrices with diagonal disorder  and nondirected graphs with random couplings }\label{sec:der2}       
We derive recursions relations for the distribution $p_R$ in the case of random matrices with diagonal disorder [the model defined in Eq.~(\ref{eq:modelxx})] and for random matrices defined on nondirected random graphs with random couplings [the model defined in Eq.~(\ref{eq:modelxxxxx})].     For the first model, we  obtain also compact expressions for the values of  $\lambda$ for which the   recursion relations admit a normalizable solution for $p_R$.  

\subsection{Random matrices with diagonal disorder} 
First, we derive the recursion Eqs.~(\ref{eq:pRecRD}) and (\ref{eq:qRecRD}) for the random matrix model Eq.~(\ref{eq:modelxx}) with diagonal elements $D_j$ drawn from a distribution $p_D$.   Using Eqs.~(\ref{eq:recursionfinal1Bxxx}) and (\ref{eq:recursionfinal2Bxxx}) for the eigenvector elements $R_j$ and $R^{(j)}_k$, and the fact that for the locally tree-like random matrices defined in Eq.~(\ref{eq:modelxx}) all  random variables on the right-hand-side of  Eqs.~(\ref{eq:recursionfinal1Bxxx}) and (\ref{eq:recursionfinal2Bxxx}) are independent, we 
readily obtain the recursion Eqs.~(\ref{eq:pRecRD}) and (\ref{eq:qRecRD}), with $p_R$ and $q_R$ as defined in Eqs.~(\ref{eq:uniform}) and (\ref{eq:out}).

Second, we determine the values of $\lambda$ for which the recursion  Eqs.~(\ref{eq:pRecRD}) and (\ref{eq:qRecRD}) admit normalizable solutions, which provide us with the deterministic outlier eigenvalues $\lambda_{\rm isol}$  and the eigenvalues $\lambda_{\rm b}$ at the boundary of the continuous part of the spectrum.  
 To this aim, we use the Eqs.~(\ref{eq:pRecRD})  to derive the set of self-consistent equations 
  \begin{eqnarray}
 \langle R \rangle_q &=& \frac{\langle K^{\rm in}K^{\rm out}\rangle}{c} \Big\langle \frac{1}{\lambda-D}\Big\rangle \langle J \rangle \langle R\rangle_q ,\label{eq:oD}\\ 
       \langle R^2 \rangle_q &=& \frac{\langle K^{\rm in}K^{\rm out}\rangle}{c} \Big\langle \frac{1}{(\lambda-D)^2}\Big\rangle \langle J^2 \rangle \langle R^2\rangle_q  \nonumber\\
      &&   +   \Big\langle\frac{\langle K^{\rm in}K^{\rm out}(K^{\rm out}-1)\rangle}{c(\lambda-D)^2}  \Big\rangle  \langle J \rangle^2 \langle R\rangle^2_q,\nonumber \\ 
    \langle |R|^2 \rangle_q &=& \frac{\langle K^{\rm in}K^{\rm out}\rangle}{c}\Big\langle \frac{1}{|\lambda-D|^2}   \Big\rangle  \langle |J|^2  \rangle \langle |R|^2\rangle_q  \nonumber\\  && + \Big\langle\frac{ \langle K^{\rm in}K^{\rm out}(K^{\rm out}-1)\rangle}{c (\lambda-D)^2}\Big\rangle  |\langle J \rangle|^2 |\langle R\rangle|^2_q, \label{eq:bD}
 \end{eqnarray}     
 in the lower order moments of $q_R$.    Solving Eq.~(\ref{eq:oD}) for $ \langle R\rangle_q\neq 0$, we obtain   Eq.~(\ref{eq:IsolD}) for the eigenvalue outliers $\lambda=\lambda_{\rm isol}$ of the random matrix ensemble.   On the other hand, setting $ \langle R\rangle_q = 0$  in Eq.~(\ref{eq:bD}), ew obtain Eq.~(\ref{eq:lambdabD}) for the eigenvalues $\lambda=\lambda_{\rm b}$ located at the boundary $\partial \sigma_{\rm ac}$ of the continuous part  $\sigma_{\rm ac}$ of the spectrum.    
 
The  moments of the  distribution $p_R$  of right eigenvector entries  associated with either $\lambda = \lambda_{\rm isol}$ or $\lambda = \lambda_{\rm b}$ solve the self-consistent equations
   \begin{eqnarray}
 \langle R \rangle &=& \Big\langle \frac{c}{\lambda+D} \Big\rangle \langle J \rangle \langle R\rangle_q, \label{eq:oxD}\\ 
       \langle R^2 \rangle &=& \Big\langle \frac{c}{(\lambda+D)^2}\Big\rangle \langle J^2 \rangle \langle R^2\rangle_q  \nonumber\\
      &&   +  \left[\langle \left(K^{\rm out}\right)^2\rangle -c\right]\Big\langle \frac{1}{ (\lambda+D)^2} \Big\rangle  \langle J \rangle^2 \langle R\rangle^2_q,  \\ 
    \langle |R|^2 \rangle &=&  \Big\langle \frac{c}{|\lambda+D|^2} \Big\rangle \langle |J|^2  \rangle \langle |R|^2\rangle_q  \nonumber\\  && +  \left[\langle \left(K^{\rm out}\right)^2\rangle -c\right]\Big\langle  \frac{1}{ |\lambda+D|^2} \Big\rangle   |\langle J \rangle|^2 |\langle R\rangle_q |^2. \nonumber\\\label{eq:bxD}
 \end{eqnarray}  
 
Note that the Eqs.~(\ref{eq:oxD}-\ref{eq:bxD}) generalize the Eqs.~(\ref{eq:ox}-\ref{eq:bx}) for the case of constant $D = d$.    

\subsection{Undirected graphs with random couplings}
We derive the recursion Eqs.~(\ref{eq:distriO2}) and (\ref{eq:distriO1}) for the random matrix model Eq.~(\ref{eq:modelxxxxx}).   Random matrices in this model are locally tree-like, and therefore, we can use the Eqs.~(\ref{eq:recursionfinal1B}), (\ref{eq:recursionfinal1xxA}), (\ref{eq:gjjB2}) and (\ref{eq:gjjzB2}) derived in Appendix~\ref{sec:locTreeRanA}.   In order to obtain a set of recursive distribution equations,  we define  the joint distributions
 \begin{eqnarray}
 p_{R,G}(r, g|\bA) = \frac{1}{n}\sum^n_{j=1} \delta(r-R_j) \delta(g-G_j)\label{eq:uniformO}
\end{eqnarray}
and 
\begin{eqnarray}
q_{R,G}(r,g|\bA) = \frac{1}{c\:n}\sum_{k=1}^n \sum_{j \in \partial_{k}} \delta(r-R_k^{(j)} )\delta(g-G^{(j)}_k). \nonumber\\ \label{eq:outP} 
\end{eqnarray} 
where $\partial_k$ is the neighborhood of node $k$, as defined in Eq.~(\ref{eq:neighbour}).    Using that $\mathbf{A}$ is locally tree-like, and therefore the random variables on the right-hand-side of Eqs.~(\ref{eq:recursionfinal1B}), (\ref{eq:recursionfinal1xxA}), (\ref{eq:gjjB2}) and (\ref{eq:gjjzB2}) are independent, we readily obtain the recursive distributional Eqs.~(\ref{eq:distriO1}) and  (\ref{eq:distriO2}).

\section*{acknowledgements}
We thank C.~Cammarota, Y.~Fyodorov, A.~Mambuca, W.~M.~Tarnowski and P.~Vivo  for fruitful discussions. 
FLM thanks London Mathematical Laboratory and CNPq/Brazil for financial support.

\section*{References}
\bibliographystyle{ieeetr} 
\bibliography{bibliography}

\end{document}